\documentclass[12pt,a4paper,reqno,noamsfonts]{amsart}
\usepackage{amsmath}
\usepackage{array}
\IfFileExists{bbm.sty}{\usepackage{bbm,eufrak}}{\typeout{ * ^^J
   * Please install the bbm bundle! I try to replace the bbm fonts by the ^^J
   * AMS blackboard font and the cm bold font ^^J *}
   \usepackage{amsfonts}\newcommand{\mathbbm}[1]{\mathbb{##1}}}
\IfFileExists{cite.sty}{\usepackage{cite}}{}

\usepackage[dvips]{graphics}

\makeatletter
\def\preprint#1{\def\@preprint{#1}} \def\@preprint{}
\def\@setaddresses{\par
  \nobreak \begingroup
\footnotesize
  \def\author##1{\nobreak\addvspace\bigskipamount}%
  \def\\{\unskip, \ignorespaces}%
  \interlinepenalty\@M
  \def\address##1##2{\begingroup \centering \renewcommand{\arraystretch}{1}
    \par\addvspace\bigskipamount\indent
    \@ifnotempty{##1}{(\ignorespaces##1\unskip) }%
    {\scshape\ignorespaces \begin{tabular}[t]{c}##2%
			   \end{tabular}}\par\endgroup}%
  \def\curraddr##1##2{\begingroup
    \@ifnotempty{##2}{\nobreak\indent{\itshape Current address}%
      \@ifnotempty{##1}{, \ignorespaces##1\unskip}\/:\space
      ##2\par\endgroup}}%
  \def\email##1##2{\begingroup\centering
    \@ifnotempty{##2}{\nobreak\indent{\itshape E-mail}%
      \@ifnotempty{##1}{, \ignorespaces##1\unskip}\/:\space
      \ttfamily##2\par\endgroup}}%
  \addresses
  \endgroup
}
\def\@maketitle{%
  \normalfont\normalsize
  \let\@makefnmark\relax  \let\@thefnmark\relax
  \ifx\@empty\@date\else \@footnotetext{\@setdate}\fi
  \ifx\@empty\@subjclass\else \@footnotetext{\@setsubjclass}\fi
  \ifx\@empty\@keywords\else \@footnotetext{\@setkeywords}\fi
  \ifx\@empty\thankses\else \@footnotetext{%
    \def\par{\let\par\@par}\@setthanks}\fi
  \@mkboth{\@nx\shortauthors}{\@nx\shorttitle}%
  {}\hfill \raisebox{1cm}{\if \@preprint \empty \else 
			{\renewcommand{\arraystretch}{1} 
			 \begin{tabular}{c}\@preprint\end{tabular}} \fi}
  \global\topskip42\p@\relax % 5.5pc   "   "   "     "     "
  \@settitle
  \ifx\@empty\authors \else \@setauthors \fi
  \ifx\@empty\addresses \else \@setaddresses \fi
  \ifx\@empty\@dedicatory
  \else
    \baselineskip18\p@
    \vtop{\centering{\footnotesize\itshape\@dedicatory\@@par}%
      \global\dimen@i\prevdepth}\prevdepth\dimen@i
  \fi
  \@setabstract
  \normalsize
  \if@titlepage
    \newpage
  \else
    \dimen@34\p@ \advance\dimen@-\baselineskip
    \vskip\dimen@\relax
  \fi
} % end \@maketitle
\let\enddoc@text\empty
\makeatother

\makeatletter
\def\eqnarr@left{\@centering}
\let\eqnarr@opts\relax
\def\equationarray{%
 \col@sep\arraycolsep
 \def\d@llarbegin{$\displaystyle}
 \def\d@llarend{$}%
 \stepcounter{equation}%
 \let\@currentlabel=\theequation
 \set@eqnsw \global\@eqcnt\z@ \global\@eqargcnt\z@
 \let\@classz\@eqnclassz
 \def\multicolumn##1##2##3{\@eqnmulticolumn{##1}{##2}{##3}%
                           \global\advance\@eqcnt##1
                           \global\advance\@eqcnt\m@ne}%
 \def\@halignto{to\displaywidth}%
 \@ifnextchar[{\@equationarray}{\@equationarray[.]}}
\let\@eqnmulticolumn=\multicolumn
\def\yesnumber{\global\@eqnswtrue}
\let\set@eqnsw=\yesnumber
\def\@amper{&}
\newcount\@eqargcnt  % counts number of columns
\def\@equationarray[#1]#2{%
     \eqnarr@opts
     \@tempdima \ht \strutbox
     \advance \@tempdima by\extrarowheight
     \setbox\@arstrutbox=\hbox{\vrule
               \@height\arraystretch \@tempdima
               \@depth\arraystretch \dp \strutbox
               \@width\z@}%
     \gdef\advance@eqargcnt{\global\advance\@eqargcnt\@ne}%
     \begingroup
     \@mkpream{#2}%
     \xdef\@preamble{%
      \if #1l\tabskip\z@ \else\if #1r\tabskip\@centering
                         \else\if #1c\tabskip\@centering
                         \else\tabskip\eqnarr@left \fi\fi\fi
      \halign \@halignto
      \bgroup \tabskip\z@ \@arstrut \@preamble
      \if #1l\tabskip\@centering \else\if #1r\tabskip\z@
                                 \else\tabskip\@centering \fi\fi
      \@amper\llap{\@sharp}\tabskip\z@\cr}%
     \endgroup
     \gdef\advance@eqargcnt{}%
     \bgroup
     \let\@sharp## \let\protect\relax
     \m@th   \let\\=\@equationcr
     \let\par\@empty
     $$                            % $$ BRACE MATCHING HACK
     \lineskip \z@
     \baselineskip \z@
     \@preamble}
\def\@eqnclassz{\@classx
   \@tempcnta \count@
   \advance@eqargcnt
   \prepnext@tok
   \@addtopreamble{%
      \global\advance\@eqcnt\@ne
      \ifcase \@chnum
      \hfil \d@llarbegin \insert@column \d@llarend\hfil \or
      \d@llarbegin \insert@column \d@llarend \hfil \or
      \hfil\kern\z@ \d@llarbegin \insert@column \d@llarend \or
      $\vcenter
      \@startpbox{\@nextchar}\insert@column \@endpbox $\or
      \vtop \@startpbox{\@nextchar}\insert@column \@endpbox \or
      \vbox \@startpbox{\@nextchar}\insert@column \@endpbox
      \fi}\prepnext@tok}
\def\endequationarray{\@zequationcr
   \egroup
   \global\advance\c@equation\m@ne $$  % $$ BRACE MATCHING HACK
   \egroup\global\@ignoretrue
   \gdef\@preamble{}}
\def\@equationcr{${\ifnum0=`}\fi\@ifstar{\global\@eqpen\@M
    \@xequationcr}{\global\@eqpen\interdisplaylinepenalty
                   \@xequationcr}}
\def\@xequationcr{%
    \@ifnextchar[{\@argequationcr}{\ifnum0=`{\fi}${}%
    \@zequationcr}}
\def\@argequationcr[#1]{\ifnum0=`{\fi}${}\ifdim #1>\z@
   \@xargequationcr{#1}\else
   \@yargequationcr{#1}\fi}
\def\@xargequationcr#1{\unskip
   \@tempdima #1\advance\@tempdima \dp \@arstrutbox
   \vrule \@depth\@tempdima \@width\z@
   \@zequationcr\noalign{\penalty\@eqpen}}
\def\@yargequationcr#1{%
   \@zequationcr\noalign{\penalty\@eqpen\vskip #1}}
\def\@zequationcr{\@whilenum\@eqcnt <\@eqargcnt
   \do{\@amper\omit\global\advance\@eqcnt\@ne}%
   \@amper
   \if@eqnsw\@eqnnum\stepcounter{equation}\fi
   \set@eqnsw\global\@eqcnt\z@\cr}
\@namedef{equationarray*}{%
   \let\set@eqnsw=\nonumber \equationarray}
\@namedef{endequationarray*}{\endequationarray}
\makeatother

\DeclareMathOperator{\id}{id}
\DeclareMathOperator{\End}{End}
\DeclareMathOperator{\Hom}{Hom}
\DeclareMathOperator{\tr}{tr}

\DeclareMathOperator{\diag}{diag}

\def\ot{\otimes}
\def\op{\oplus}
\def\t{\tilde}
\def\h{\hat}
\def\iu{\mathrm{i}}
 
\def\bsj{\boldsymbol{\psi}}
\def\bsu{\boldsymbol{u}}
\def\bsd{\boldsymbol{d}}

\def\bfJ{\boldsymbol{\Psi}}       \def\bftJ{\tilde{\boldsymbol{\Psi}}}       
\def\bfF{\boldsymbol{\Phi}}	  \def\bftF{\tilde{\boldsymbol{\Phi}}}	     
\def\bY{\boldsymbol{\Upsilon}}	  \def\bftY{\tilde{\boldsymbol{\Upsilon}}}   
\def\bX{\boldsymbol{\Xi}}	  \def\bftX{\tilde{\boldsymbol{\Xi}}}	
\def\itJ{\varPsi}		  \def\ittY{\tilde{\varUpsilon}}  
\def\itF{\varPhi}                 \def\ittJ{\tilde{\varPsi}}      
\def\itY{\varUpsilon}		  \def\ittF{\tilde{\varPhi}}	  
\def\itX{\varXi}		  \def\ittX{\tilde{\varXi}}

\def\bfW{\mathbf{W}}		  \def\bfG{\mathbf{G}}		    
\def\bfX{\mathbf{X}}		  	
\def\bfn{\mathbf{n}}		  \def\bfd{\mathbf{d}}	
\def\bfm{\mathbf{m}}		  \def\sfD{\mathsf{D}}	
\def\vre{\varepsilon} 

		  \def\M{\mathcal{M}}  
\def\C{\mathbbm{C}}		  \def\R{\mathbbm{R}}  
		  \def\CX{{C^\infty (X)}}
\def\Mu{M_{\t{u}}}		  \def\Mn{M_{\t{n}}}   
\def\Mud{M_{ud}}		  \def\Men{M_{en}}

\def\hs{\hspace}		  \def\vs{\vspace}
\def\ds{\displaystyle}		  \def\tsum{\textstyle \sum}	
\def\th{\tfrac{1}{2}}		   

\def\SU#1{\mathrm{SU(#1)}}	  \def\su#1{\mathrm{su(#1)}} 
\def\U#1{\mathrm{U(#1)}}	  \def\u#1{\mathrm{u(#1)}}   
\def\ul#1{\underline{#1}}	  \def\ol#1{\,\overline{\!#1\!}\,}

\def\yn{\yesnumber}
\def\npb{\nopagebreak}
\newskip\eqnskipamount \eqnskipamount=0ex plus 0.3ex minus 0.1ex
\def\eqnskip{\vspace\eqnskipamount}
\newcommand\eq[2][]{\begin{equation} \label{#1} #2 \end{equation}}
\newcommand\eqa[2]{\begin{equationarray*}{#1} #2 \end{equationarray*}}
\newcommand\eqas[3][]{\begin{equation}\label{#1} \begin{array}{#2} #3 
		      \end{array} \end{equation}}
\newcommand\al[1]{\begin{align} #1 \end{align}}
\newcommand\seq[1]{\begin{subequations} #1 \end{subequations}}
\newcommand{\mb}[3][\!\!\!]{\left( #1 \begin{array}{#2} #3 
			    \end{array} #1 \right)} 
\newcommand{\mpa}[3][\!]{\left[ #1 \begin{array}{#2} #3 
			    \end{array} #1 \right]} 
\def\mc{\multicolumn}
\newcolumntype{D}[1]{>{$\hfil}m{#1mm}<{$\hfil}>{$\hfil}%
			      m{#1mm}<{$\hfil}>{$\hfil}m{#1mm}<{$\hfil}}
\newcolumntype{Z}[1]{>{$\hfil}m{#1mm}<{$\hfil}>{$\hfil}m{#1mm}<{$\hfil}}
\newcolumntype{E}[1]{>{$\hfil}m{#1mm}<{$\hfil}}

\newcommand{\W}[2][g]{\Omega^{#2}{\mathfrak #1}}
\newcommand{\p}[2][a]{\hat{\pi}(\Omega^{#2}{\mathfrak #1})}

\newcommand{\cJ}[2][g]{\mathbbm{J}^{#2}\!\mathfrak{#1}}
\newcommand{\jj}[2][a]{\hat{\pi}(\mathcal{J}^{#2} \mathfrak{#1})}
\newcommand{\pf}[1][a]{\hat{\pi}(\mathfrak{#1})}
\IfFileExists{bbm.sty}%
     {\newcommand{\cj}[2][a]{\mathbbm{j}^{##2}\!\mathfrak{##1}}}%
     {\newcommand{\cj}[2][a]{ \mathbf{j}^{##2}\!\mathfrak{##1}}}
\IfFileExists{bbm.sty}%
     {\newcommand{\cc}[2][a]{\mathbbm{c}^{##2}\!\mathfrak{##1}}}%
     {\newcommand{\cc}[2][a]{ \mathbf{c}^{##2}\!\mathfrak{##1}}}
\IfFileExists{bbm.sty}%
     {\newcommand{\bbr}[2][a]{\mathbbm{r}^{##2}\!\mathfrak{##1}}}%
     {\newcommand{\bbr}[2][a]{ \mathbf{r}^{##2}\!\mathfrak{##1}}}
\IfFileExists{bbm.sty}%
     {\def\one{\mathbbm{1}}}%
     {\def\one{\mathrm{I}}}

\newcommand{\la}[1][]	 {\mu^{\mathrm{#1}}}
\newcommand{\lat}[1][]	 {\tilde{\mu}^{\mathrm{#1}}}
\newcommand{\lac}[1][]	 {\check{\mu}^{\mathrm{#1}}}
\newcommand{\lah}[1][]   {\hat{\mu}^{\mathrm{#1}}}
\newcommand{\ru}[2]{\vrule width0em height#1ex depth#2ex} 
\newcommand{\prt}[1][]{\partial_{#1}} 
\newcommand{\Lgr}[1][]{\mathcal{L}_{#1}}
\newcommand{\gh}[1][5]{{\hat{\gamma}}^{#1}}
\newcommand{\f}[1][g]{\mathfrak{#1}}
\newcommand{\mat}[2][\C]{{\mathrm{M}}_{#2}{#1}}
\newcommand{\rf}[1][]{\textup{\eqref{#1}}}
\newcommand{\g}[1][5]{\gamma^{#1}}
\newcommand{\hsg}[1][\mathfrak{g}]{\h{\sigma}_{#1}} 

\def\cbftJ{\check{\tilde{\boldsymbol{\Psi}}}}
\def\cbftF{\check{\tilde{\boldsymbol{\Phi}}}}
\def\hbftY{\hat  {\tilde{\boldsymbol{\Upsilon}}}}
\def\hbftF{\hat  {\tilde{\boldsymbol{\Phi}}}}
\def\cittF{\check{\tilde{\varPhi}}}
\def\hittF{\hat  {\tilde{\varPhi}}}

\renewcommand{\arraystretch}{1.2}
\numberwithin{equation}{section}

\begin{document}

\title {   Grand Unification in Non--Associative Geometry}
\author{   Raimar Wulkenhaar} 
\address{  Institut f\"ur Theoretische Physik, Universit\"at Leipzig \\ 
	   Augustusplatz 10/11, D--04109 Leipzig, Germany} 
\email{    raimar.wulkenhaar@itp.uni-leipzig.de} 
\date{	   February 10, 1997}
\preprint{ \texttt{hep-th/9607237}\\ revised version}

\begin{abstract}
We formulate the flipped $\mathrm{SU(5)} \times \mathrm{U(1)}$--GUT within the 
framework of non--associative geometry. It suffices to take the matrix Lie 
algebra $\mathrm{su(5)}$ as the input; the $\mathrm{u(1)}$--part with its 
representation on the fermions is an algebraic consequence. The occurring 
Higgs multiplets ($\underline{24},\underline{5},\underline{45},
\underline{50}$--representations of $\mathrm{su(5)}$) are uniquely determined 
by the fermionic mass matrix and the spontaneous symmetry breaking pattern to 
$\mathrm{SU(3)}_C \times \mathrm{U(1)}_{EM}$. We find the most general gauge 
invariant Higgs potential that is compatible with the given Higgs vacuum. Our 
formalism yields tree--level predictions for the masses of all 
gauge and Higgs bosons. It turns out that the low--energy sector is identical 
with the standard model. In particular, there exists precisely one light 
Higgs field, whose upper bound for the mass is $1.45\,m_t$. All remaining 
$207$ Higgs fields are extremely heavy.
\begin{center}\hs*{-2em}\renewcommand{\arraystretch}{1}\begin{tabular}{ll}
PACS:	  & 02.40.--k; 12.10.Kt; 12.60.--i; 14.80.Cp	\\
keywords: & non--associative geometry; grand unification; 
	    masses of Higgs bosons \hs*{-3em}
\end{tabular}\end{center}

\end{abstract}

\maketitle 

\section{Introduction}

One of the most important applications of non--commutative geometry (NCG) 
to physics is a unified description of the standard model. The most elegant 
version rests upon a K--cycle \cite{cl, ac} with real structure \cite{acr}, see 
\cite{iks, mgv} for details of the construction. The standard model is the only 
realistic physical model that one can formulate within the most elegant 
NCG--prescription \cite{lmms}. On the other hand, there exist good reasons 
\cite{l} why one could be interested in Grand Unified Theories (GUT's): 
GUT's explain the quantization of electric charge, yield a 
fairly well prediction for the Weinberg angle, explain the convergence of 
running coupling constants at high energies, include massive neutrinos to 
solve the solar neutrino problem, produce the observed baryon asymmetry of the 
universe, etc. However, the results of \cite{lmms} imply that one needs 
additional structures or different methods for a NCG--formulation of these 
models.

The perhaps most successful NCG--approach towards grand unification was 
proposed by Chamseddine, Felder and Fr\"ohlich. In the $\SU5$--model 
\cite{cff1, cff2}, the authors start to construct an auxiliary K--cycle. Within 
this framework they construct the bosonic sector. Then they interpret some of 
these bosonic quantities as Lie algebra valued and consider Lie algebra 
representations on the physical Hilbert space to obtain the fermionic sector. 
An aesthetic shortcoming of that approach is the auxiliary character of the 
K--cycle, which of course is inevitable in view of \cite{lmms}. The 
$\mathrm{SO(10)}$--model \cite{cf} by Chamseddine and Fr\"ohlich fits
well\footnote{Nevertheless, the use of Lie algebras instead of algebras 
could probably justify certain assumptions made in \cite{cf}.} into the 
NCG--scheme. The reason why this model was excluded in \cite{lmms} is that 
only models possessing complex fundamental irreducible representations were 
admitted in that article. 

The author of this paper has proposed in \cite{rw2} a modification of 
non--commutative geometry. In that approach one uses skew--adjoint Lie algebras 
instead of unital associative $*$--algebras. Lie algebras are non--associative 
algebras~-- this is the motivation for the working title ``non--associative 
geometry''. The advantage of non--associative geometry is that a larger class 
of physical models can be constructed from the same amount of structures as in 
the most elegant NCG--formulation. That class includes the standard model 
\cite{rw3} and the flipped $\SU5 \times \U1$--GUT as well, as we show in this 
paper. The $\SU5$--model can be obtained as a special case. For the classical 
treatment of the flipped $\SU5 \times \U1$--model see \cite{dkn}.

We give in Section~\ref{nag} a recipe how to construct classical gauge field 
theories within non--associative geometry. The arguments why this recipe 
works can be found in \cite{rw2}. In Section~\ref{tum} we construct the matrix 
part of the $\SU5 \times \U1$--model: In Section~\ref{rema} we consider 
relevant $\su5$--representations. The remaining ingredients of non--associative 
geometry are defined in Section~\ref{gdo}. Then it is not difficult to derive 
in Section~\ref{spp} the matrix part of the connection form. Finally, we 
perform in Section~\ref{fact} the factorization of the curvature with respect 
to a canonically given ideal, which we construct before in Section~\ref{tide}. 

In Section~\ref{aum} we include the space--time part and derive the action for 
our model: Out of the curvature obtained in Section~\ref{cum} we build in 
Section~\ref{baum} the bosonic action. To compare it with usual formulae of 
gauge field theory we write down this action in terms of local coordinates, 
see Section~\ref{bllc}. The fermionic action 
is derived in Section~\ref{fa}. Comparing it with phenomenology we can identify 
certain parameters of the generalized Dirac operator with fermion masses 
and Kobayashi--Maskawa mixing angles. 

This information plays an essential r\^ole in deriving the masses of the 
Higgs bosons in Section~\ref{thp}. Finally, we comment on a formal derivation 
of the $\SU5$--GUT in Section~\ref{su5m}. 

\section{The Recipe of Non--associative Geometry}
\label{nag}

The basic object in non--associative geometry is an L--cycle 
$(\f[g],h,D,\pi,\Gamma)\,,$ which consists of a $*$--representation $\pi$ of a 
skew--adjoint Lie algebra $\f[g]$ as bounded operators on a Hilbert space 
$h\,,$ together with a selfadjoint operator $D$ on $h$ with compact resolvent 
and a selfadjoint operator $\Gamma$ on $h\,,$ $\Gamma^2=\id_h\,,$ which 
commutes with $\pi(\f[g])$ and anticommutes with $D\,.$ The operator $D$ may 
be unbounded, but such that $[D,\pi(\f[g])]$ is bounded. L--cycles are 
naturally related to physical models on a space--time manifold $X$ if the 
following input data are given:
\begin{enumerate}
\vs{-\topsep} 
\item
A unitary matrix Lie group $G$ and its associated gauge group $\mathcal{G} 
=\CX \ot G\,.$ Here, $\CX$ denotes the algebra of real--valued smooth functions 
on $X$.  \label{sybr1}
\vs{-\itemsep} \vs{-\parsep}

\item 
Chiral fermions $\bsj$ transforming under a representation $\t{\pi}_0$ 
of $G\,.$ The induced representation of the gauge group $\mathcal{G}$ is 
$\t{\pi}=\id \ot \t{\pi}_0\,.$ 
\vs{-\itemsep} \vs{-\parsep}

\item
The fermionic mass matrix $\widetilde{\M}\,,$ i.e.\ fermion masses plus 
generalized Kobayashi--Maskawa matrices. 
\label{sybr3}
\vs{-\itemsep} \vs{-\parsep}

\item
Possibly the spontaneous symmetry breaking pattern of $G\,.$  
\label{sybr4}
\end{enumerate}
\vs{-\topsep}
For technical reasons we pass to a compact Euclidian spin manifold $X$. 
We take $\f[g]=\CX \ot \f[a]$ as the Lie algebra of $\mathcal{G}\,.$ Here, 
$\f[a] = \f[a]' \op \f[a]''$ is a skew--adjoint matrix Lie algebra, where 
$\f[a]'$ is semisimple and $\f[a]''$ Abelian. We shall only consider the case 
that the Abelian part is not present, i.e.\ $\f[a]=\f[a]'\,.$ We choose 
$h=L^2(X,S) \ot \C^F$ as the space where the Euclidian fermions $\bsj$ live. 
Here, $L^2(X,S)$ is the Hilbert space of square integrable bispinors. We take
$\pi=1 \ot \h{\pi}$ as the differential $\t{\pi}_*\,,$ where $\h{\pi}$ is a 
representation of $\f[a]$ in $\mat{F}\,.$ We define $D=\sfD \ot \one_F 
+ \g \ot \M\,,$ where $\sfD$ is the Dirac operator associated to the spin 
connection and $\M \in \mat{F}\,.$ Here, $\g \ot \M$ has to coincides with 
$\widetilde{\M}$ on chiral fermions. The chirality properties of the fermions 
are encoded in $\Gamma=\g \ot \h{\Gamma},$ where $\{\h{\Gamma},\M\}=0$ and 
$[\h{\Gamma},\pf]=0\,.$ 

The recipe towards the (classical) gauge field theory associated to the 
L--cycle is the following: Let $\W[a]{1}$ be the space of formal commutators
\eq[udla]{
\omega^1=\tsum_{\alpha,z \geq 0} [a^z_\alpha,[\dots [a^1_\alpha,d a^0_\alpha] 
\dots ]]~, \quad a^i_\alpha \in \f[a]~.
   }
Apply linear mappings $\h{\pi}: \W[a]{1} \to \mat{F}$ and $\h{\sigma}: 
\W[a]{1} \to \mat{F}$ defined by
\seq{
\al{
\h{\pi}(\omega^1) &:= \tsum_{\alpha,z \geq 0} [\h{\pi}(a^z_\alpha),[\dots 
[\h{\pi}(a^1_\alpha), [-\iu \M, \h{\pi}(a^0_\alpha)]] \dots ]] ~, 
\label{hpi} \\
\h{\sigma}(\omega^1) &:= \tsum_{\alpha,z \geq 0} [\h{\pi}(a^z_\alpha),[\dots 
[\h{\pi}(a^1_\alpha), [\M^2, \h{\pi}(a^0_\alpha)]] \dots ]] ~.
\label{hsig}
    }}
Define $\W[a]{n} \ni \omega^n=\tsum_\alpha [\omega^1_{n,\alpha}, 
[ \omega^1_{n-1,\alpha}, \dots [\omega^1_{2,\alpha}, \omega^1_{1,\alpha}] 
\dots ]]\,,$ where $\omega^1_{i,\alpha} \in \W[a]{1}\,.$ 
Extend $\h{\pi}$ and $\h{\sigma}$ recursively to $\W[a]{n}$ by 
\eqa{rcl}{
\h{\pi}([\omega^1,\omega^k]) &:=& \h{\pi}(\omega^1) \h{\pi}(\omega^k) 
- (-1)^k \h{\pi}(\omega^k) \h{\pi}(\omega^1) ~,~~ \label{on} \yn \\
\h{\sigma}([\omega^1,\omega^k]) &:=& \h{\sigma}(\omega^1) \h{\pi}(\omega^k) 
- \h{\pi}(\omega^k) \h{\sigma}(\omega^1) 
- \h{\pi}(\omega^1) \h{\sigma}(\omega^k) - (-1)^k \h{\sigma}(\omega^k) 
\h{\pi}(\omega^1) ~. 
    }
Define for $n \geq 2$ 
\begin{equation}
\jj{n} :=\{~ \h{\sigma}(\omega^{n-1})~,~~ \omega^{n-1} \in \W[a]{n-1} \cap \ker 
\h{\pi}~\} ~. \label{js}
\end{equation}
Define spaces $\bbr{0} \subset \mat{F}$ and $\bbr{1} \subset \mat{F}$ 
elementwise by 
\begin{align}
\bbr{0} &= -(\bbr{0})^* = \h{\Gamma} (\bbr{0}) \h{\Gamma} \;, &
\bbr{1} &= -(\bbr{1})^* = -\h{\Gamma} (\bbr{1}) \h{\Gamma} \;, \notag \\{}
[\bbr{0} , \pf] &\subset \pf\;, & [\bbr{0} , \p{1}] &\subset \p{1} \;, 
\label{rlh}\\
\{\bbr{0} , \pf\} &\subset \{ \pf, \pf\} + \p{2}\;, &
\{\bbr{0} , \p{1}\} &\subset \{ \pf, \p{1}\} + \p{3}\;,  \notag\\{}
[\bbr{1} ,\pf] &\subset \p{1}\;, &
\{\bbr{1} ,\p{1}\} &\subset \p{2} + \{ \pf, \pf\} \;.  \notag
\end{align}
Define spaces $\cj{0},\cj{1},\cj{2} \subset \mat{F}$ elementwise by 
\eqa{rclrcl}{
\cj{0} &:=& \cc{0} ~, &  \cj{1} &:=& \cc{1} ~, \npb \\ 
\cj{2} &:=& \cc{2} + \jj{2} + \{\pf, \pf\} ~, & && \mbox{where} 
\label{cj} \yn \\
\cc{0} &=& -(\cc{0})^* = \h{\Gamma} (\cc{0}) \h{\Gamma}\;, &
\cc{1} &=& -(\cc{0})^* =-\h{\Gamma} (\cc{0}) \h{\Gamma}\;, \npb \\
\cc{2} &=& (\cc{0})^* = \h{\Gamma} (\cc{0}) \h{\Gamma}\;, \\{}
\cc{0} \cdot \pf &=& 0~, & \cc{0} \cdot \p{1} &=& 0 ~, \npb \\
\cc{1} \cdot \pf &=& 0~, & \cc{1} \cdot \p{1} &=& 0 ~, \npb \\{}
[\cc{2}, \pf] &=& 0~, & [\cc{2}, \p{1} ] &=& 0~.
    }
The connection form $\rho$ has the structure
\eqas[rho]{rcl}{
\rho &=& \tsum_\alpha (c^1_\alpha \ot m^0_\alpha + c^0_\alpha \g \ot 
m^1_\alpha) ~,~~ \\ 
&& c^1_\alpha \in \Lambda^1~,\quad c^0_\alpha \in \Lambda^0~, \quad 
m^0_\alpha \in \bbr{0}~, \quad m^1_\alpha \in \bbr{1}~,
	 }
where $\Lambda^k$ is the space of differential $k$--forms represented by gamma 
matrices. The curvature $\theta$ is computed from the connection form $\rho$ by 
\begin{equation}
\begin{split}
\label{th}
\theta &= \bfd \rho + \rho^2 - \iu \{ \g \ot \M, \rho\} 
+ \h{\sigma}_{\f[g]} (\rho) \g + \cJ{2}~, \\
\cJ{2} &= (\Lambda^2 \ot \cj{0}) \op (\Lambda^1 \g \ot \cj{1}) \op 
(\Lambda^0 \ot \cj{2})~, 
\end{split}
\end{equation}
where $\bfd$ is the exterior differential and $\h{\sigma}_{\f[g]}$ the 
extension of $\id \ot \h{\sigma}$ to elements of the form \rf[rho]. Select the 
representative $\f[e](\theta)$ 
orthogonal to $\cJ{2}\,,$ i.e.\ find $\mathrm{j} \in \cJ{2}$ such that
\eqas[eoft]{l}{
\f[e](\theta) = \bfd \rho + \rho^2 - \iu \{ \g \ot \M, \rho\} 
+ \h{\sigma}(\rho) \g + \mathrm{j}~,~~ \\
{\ds \int_X \!\! dx} \;\tr_c(\f[e](\theta) \,\mathrm{j}_2) = 0~, \quad
\forall \,\mathrm{j}_2 \in \cJ{2}~.
   }
The trace $\tr_c$ includes the trace in $\mat{F}$ and over gamma matrices. 
Compute the bosonic and fermionic actions 
\al{
S_B &= \int_X \!\! dx \; \frac{1}{g_0^2\,F} \, \tr_c ( \f[e](\theta)^2 )~, &
S_F &= \int_X \!\! dx \; \bsj^* (D+\iu \rho) \bsj ~, \label{sb} 
   }
where $g_0$ is a coupling constant and $\bsj \in h\,.$ Finally, 
perform a Wick rotation to Minkowski space. 

\section{The Matrix Part of the Unification Model}
\label{tum}

\subsection{The Representations under Consideration}
\label{rema}

We shall adapt our notations to the $\SU5 \times \U1$--model. In contrast to 
what one could expect from the classical treatment \cite{dkn} of that model, 
the matrix Lie algebra is not $\su5 \op \u1$ but $\f[a]=\su5\,.$ In our 
approach, the $\u1$--part is not an input of the model but an algebraic 
consequence. The internal Hilbert space is 
\eq[c192]{
\C^{192}=\big(\ul{10} \op \ul5^* \op \ul1 \op \ul{10}^* \op \ul5 \op \ul1 \big) 
\ot \C^2 \ot \C^3~,
          }
where $\ul{10},\ul{10}^*,\ul5,\ul5^*,\ul1$ are representations of $\su5\,.$ 
Since we consider linear operators on $\C^{192}\,,$ we need the decomposition 
rules for homomorphisms between the $\su5$--representations occurring in 
\rf[c192]: 
\seq{
\label{irrep}
\eqa{lclcccccccccc}{
\End(\ul{10}) &=& \End(\ul{10}^*) &=& \ul{10} & \ot & \ul{10}^* 
&=& \ul1 & \op & \ul{24} & \op & \ul{75} \yn \label{irrea} \npb \\
\End(\ul5) &=& \End(\ul5^*) &=& \ul5 & \ot & \ul5^*	 
&=& \ul1 & \op & \ul{24}  \yn \label{irreb} \\
\End(\ul1) & & & & & & &=& \ul1 
\yn \\
\Hom(\ul5,\ul{10}) &=& \Hom(\ul{10}^*,\ul5^*) &=& \ul5^* & \ot & \ul{10} 
&=& \ul5 & \op & \ul{45}^* \yn \label{irred} \\
\Hom(\ul5,\ul{10}^*) &=& \Hom(\ul{10},\ul5^*) &=& 
\ul5^* & \ot & \ul{10}^* &=& \ul{10} & \op & \ul{40}^*	\yn \label{irree} \\
\Hom(\ul5^*,\ul5) && &=& \ul5 & \ot & \ul5
&=& \ul{10} & \op & \ul{15} \yn \\
\Hom(\ul{10}^*,\ul{10}) \hs*{-1em} && &=& \ul{10} & \ot & \ul{10} 
&=& \ul5^* & \op & \ul{45} & \op & \ul{50}
\yn \label{irreg} \\
\Hom(\ul1,\ul5) &=& \Hom(\ul5^*,\ul1) &&&& &=& \ul5 \yn \\
\Hom(\ul1,\ul{10}) &=& \Hom(\ul{10}^*,\ul1) &&&& &=& \ul{10} \yn
       }}
We identify the matrix Lie algebra $\su5$ with its 
$\ul{24}$--representation. Then, we get a natural representation $\h{\pi}$ of 
$\su5$ in $\End(\C^{192})$ by selecting the $\ul{24}$--representations in 
\rf[irrep]: 
\eq[piofa]{
\h{\pi}(a) := \mb{Z{15}E{15}|Z{15}E{15}}{
\pi_{10}(a) & 0 & 0 & & & \\
0 & \ol{\pi_5(a)} & 0 & \mc{3}{c}{\mathrm{O}} \\
0 & 0 & 0_3 & & &  \\  \hline 
& & & \; \ol{(\pi_{10}(a)} & 0 & 0 \ru{3}{0}\\
\mc{3}{c}{\mathrm{O}} \vline & 0 & \pi_5(a) & 0 \\ 
& & & 0 & 0 & 0_3 } \ot \one_6~ .
  }
Here, $\pi_{10}$ and $\pi_5$ denote the embeddings of $\ul{24}$ into 
\rf[irrep]. 

We define the $\ul{75}$--representation of $\su5$ occurring in 
the decomposition \rf[irrea] as the set $\f[v]$ of $10 \times10$--matrices 
of the form
\eq{
\f[v]:=\{~v \in \mathrm{su}(10)~,~~ \tr(v \,\pi_{10}(a))=0~~
\forall a \in \f[a]~ \}~.
   }

Next, we consider the $\ul5$--representations occurring on the r.h.s.\ of 
\rf[irrep]. Let $\f[b]=\C^5$ be the vector space of matrices represented in the 
form
\eq[bbb]{
b = \iu (b_1, b_2, b_3, b_4, b_5)^T ~, \quad b_i \in \C~.
	}
We define a linear map $\h{\pi}$ of $\f[b]$ in $\End (\C^{192})\,,$ putting
\eq{
\h{\pi}(b) = \mbox{\small{$ \mb{D{19}|D{18}}{
& & & \pi_{10,10}(b) & \pi_{10,5}(b)  & 0 \\
\mc{3}{c}{\mathrm{O}} \vline & \pi_{10,5}(b)^T & 0 & \pi_{5,1}(b) \\
& & & 0 & \pi_{5,1}(b)^T & 0 \\ \hline
-\pi_{10,10}(b)^* & - \ol{\pi_{10,5}(b)} & 0 & & & \\
-\pi_{10,5}(b)^* & 0 & -\ol{\pi_{5,1}(b)} & \mc{3}{c}{\mathrm{O}} \\
0 & -\pi_{5,1}(b)^* & 0 & & & } $}}\! \ot \one_6\,.  \raisetag{3ex}
  }
The matrices $\pi_{10,10}(b), \pi_{10,5}(b)$ and $\pi_{5,1}(b)$ are the 
embeddings of $b \in \ul5$ into $\ul{10} \ot \ul{10}\,,$ $\ul5^* \ot \ul{10}$ 
and $\ul1 \ot \ul5^*\,,$ see \rf[irrep]. Observe that 
\eq[ppa]{
[\h{\pi}(a),\h{\pi}(b)] = \h{\pi}(a b) \in \h{\pi}(\f[b])~, \quad  
a \in \f[a]~,~~ b \in \f[b]~.
	}

Due to the first three formulae in \rf[irrep], the $\ul{24}$--parts and the 
$\ul1$--parts of $\pi_{i,j}(b) \pi_{i,j}(b)^*\,,$ respectively, must be 
correlated. Indeed, we find with 
\eq[btb]{
(b,b)':= b b^* - \tfrac{1}{5} \tr(b b^*) \one_5 \in \iu \f[a]
   }
the identities \cite{phd}
\eqas[ppb]{rcl}{
\pi_{10,10}(b) \pi_{10,10}(b)^* &=& \iu \pi_{10}(\iu (b,b)') 
+\tfrac{3}{5} (b^* b) \one_{10}~,~~  \\
\pi_{10,5}(b) \pi_{10,5}(b)^* &=& 
-\iu \pi_{10}(\iu (b,b)') +\tfrac{2}{5} (b^* b) \one_{10}~,~~ \\
\pi_{10,5}(b)^* \pi_{10,5}(b) &=& \iu \pi_5(\iu (b,b)') + \tfrac{4}{5} (b^* b) 
\one_{5}~,~~ \\
\ol{\pi_{5,1}(b)} \pi_{5,1}(b)^T &=& -\iu \pi_5(\iu (b,b)') 
+ \tfrac{1}{5} (b^* b) \one_{5}~,~~\\
\pi_{5,1}(b)^T \,\ol{\!\pi_{5,1}(b)} &=& (b^* b) ~.~~  
    }

Moreover, we consider the \ul{45}--representation of $\su5$ occurring in 
\rf[irred]. It is the vector space $\f[w]$ of $10 \times 5$--matrices 
determined by 
\seq{
\eq{
\f[w]:=\{ ~w \in \Hom(\C^5,\C^{10})~,~~ \tr(w \, \pi_{10,5}(b)^*)=0~~ 
\forall b \in \ul{5}~\}~.
    }
One has 
\eq[awwa]{
[a,w] := \pi_{10}(a) w - w \pi_5(a) \in \f[w]~, \quad 
w \in \f[w]\,,~ a \in \f[a]\,.
    }}

Finally, we consider the \ul{50}--representation of $\su5$ occurring in 
\rf[irreg]. It is the vector space $\f[c]$ of symmetric complex 
$10 \times 10$--matrices determined by 
\seq{
\eq{
\f[c]:=\{ ~c \in \mat{10}~,~~ c=c^T~,~~\tr(c \, \pi_{10,10}(b)^*)=0~~ 
\forall b \in \ul{5}~\}~.
    } 
One has
\eq[acc]{
[a,c] := \pi_{10}(a) c - c \overline{\pi_{10}(a)} \equiv 
\pi_{10}(a) c + c \pi_{10}(a)^T \in \f[c] ~, \quad 
a \in \f[a]\,,~ c \in \f[c]~.
	}
        }

\subsection{The Mass Matrix}
\label{gdo}

Now we define the mass matrix $\M$ of the L--cycle. Let 
\eqa{rcl}{
m &\equiv& \pi_5(m) := \iu\, \diag( {-} \tfrac{2}{5}, {-} \tfrac{2}{5}, 
{-} \tfrac{2}{5}, \tfrac{3}{5}, \tfrac{3}{5}) \in \f[a]~, \npb \\
&& \pi_{10}(m) \equiv  \iu\, \diag(\tfrac{1}{5},\tfrac{1}{5}, 
\tfrac{1}{5}, \tfrac{1}{5}, \tfrac{1}{5},\tfrac{1}{5}, {-} \tfrac{4}{5}, 
{-} \tfrac{4}{5}, {-}\tfrac{4}{5}, \tfrac{6}{5})~, \label{mnnp} \yn \\
n &:=& \iu (0,0,0,1,0)^T \in \f[b]~, \qquad \\
m' &:=& \iu \mb{Z{11}Z{11}}{
0_{3 \times 3} & 0_{3 \times 3} & 0_{3 \times 3} & 0_{3 \times 1} \\ 
0_{3 \times 3} & 0_{3 \times 3} & 0_{3 \times 3} & 0_{3 \times 1} \\ 
0_{3 \times 3} & 0_{3 \times 3} & 0_{3 \times 3} & 0_{3 \times 1} \\ 
0_{1 \times 3} & 0_{1 \times 3} & 0_{1 \times 3} & -1 } \in \f[c] \;, \quad{} 
n' := \iu \mb{D{11}}{
\one_3	   & 0_{3 \times 1} & 0_{3 \times 1} \\ 
0_{3 \times 3} & 0_{3 \times 1} & 0_{3 \times 1} \\ 
0_{3 \times 3} & 0_{3 \times 1} & 0_{3 \times 1} \\ 
0_{1 \times 3} & 0_{1 \times 1} & 3 } \in \f[w] \;.
    }
Then we put
\seq{
\label{M}
\eqa{rclrcl}{
\M &=& \multicolumn{4}{l}{ \mb{E{14}E{16}E{12}|E{16}E{12}E{16}}{
\M_{10} & 0 & 0 & \M_{10,10} & \M_{10,5} & 0 \\  
0 & \overline{\M_5} & 0 & \M_{10,5}^T & 0 & \M_{5,1} \\  
0 & 0 & 0 & 0 & \M_{5,1}^T & 0 \\  \hline 
\M_{10,10}^* & \overline{\M_{10,5}} & 0 & \overline{\M_{10}} & 0 & 0 \\  
\M_{10,5}^* & 0 & \overline{\M_{5,1}} & 0 & \M_5 & 0  \\  
0 & \M_{5,1}^* & 0 & 0 & 0 & 0	}~,~~  \mbox{where} \hs*{3em} }  \yn
\\
\M_{10} & = & \iu \pi_{10}(m) \ot M_{10}' \;,\qquad{} & 
\M_5 & = & -\iu \pi_5(m) \ot M_5' \;, \npb \\
\M_{10,10}  & = & \iu \pi_{10,10}(n) \ot M_d' + \iu m' \ot M_N' \;, \qquad{} &
\M_{5,1} & = & \iu \pi_{5,1}(n) \ot M_e'  \;, \yn \npb \\
\M_{10,5} & = & \iu \pi_{10,5}(n) \ot \Mu' + \iu n' \ot \Mn' \;.  
    }}
Here, $M_{10}',M_5',M_N',\Mu',M_d',M_e',\Mn'$ are $6 \times 6$--matrices of the 
following block structure:
\eq[mimf]{
M_i'=\mb{Z{12}}{ 0_3 & M_i \\ M_i^* & 0_3 }~, \qquad
M_f'=\mb{Z{12}}{ M_f & 0_3 \\ 0_3 & M_f }~,~~ 
   }
for $i \in \{5,10\}$ and $f \in \{\t{u},d,e,\t{n},N\}\,.$ The only condition to 
the $3 \times 3$--mass matrices $M_{10},M_5,\Mu,M_d,M_e,\Mn$ and $M_N$ is 
\al{
M_d &= M_d^T~, &  M_N &= M_N^T~.
  }

The final input of our L--cycle is the grading operator $\h{\Gamma}$, which we 
choose as
\al{
\label{hG}
\h{\Gamma}&=\mb{Z{22}}{
-\one_{16} \ot \h{\Gamma}' & 0_{96} \\ 0_{96} & \one_{16} \ot \h{\Gamma}' }~, &
\h{\Gamma}'  &=\mb{Z{15}}{ \one_{3} & 0 \\ 0 & -\one_{3} }~.
   }
From \rf[piofa] and \rf[hG] there follows that $\h{\Gamma}$ commutes with 
$\pf\,.$ The fact that $\h{\Gamma}'$ commutes with $M'_{\t{u},d,e,\t{n},N}$ 
and anticommutes with $M_{10,5}'$ implies that $\h{\Gamma}$ anticommutes with 
$\M\,.$ Therefore, the tuple $(\f[a],\C^{192},\M,\h{\pi},\h{\Gamma})$ is an 
L--cycle. 

Let us summarize the block structure of this L--cycle, for instance 
in terms of $4 \times 4$--block matrices with entries in 
$48 \times 48$--matrices:
\eqa{rcl}{
\mc{3}{l}{
\h{\pi}(a) \!=\!\! \mbox{\footnotesize{$\mb{Z{6}Z{6}}{
A & 0 & 0 & 0 \\ 0 & A & 0 & 0 \\ 0 & 0 & \bar{A} & 0 \\ 0 & 0 & 0 & \bar{A} }
$}},~~
\M=\!\! \mbox{\footnotesize{$ \mb{Z{8.4}Z{8.4}}{
0 & \M_i & \M_f & 0 \\ \M_i^* & 0 & 0 & \M_f \\ \M_f^* & 0 & 0 & 
\ol{\M_i} \\ 0 & \M_f^* & \M_i^T & 0 } $}},~~ 
\h{\Gamma} \!=\!\! \mbox{\footnotesize{$ \mb[\!\!]{E{8.5}Z{7.5}E{8.5}}{
-\one_{48} & 0 & 0 & 0 \\ 0 & \one_{48} & 0 & 0 \\ 0 & 0 & \one_{48} & 0 \\ 
0 & 0 & 0 & -\one_{48} } $}},} \npb \eqnskip \\
\mbox{with} \yn  \label{AMM} \eqnskip \eqnskip
\\
A &:=& \diag \big( \pi_{10}(a) \ot \one_3~,~ \overline{\pi_5(a) } \ot \one_3~,~ 
0_3 \big) ~,  \eqnskip \\
\M_i &:=& \diag \big( \iu \pi_{10}(m) \ot M_{10}~,~ 
\overline{ -\iu \pi_5(m) \ot M_5} ~,~ 0_3 \big) ~, \eqnskip \eqnskip\\
\M_f &:=& \mbox{\small{$ \!\! \mb{ccc}{
{}~\iu \pi_{10,10}(n) \!\ot\! M_d + \iu m' \!\ot\! M_N & {}~~
\iu \pi_{10,5} (n) \!\ot\! \Mu + \iu n' \ot \Mn & 0 \\ 
{}~\iu \pi_{10,5}(n)^T \!\ot\! \Mu^T + \iu n'{}^T \!\ot\! \Mn^T & 
0 & \hs*{-0.7em} \iu \pi_{5,1}(n) \!\ot\! M_e \\  
0 & \iu \pi_{5,1}(n)^T \!\ot\! M_e^T & 0 } \!\! $}} \equiv \M_f^T \,.
    }

\subsection{The Structure of $\p{1}$ and $\p{2}$}
\label{spp}

We recall \rf[hpi] that elements $\tau^1 \in \p{1}$ are of the form 
\eq[eip]{
\tau^1=\tsum_{\alpha,z \geq 0} [\h{\pi}(a_\alpha^z), 
[ \dots [\h{\pi}(a_\alpha^1), [-\iu \M,\h{\pi}(a_\alpha^0)]] \dots ]] ~.~~
	}
Using \rf[ppa], \rf[awwa] and the fact that $\h{\pi}$ is a representation we 
obtain the explicit structure of elements $\tau^1 \in \p{1}$: 

\seq{
\label{tau1}
\eqa{rcl}{
\mc{3}{l}{\tau^1=  } \label{impb} \yn \npb \\
\mc{3}{l}{\arraycolsep0.02em
\mbox{\footnotesize{$ 
\mb{E{28}E{27}E{17}|E{25}E{26}E{15}}{
\pi_{10}(a) \ot M_{10}' & 0 & 0 & \mpa{c}{ \pi_{10,10}(b) \ot M_d' \\ 
+ c \ot M_N' } \hs*{-0.4em} & \mpa{c}{
\pi_{10,5} (b) \ot \Mu' \\ + w \ot \Mn' } \hs*{-0.8em} 
& 0 \\ 
0 & \overline{\pi_5(a) \ot M_5'} & 0 & \mpa{c}{ \pi_{10,5}(b)^T \!\ot\! 
\Mu'{}^T \\ + w^T \ot \Mn'{}^T } \hs*{-1.2em} & 0 & \hs*{-1.6em} 
\pi_{5,1}(b) \ot M_e' \\ 
0 & 0 & 0 & 0 & \hs*{-0.5em} \pi_{5,1}(b)^T \ot M_e'{}^T \hs*{-0.5em} & 0 
\\  \hline 
\mpa{c}{ -\pi_{10,10}(b)^* \!\ot\! M_d'{}^* \\ - c^* \ot M_N' } \hs*{-0.9em} & 
\mpa{c}{ -\overline{\pi_{10,5}(b) \ot \Mu'} \\ - \overline{w \ot \Mn'} }
\hs*{-1.3em} & 0 & -\overline{\pi_{10}(a) \ot M_{10}'} \hs*{-0.5em} & 0 & 0 \\  
\mpa{c}{ -\pi_{10,5}(b)^* \ot \Mu'{}^* \\ - w^* \ot \Mn'{}^* } \hs*{-0.5em} & 
0 & \hs*{-1.6em} -\overline{\pi_{5,1}(b) \ot M_e'} & 
0 & -\pi_5(a) \ot M_5' & 0  \\  
0 & -\pi_{5,1}(b)^* \ot M_e'{}^* & 0 & 0 & 0 & 0
}  \! $}}, \eqnskip \eqnskip } 
\\
a &=& \tsum_{\alpha,z \geq 0} [a_{\alpha}^z,[ \dots [a_{\alpha}^1, 
[m,a_{\alpha}^0]] \dots ]] \in \f[a] ~,~~ \yn \\
b &=& -\tsum_{\alpha,z \geq 0} a_{\alpha}^z \cdots a_{\alpha}^1 a_{\alpha}^0 n 
\in \f[b]~, \yn \\
w &=& \tsum_{\alpha,z \geq 0} [a_{\alpha}^z,[ \dots 
[a_{\alpha}^1,[n',a_{\alpha}^0]] \dots ]] \in \f[w]~,  \yn \npb \label{taw} \\
c &=& \tsum_{\alpha,z \geq 0} [a_{\alpha}^z,[ \dots 
[a_{\alpha}^1,[m',a_{\alpha}^0]] \dots ]] \in \f[c]~.  \yn \label{tac}
    }}
Here, the commutators \rf[taw] and \rf[tac] are understood in the sense 
\rf[awwa] and \rf[acc]. It is obvious that $a,b,c,w$ are independent as 
elements of different irreducible representations of $\su5\,.$ 

Next, we are going to construct $\p{2}\,.$ According to \rf[on], elements 
$\tau^2 \in \p{2}$ are obtained by summing up elements of the type 
\eq[pcw]{
\tau^2:=\th \{\tau^1,\tau^1 \}~,\qquad \tau^1 \in \p{1}~.
    }
Thus, using \rf[ppb] we get from \rf[impb] the structure 
\seq{
\label{tau2}
\eqa{rclrcl}{
\tau^2 &=& \mb{D{14}|D{14}}{
\tau_{10} & \tau_{\widetilde{10,5}} & \tau_{10,1} & \tau_{10,10} & 
\tau_{10,5} & 0 \\  
\tau_{\widetilde{10,5}}^* & \tau_5^T & 0 & \tau_{10,5}^T & 0 & \tau_{5,1} \\  
\tau_{10,1}^* & 0 & \tau_1 & 0 & \tau_{5,1}^T & 0 \\  \hline 
\tau_{10,10}^* & \overline{\tau_{10,5}} & 0 & \tau_{10}^T & 
\overline{\tau_{\widetilde{10,5}}} & \overline{\tau_{10,1}} \\	
\tau_{10,5}^* & 0 & \overline{\tau_{5,1}} & \tau_{\widetilde{10,5}}^T & 
\tau_5 & 0  \\  
0 & \tau_{5,1}^* & 0 & \tau_{10,1}^T & 0 & \tau_1^T }  ~,~~ \mbox{where}  \yn
\label{poaw}
\eqnskip
\\
\tau_{10} &=& \iu \pi_{10}(\iu (b,b)') 
\ot (\Mu' \Mu'{}^* - M_d' M_d'{}^*) - (b^* b) \one_{10} \ot 
(\tfrac{2}{5} \Mu' \Mu'{}^* + \tfrac{3}{5} M_d' M_d'{}^*) \hs*{-3em} \npb \\ && 
+ \th \{\pi_{10}(a),\pi_{10}(a)\} \ot M_{10}'{\!}^2  \npb \\ &&
- w w^* \!\ot\! \Mn' \Mn'{}^* - w \pi_{10,5}(b)^* \ot \Mn' \Mu'{}^* 
- \pi_{10,5}(b) w^* \ot \Mu' \Mn'{}^* 
\npb \\ &&
- c c^* \ot M_N' M_N'{}^* - c \pi_{10,10}(b)^* \ot M_N' M_d'{}^* 
- \pi_{10,10}(b) c^* \ot M_d' M_N'{}^* 
\\ 
\tau_5 &=& \iu \pi_5(\iu (b,b)') \ot (\bar{M}_e' M_e'{}^T - \Mu'{}^* \Mu' ) 
- (b^* b) \one_{5} \ot (\tfrac{4}{5} \Mu'{}^* \Mu' 
+ \tfrac{1}{5} \bar{M}_e' M_e'{}^T) \hs*{-3em} \npb \\ && 
+ \th \{\pi_5(a),\pi_5(a)\} \ot M_5'{}^2 \npb \yn  \label{tdev}\\ &&
- w^* w \ot \Mn'{}^* \Mn' 
- w^* \pi_{10,5}(b) \ot \Mn'{}^* \Mu' - \pi_{10,5}(b)^* w \ot M_{u}'{}^* \Mn'~,  
\npb \\ 
\tau_1 &=& - b^* b \ot M_e'{}^T \bar{M}_e' ~, 
\\ 
\tau_{10,10} &=& \pi_{10,10}(a b) \ot \th (M_{10}' M_d' + M_d' M_{10}'{\!}^T) 
\npb \\ && 
+ (\pi_{10}(a) \pi_{10,10}(b) - \pi_{10,10}(b) \pi_{10}(a)^T) 
\ot \th (M_{10}' M_d' - M_d' M_{10}'{\!}^T)  \npb \\ &&
+ (\pi_{10}(a) c + c \pi_{10}(a)^T) \ot \th (M_{10}' M_N' + M_N' M_{10}'{\!}^T) 
\npb \\ && 
+ (\pi_{10}(a) c - c \pi_{10}(a)^T) \ot \th (M_{10}' M_N' - M_N' M_{10}'{\!}^T)  
{}~, \\
\tau_{10,5} &=& \pi_{10}(a) \pi_{10,5}(b) \ot M_{10}' \Mu' 
- \pi_{10,5}(b) \pi_5(a) \ot \Mu' M_5' \yn \label{tdow} \\ && 
+ \pi_{10}(a) w \ot M_{10}' \Mn' - w \pi_5(a) \ot \Mn' M_5' \,, \\  
\tau_{\widetilde{10,5}} &=& -\pi_{10,10}(b) \overline{w} 
\ot M_d' \bar{M}_{\t{n}}'- c \overline{w} \ot M_N' \bar{M}_{\t{n}}' 
- c \overline{\pi_{10,5}(b)} \ot M_N' \bar{M}_{\t{u}}' ~,\\
\tau_{5,1} &=& \pi_{5,1}(a b) \ot M_5'{}^T M_e' ~, \qquad 
\tau_{10,1} = -w \overline{\pi_{5,1}(b)} \ot \Mn' \bar{M}_e' ~. 
     }}

\subsection{The Structure of the Connection Form}

We know from \rf[rho] that for constructing the connection form $\rho$ we need 
knowledge of the spaces $\bbr{0}$ and $\bbr{1}$ determined by the equations 
\rf[rlh]. To compute the structure of elements $\eta^0 \in \bbr{0}$ we first 
decompose $\eta^0$ according to \rf[irrep] into irreducible 
$\su5$--representations, each of them tensorized by $\mat{6}\,.$ Then, the 
condition $[\bbr{0} , \pf] \subset \pf$ yields the block structure 
\[
\eta^0 = \h{\pi}(a) + \iu \,\diag (\one_{10} \ot m_{10}'\,,\, 
\one_{5} \ot m_{\t{5}}' \,,\, m_1'\,,\, \one_{10} \ot m_{\widetilde{10}}'\,,\, 
\one_{5} \ot m_5'\,,\, m_{\t{1}}')\,,
\]
where $a \in \f[a]$ and $m_{10,\t{5},1, \widetilde{10},5, \t{1}}'$ are 
selfadjoint elements of $\mat{6}\,.$ The condition $\bbr{0} = \h{\Gamma} 
(\bbr{0}) \h{\Gamma}$ implies $m_i'=\diag(m_i,\h{m}_i)\,,$ for 
$m_i,\h{m}_i \in \mat{3}\,.$ 

We insert this structure into the condition $[\bbr{0}, \p{1}] \subset \p{1}\,.$ 
Using \rf[impb], \rf[ppa], \rf[awwa] and \rf[acc] we obtain from the 
off--diagonal blocks the equations 
\seq{
\eqas[mh0]{rclrcl}{
m_{10} M_d - M_d m_{\widetilde{10}} &=& -\iu \bar{\alpha} M_d ~, \qquad{} &
m_{10} M_N - M_N m_{\widetilde{10}} &=& -\iu \bar{\alpha}' M_N ~, \\
m_{10} \Mu - \Mu m_5 &=& -\iu \alpha \Mu ~, \qquad{} &
m_{10} \Mn - \Mn m_5 &=& -\iu \alpha'' \Mn ~, \\
m_{\t{5}} \Mu^T - \Mu^T m_{\widetilde{10}} &=& -\iu \alpha \Mu^T ~, \qquad{} &
m_{\t{5}} \Mn^T - \Mn^T m_{\widetilde{10}} &=& -\iu \alpha'' \Mn^T~,  \\
m_{\t{5}} M_e - M_e m_{\t{1}} &=& -\iu \bar{\alpha} M_e ~, \qquad{} &
m_1 M_e^T - M_e^T m_5 &=& -\iu \bar{\alpha} M_e^T ~,~~ 
    }
for $\alpha,\alpha',\alpha'' \in \C\,.$ The same equations hold for 
$\h{m}_i\,,$ with the same parameters $\alpha,\alpha',\alpha''\,.$ Multiplying 
the first equation by $M_d^*$ from the right and subtracting the Hermitian 
conjugate of the resulting equation we get for instance 
\[
[m_{10},M_d M_d^*] = -\iu(\alpha+\bar{\alpha}) M_d M_d^*\,.
\]
Applying the trace and respecting $\tr(M_d M_d^*) >0$ we get 
$\alpha=\iu \lambda\,,$ for $\lambda \in \R\,.$ Analogously, we have 
$\alpha'=\iu \lambda'$ and $\alpha''=\iu \lambda''\,.$ Thus, we find the 
equations
\eq[m1dd]{
[m_{10},M_d M_d^*] = [m_{10},M_N M_N^*] = [m_{10},\Mu \Mu^*] 
= [m_{10},\Mn \Mn^*] = 0~.
	 }
For generic mass matrices $M_{d,N,\t{u},\t{n}}\,,$ these equations can only be 
satisfied for $m_{10}=\mbox{$(\nu {-} \th \lambda) \one_3$}\,,$ for 
$\nu \in \R\,.$ We assume that $M_{d,\t{u},e}$ are invertible and find the 
solution
\eqas[mh01]{rclrclrcl}{
m_{10} &=& (\nu-\th \lambda) \one_3~,\qquad{} & 
m_5 &=& (\nu-\tfrac{3}{2} \lambda) \one_3~, \qquad{} & 
m_1 &=& (\nu-\tfrac{5}{2} \lambda) \one_3~, \\
m_{\widetilde{10}} &=& (\nu + \th \lambda) \one_3~, \qquad{} &
m_{\t{5}} &=& (\nu+\tfrac{3}{2} \lambda) \one_3~, \qquad{}& 
m_{\t{1}} &=& (\nu+\tfrac{5}{2} \lambda) \one_3~,
     }
     }
where $\nu,\lambda \in \R\,.$ For $\h{m}_i$ we get the same equations, with the 
same $\lambda$ but possibly a different $\h{\nu}$ instead of $\nu\,.$ Inserting 
this result into the $\pi_{10}$--block we get the equations
\al{
(\nu-\h{\nu}) M_{10} &= \beta M_{10}~, &
(\nu-\h{\nu}) M_{10}^* &= -\beta M_{10}^*~, \notag
   }
which are only compatible with $\nu=\h{\nu}\,.$ Thus, we obtain with \rf[piofa] 
the preliminary result
\seq{
\label{apru1}
\eqa{rrl}{
\eta^0 &=& \h{\pi}(a) + \h{\pi}(\u1) + \iu \nu \one_{192} ~,~~ 
\label{apr} \npb \yn \\
\h{\pi}(\iu \lambda) &:=& \iu \lambda \, \diag (\, {-} \th \one_{10} \;,~ 
\tfrac{3}{2} \one_5 \;,~ {-}\tfrac{5}{2} \;,~ 
\th \one_{10} \;,~ {-}\tfrac{3}{2} \one_5 \;,~ 
\tfrac{5}{2} \,) \ot \one_6~. \quad{} \yn \label{u1}
	}}

Now, one finds \cite{phd} that the $\u1$--part $\h{\pi}(\u1)$ is compatible 
with the two conditions $\{\bbr{0} , \pf\} \subset \{ \pf, \pf\} + \p{2}$ and 
$\{\bbr{0} , \p{1}\} \subset \{ \pf, \p{1}\} + \p{3}\,,$ whereas the 
identity part $\iu \nu \one_{192}$ is not. Here, one has to use the following 
identities:
\seq{
\label{idaa}
\eqa{c}{
\begin{array}{rclcr}
\tr (\pi_{10}(a)\,\pi_{10}(a)) &=& \tr (\overline{\pi_{10}(a)}\, 
\overline{\pi_{10}(a)}) &=& 3 \, \tr (a a)~, \\  
\tr (\pi_5(a)\,\pi_5(a)) &=& \tr (\overline{\pi_5(a)}\, \overline{\pi_5(a)}) 
&=&  \tr (a a)~, 
\end{array}   \yn 
\\
\iu \{\pi_{10}(a),\pi_{10}(a)\}_{\ul{24}} 
= \tfrac{1}{3} \pi_{10}\big(\iu \{\pi_5(a),\pi_5(a)\}_{\ul{24}}\big)~, \yn
\\
\begin{array}{rclrcl}
(\pi_{10}(a) \pi_{10,5}(b))_{\ul5} &=& \tfrac{3}{4} \pi_{10,5} (ab)~, 
\quad{} &
(\pi_{10,5}(b) \pi_5(a))_{\ul5} &=& -\tfrac{1}{4} \pi_{10,5} (ab)~, 
\quad{} \npb \\
(\pi_{10}(a) \pi_{10,5}(b))_{\ul{45}} &=& (\pi_{10,5}(b) \pi_5(a))_{\ul{45}} ~, 
\quad{} & (\pi_{10}(a) w)_{\ul5} &=& (w \pi_5(a))_{\ul5} ~,
\end{array} \quad{} \yn
  }}	
for $a \in \f[a]\,,$ $b \in \f[b]$ and $w \in \f[w]\,.$ 

The evaluation of the formulae for $\bbr{1}$ in \rf[rlh] yields for a generic 
choice of the mass matrices $M_{\t{u},d,e,\t{n},10,5}$ the simple result 
$\bbr{1} = \p{1}\,.$ Therefore, the connection form has the structure
\eq[gc]{
\rho \in \big( \Lambda^1 \otimes (\pf + \h{\pi}(\u1)) \big) \op \big( 
\Lambda^1 \g[5] \ot \p{1} \big) ~.
   }
We see that our formalism generates an additional $\u1$--part for the 
connection form and determines uniquely its representation \rf[u1] on the 
fermionic Hilbert space. Remarkably, this representation is realized in nature! 

\subsection{The Ideal $\cj{2}$} 
\label{tide}

We recall \rf[js] that for the analysis of $\jj{2}$ we must find the space of 
elements $\h{\sigma} (\omega^1)\,,$ where $\omega^1 \in \W[a]{1} \cap \ker 
\h{\pi}\,.$ For the computation of $\h{\sigma} (\omega^1)$ we need knowledge of 
$\M^2\,,$ see \rf[hsig]. We define
\eqas{rcl}{
v_0 &:=& \iu \,\diag({-}\tfrac{1}{3},{-}\tfrac{1}{3},{-}\tfrac{1}{3},
{-}\tfrac{1}{3},{-}\tfrac{1}{3},{-}\tfrac{1}{3},
\tfrac{1}{3},\tfrac{1}{3},\tfrac{1}{3}, 1\,) \in \f[v]~,\\
I_3 &\equiv& \pi_5(I_3) := \iu \, \diag\,(0, 0, 0, \tfrac{1}{2}, 
{-}\tfrac{1}{2} )~, \\
\pi_{10}(I_3) &\equiv& \iu \, \diag\,(\tfrac{1}{2}, \tfrac{1}{2}, \tfrac{1}{2}, 
{-}\tfrac{1}{2},{-}\tfrac{1}{2},{-}\tfrac{1}{2}, 0,0,0,0)
    }
and abbreviate
\al{
\label{mtu}
M_u' &:= \Mu' + \Mn'~, &  M_n' &:= \Mu' - 3 \Mn'~,
	}
analogously for the primeless matrices $M_{u,\nu,\t{u},\t{n}}\,.$ 
Then, using \rf[mnnp] and \rf[M], we find the following formula for $\M^2\,:$ 
\seq{
\label{Msq}
\eq[M2]{
\M^2 = \mbox{\small{$ \mb{Z{18}E{15}|Z{18}E{15}}{
(\M^2)_{10} & (\M^2)_{\widetilde{10,5}} & 0 & 
(\M^2)_{10,10} & (\M^2)_{10,5} & 0 \\   
(\M^2)_{\widetilde{10,5}}^*  & (\M^2)_{5}^T & 0 & 
(\M^2)_{10,5}^T & 0 & (\M^2)_{5,1} \\  
0 & 0 & (\M^2)_{1} & 0 & (\M^2)_{5,1}^T & 0 \\	\hline 
(\M^2)_{10,10}^* & \overline{(\M^2)_{10,5}} & 0 & 
(\M^2)_{10}^T & \overline{(\M^2)_{\widetilde{10,5}}} & 0 \\ 
(\M^2)_{10,5}^* & 0 & \overline{(\M^2)_{5,1}} & 
(\M^2)_{\widetilde{10,5}}^T & (\M^2)_{5} & 0 \\  
0 & (\M^2)_{5,1}^* & 0 & 0 & 0 & (\M^2)_{1}^T  
} , $}}
     }
where
\eqa{rcl}{
(\M^2)_{10} &=& \one_{10} \ot (\tfrac{9}{25} M_{10}'{\!}^2 
+ \tfrac{4}{10} \Mu' \Mu'{}^* + \tfrac{6}{10} M_d' M_d'{}^* 
+ \tfrac{12}{10} \Mn' \Mn'{}^* + \tfrac{1}{10} M_N' M_N'{}^* ) \npb \\ &&  
- \iu v_0 \ot (M_{10}'{\!}^2 - 2 (\Mu' \Mn'{}^* + \Mn' \Mu'{}^*) 
+ 4 \Mn' \Mn'{}^* + \th M_N' M_N'{}^*) \npb \\ &&
-\iu  \pi_{10}(\tfrac{Y'}{2} +I_3) \ot (M_u' M_u'{}^*-M_d' M_d'{}^*) \npb \\ && 
- \tfrac{1}{3} \iu \pi_{10}(m) (\tfrac{1}{5} M_{10}'{\!}^2 
- 4 (\Mu' \Mn'{}^* + \Mn' \Mu'{}^*) + 8 \Mn' \Mn'{}^* + M_N' M_N'{}^*) \,,
\\
(\M^2)_5 &=& \one_{5} \ot ( \tfrac{6}{25} M_5'{}^2 + \tfrac{12}{5} 
\Mn'{}^* \Mn' + \tfrac{4}{5} \Mu'{}^* \Mu' + \tfrac{1}{5} \bar{M}_e' M_e'{}^T 
) \npb \\ && \yn \label{Msqe} 
- \iu \pi_5(\tfrac{Y'}{2}+I_3) \ot (\bar{M}_e' M_e'{}^T-M_n'{}^* M_n') 
\npb \\ &&
- \iu \pi_5(m) (\tfrac{1}{5} M_5'{}^2 - 4 (\Mu'{}^* \Mn' + \Mn'{}^* \Mu' )
+ 8 \Mn'{}^* \Mn') \,, \npb 
\\ 
(\M^2)_1 &=& M_e'{}^T \bar{M}_e' ~,~~ 
\\
(\M^2)_{10,10} & = & \tfrac{3\iu}{5} \pi_{10,10}(n) \ot \th (M_{10}' M_d' + 
M_d M_{10}'{\!}^T) \npb \\ &&
+ \tfrac{\iu}{2} \pi_{10,10}(n') \ot \th (M_{10}' M_d' {-} M_d M_{10}'{\!}^T) 
- \tfrac{12\iu}{5} m' \ot \th (M_{10}' M_N' {+} M_N' \overline{M_{10}'})\;, \\
(\M^2)_{5,1} &=& \tfrac{3\iu}{5} \pi_{5,1}(n) \ot M_5'{}^T M_e' ~,~~  \npb \\
(\M^2)_{10,5} & = & -\iu \pi_{10,5}(n) \ot (\tfrac{9}{20} M_{10}' \Mu' 
+ \tfrac{3}{20} \Mu' M_5' - \tfrac{3}{4} M_{10}' \Mn' 
+ \tfrac{3}{4} \Mn' M_5') \npb \\ && 
-\iu n' \ot (-\tfrac{1}{4} M_{10}' \Mu' + \tfrac{1}{4} \Mu' M_5' 
+ \tfrac{19}{20} M_{10}' \Mn' - \tfrac{7}{20} \Mn' M_5') ~, \npb \\
(\M^2)_{\widetilde{10,5}} &=& -\iu n'' \ot M_N' \overline{M_n'}~.
      }
      }
Here, $n''$ is a generator of the $\ul{40}^*$--representation of $\su5$ 
occurring in the decomposition \rf[irree]:
\enlargethispage{5mm}
\eq{
n'' := \iu \mb{D{12}}{
0_{3 \times 3} & 0_{3 \times 1} & 0_{3 \times 1} \\ 
0_{3 \times 3} & 0_{3 \times 1} & 0_{3 \times 1} \\ 
0_{3 \times 3} & 0_{3 \times 1} & 0_{3 \times 1} \\ 
0_{1 \times 3} & 0_{1 \times 1} & 1 } \in \ul{40}^* \;.
    }

Due to \rf[js], the ideal $\jj{2}$ is given as the set of elements $j_2$ of 
the form 
\seq{
\label{j23}
\eqa{l}{
j_2={\sum}_{\alpha,z \geq 0} [\h{\pi}(a_\alpha^z),[ \dots [\h{\pi}(a_\alpha^1), 
[\M^2,\h{\pi}(a_\alpha^0)]] \dots ]] ~,~~\mbox{where} \label{j2} \yn \npb \\ 
0={\sum}_{\alpha,z \geq 0} [\h{\pi}(a_\alpha^z),[ \dots [\h{\pi}(a_\alpha^1),
[-\iu \M,\h{\pi}(a_\alpha^0)]] \dots ]] ~.~ \label{j3} \yn
       }}
Obviously, terms in $\M^2$ proportional to the identities $\one_{10},\one_{5}, 
1$ do not contribute to $j_2\,.$ Next, the term $(\M^2)_{5,1}=\tfrac{3\iu}{5} 
\pi_{5,1}(n) \ot M_5'{}^T M_e'$ gives a contribution to $j_2\,,$ which is 
$\tfrac{3\iu}{5} \ot M_5'{}^T$ times (from the left) the contribution of 
$-\iu \M_{5,1}=\pi_{5,1}(n) \ot M_e'$ to \rf[j3], and hence equals zero. For 
the same argument, all terms in $(\M^2)_{10,10}$ and $(\M^2)_{10,5}$ 
do not contribute to $j_2\,.$ The same is true for the terms 
proportional to $\pi_{10}(m)$ and $\pi_5 (m)\,.$ Thus, there remain only 
contributions from the terms $- \iu \pi_{10} (\tfrac{m}{2}+I_3) \ot 
M^2_{A,10}\,,$ $- \iu \pi_5 (\tfrac{m}{2}+I_3) \ot M^2_{A,5}\,,$ 
$- \iu v_0 \ot M_V^2$ and $- \iu n'' \ot M_N' \overline{M_n'}\,,$ where 
\seq{
\eqa{rcl}{
M_V^2 &:=& M_{10}'{\!}^2 - 2(\Mu' \Mn'{}^* + \Mn' \Mu'{}^* 
- 2 \Mn' \Mn'{}^*)~,~~ \yn \label{MVa} \npb \\
M_{A,10}^2 &:=& M_u' M_u'{}^*-M_d' M_d'{}^* ~, \qquad
M_{A,5}^2:= \bar{M}_e' M_e'{}^T -M_n'{}^* M_n'~.~~  \yn 
     }}
Since the irreducible representations $\ul{24},\ul{75},\ul5,\ul{45}^*,\ul{50}, 
\ul{40}^*$ are independent, it is always possible to fulfil \rf[j3] and to 
generate by the commutators \rf[j2] representations of arbitrary elements of 
$\ul{75}$ and $\ul{40}^*.$ Moreover, it can be checked that the generator 
$\tfrac{m}{2} + I_3$ occurring in $\M^2$ generates independent elements of the 
$\ul{24}$--representation. Hence, $j_2 \in J_2:=\jj{2}$ takes the form
\eqa{l}{
j_2= \yn \label{j1m} \npb \\
\mbox{\small{$ \mb{ccE{10}|ccE{10}}{~
\mpa{c}{ \iu \pi_{10}(a) \ot M_{A,10}^2 \\ + \iu v \ot M_V^2 } & 
\iu c'' \ot M_N' \overline{M_n'} & 0 & & & \\ 
(\iu c'' \ot M_N' \overline{M_n'})^* & 
(\iu \pi_5(a) \ot M_{A,5}^2)^T & 0 & 
\mc{3}{c}{\raisebox{2ex}{O}} \\ 
0 & 0 & 0 & & & \\ \hline 
& & & \mpa{c}{ (\iu \pi_{10}(a) \ot M_{A,10}^2 \\
+ \iu v \ot M_V^2)^T } & \overline{\iu c'' \ot M_N' \overline{M_n'}} & 0 \\ 
\mc{3}{c}{\raisebox{2ex}{O}} \vline & (\iu c'' \ot M_N' \overline{M_n'})^T & 
\iu \pi_5(a) \ot M_{A,5}^2 & 0 \\ 
& & & 0 & 0 & 0 
} $}},
    }
where $a \in \f[a]\,,\ v \in \f[v]$ and $c'' \in \ul{40}^*\,.$ 
\enlargethispage{-1mm}

Let $J_0:=\{\pf,\pf \}\,.$ From \rf[piofa] and \rf[idaa] we conclude 
that elements $j_0 \in J_0$ are of the form 
\eqa{l}{
j_0 = \yn \label{j3m} \npb \\ 
\mbox{\footnotesize{$ \mb{ccE{7.5}|ccE{7.5}}{
\; \tfrac{3}{5} \alpha \one_{10} + \tfrac{1}{3} \iu \pi_{10}(a) + \iu v  
\hs*{-0.6em} & 0 & 0 & & & \\
0 & \hs*{-0.6em} \tfrac{2}{5} \alpha \one_5 + \iu \pi_5(a)^T & 0 & 
\mc{3}{c}{\mathrm{O}} \\ 
0 & 0 & 0 & & & \\ \hline
& & & \;\tfrac{3}{5}\alpha \one_{10} + \tfrac{1}{3} \iu \pi_{10}(a)^T + \iu v^T 
\hs*{-0.6em} & 0 & 0 \\ 
\mc{3}{c}{\mathrm{O}} \vline & 0 & \hs*{-0.6em} \frac{2}{5}\alpha \one_5 
+ \iu \pi_5(a) & 0 \\  
& & & 0 & 0 & 0 } $}} \!\! \ot \one_6\,,  
     }
where $\alpha \in \R\,,~a \in \f[a]$ and $v \in \f[v]\,.$ 

It remains to find the spaces $\cj{0},\cj{1},\cj{2}$ occurring in \rf[cj]. 
For generic mass matrices $M_{\t{u},d,e,\t{\nu},10,5}$ the result is \cite{phd} 
\eq[lcj]{
\cj{0}=\{0\}~, \qquad  \cj{1}=\{0\}~, \qquad  
\cj{2} = \jj{2} + \{\pf,\pf\} + \R \one_{192} ~.~~  
  }
Let $J_3:=\R \one_{192}\,.$ It is advantageous to construct an orthogonal 
decomposition $J_0+J_2+J_3=(J_0+J_3) \op J_2\,.$ The result is 
that elements $j_2' \in J_2'$ are of the form

\eqa{l}{
j_2' = \yn \label{jzp} \npb \\
\mbox{\small{$ \mb{ccE{10}|ccE{10}}{~
\mpa{c}{ \iu \pi_{10}(a) \ot \Mud^2 \\ + \iu v \ot \t{M}_V^2 } & 
\iu c'' \ot M_N' \overline{M_n'} & 0 & & & \\ 
(\iu c'' \ot M_N' \overline{M_n'})^* & 
{}~~(\iu \pi_5(a) \ot \Men^2)^T & 0 & 
\mc{3}{c}{\raisebox{2ex}{O}} \\ 
0 & 0 & 0 & & & \\ \hline 
& & & \mpa{c}{ (\iu \pi_{10}(a) \ot \Mud^2 \\
+ \iu v \ot \t{M}_V^2)^T } & \overline{\iu c'' \ot M_N' \overline{M_n'}} & 0 
\\ \mc{3}{c}{\raisebox{2ex}{O}} \vline & 
(\iu c'' \ot M_N' \overline{M_n'})^T & {}~~\iu \pi_5(a) \ot \Men^2 & 0 \\ 
& & & 0 & 0 & 0 
} $}},
    }
where
\seq{
\eqa{rcl}{
{}\!\begin{array}{r} \Mud^2 \!{} \\ \Men^2 \!{} \end{array} & 
\begin{array}{c} := \\ := \end{array} & 
\begin{array}{l} 
\! (M_u' M_u'{}^* {-} M_d' M_d'{}^*) -\tfrac{1}{24} \tr(M_u' M_u'{}^* 
{-} M_d' M_d'{}^* {+} \bar{M}_e' M_e'{}^T {-} M_n'{}^* M_n') \one_6 \,, \\
\! (\bar{M}_e' M_e'{}^T {-} M_n'{}^* M_n') -\tfrac{1}{8} 
\tr(M_u' M_u'{}^* {-} M_d' M_d'{}^* {+} \bar{M}_e' M_e'{}^T {-} 
M_n'{}^* M_n') \one_6 \,, \end{array} \hs*{2.8em} \yn \label{mudz} \\  
\t{M}_V^2 &:=& M_V^2 -\tfrac{1}{6} \tr(M_V^2) \one_6~.~~  \label{tMV} \yn 
     }}

\subsection{The Factorization}
\label{fact}

Due to \rf[lcj], the problem of solving \rf[eoft] is equivalent to finding 
for each given $\tau^2 \in \p{2}$ an element $j \in J$ such that  
\eq[trt]{
\tr (\t{j}^{\,*} (\tau^2+j))=0~, \quad \forall \,\t{j} \in J ~.~~  
	}
Since $J$ is block--diagonal, the off--diagonal blocks $\tau_{i,j}$ do not 
contribute to the trace \rf[trt]. Next, in the parts 
$\pi_{10;5}(\iu (b,b)')$ we can (and must) modulo $J_2$ replace 
\eqas{l}{
\hs*{-0.7em} \Mu' \Mu'{}^*{-}M_d' M_d'{}^* \mapsto \Mu' \Mu'{}^* 
{-} M_u' M_u'{}^* = {-} \Mu' \Mn'{}^* {-} \Mn' \Mu'{}^* {-} \Mn' \Mn'{}^* \;, 
\\
\hs*{-0.7em} \bar{M}_e' M_e'{}^T {-} \Mu'{}^* \Mu' \mapsto M_n'{}^* M_n' 
{-} \Mu'{}^* \Mu' = {-} 3 \Mu'{}^* \Mn' {-} 3 \Mn'{}^* \Mu' 
{+} 9 \Mn'{}^* \Mn' \;,
       }
see \rf[mtu]. In the diagonal part \rf[tdev] of $\tau^2$ let us define 
\eqas{rclrcl}{
A^{10} &:=& \th \{\pi_{10}(a), \pi_{10}(a)\} ~,~~ & 
A^5 &:=& \th \{\pi_5(a), \pi_5(a)\} ~,~~ \npb \\
B &:=& -b^* b ~,~~ & 
(b,b)' &:=& b b^* - \tfrac{1}{5} \tr(b b^*) \one_5 ~,~~ \npb \\
U^{10} &:=& - c c^* ~,~~ & \t{U}^{10} &:=& -c \pi_{10,10}(b)^* \\
\t{V}^{10} &:=& - w w^* ~,~~ &
V^{10} &:=& \t{V}^{10} - \iu \pi_{10}(\iu (b,b)') ~,~~ \npb \\
\t{V}^5 &:=& - w^* w ~,~~ &
V^5 &:=& \t{V}^5 + 9 \iu \pi_5(\iu (b,b)') ~,~~ 
\npb \\
\t{W}^{10} &:=& - \pi_{10,5}(b) w^* ~,~~ &
W^{10} &:=& \t{W}^{10} - \iu \pi_{10}(\iu (b,b)') ~,~~ \npb \\
\t{W}^5 &:=& - w^* \pi_{10,5}(b) ~,~~ &
W^5 &:=& \t{W}^5 - 3 \iu \pi_5(\iu (b,b)') ~.~~
         }
It is necessary to split $A^{10},U^{10},\t{U}^{10},V^{10}$ and $W^{10}$ 
according to \rf[irrea] and $A^5,V^5$ and $W^5$ according to \rf[irreb] into 
irreducible components. It turns out that the non--vanishing components are 
\eqas{cccccccccccl}{
A^{10} &=& A^{10}_{\ul1} &\op& A^{10}_{\ul{24}} &\op& A^{10}_{\ul{75}}~, 
\qquad{} & A^5 &=& A^5_{\ul1} &\op& A^5_{\ul{24}} ~, \\
U^{10} &=& U^{10}_{\ul1} &\op& U^{10}_{\ul{24}} &\op& U^{10}_{\ul{75}}~, 
\qquad{} & \t{U}^{10} &=& \t{U}^{10}_{\ul{75}} &\,, \\
V^{10} &=& V^{10}_{\ul1} &\op& V^{10}_{\ul{24}} &\op& V^{10}_{\ul{75}}~, 
\qquad{} & V^5 &=& V^5_{\ul1} &\op& V^5_{\ul{24}} ~, \\
W^{10} &=& W^{10}_{\ul{24}} &\op& W^{10}_{\ul{75}} &\,,&  & 
W^5 &=& W^5_{\ul{24}} &\,.
	     }
For these components we find
\eqa{rclrclrcl}{
A^{10}_{\ul1} &=& \tfrac{3}{10} \, \tr(A^5) \one_{10}~, \qquad{} &
A^5_{\ul1} &=& \tfrac{1}{5} \, \tr(A^5) \one_{5}~, \quad{} & 
A^{10}_{\ul{75}} &=& A^{10}{-}A^{10}_{\ul{24}}{-}A^{10}_{\ul1} ~,  \npb \\
A^5_{\ul{24}} &=& A^5 {-} \tfrac{1}{5} \tr(A^5) \one_{5}~, &&&&
A^{10}_{\ul{24}} &=& {-}\tfrac{1}{3} \iu \pi_{10}(\iu A^5_{\ul{24}}) ~, \\
U^{10}_{\ul1} &=& \tfrac{1}{10} \, \tr(U^{10}) \one_{10}~, \qquad{} &
W^{10}_{\ul{75}} &=& W^{10}{-}W^{10}_{\ul{24}}~, & 
U^{10}_{\ul{75}} &\equiv& U_{10}-U^{10}_{\ul1}-U^{10}_{\ul{24}} ~, 
\hs*{3em} \npb \\
V^{10}_{\ul1} &=& \tfrac{1}{10} \, \tr(\t{V}^5) \one_{10}~, \qquad{} &
V^5_{\ul1} &=& \tfrac{1}{5} \, \tr(\t{V}^5) \one_{5}~, \quad{} &
V^{10}_{\ul{75}} &=& V^{10}{-}V^{10}_{\ul{24}}{-}V^{10}_{\ul1} ~, 
\npb \yn \label{spla} \\
V^5_{\ul{24}} &=& \mc{4}{l}{ \t{V}^5 {-} \tfrac{1}{5} \tr(\t{V}^5) \one_{5} 
{+} 9 \iu \pi_5(\iu (b,b)') ~, } & 
V^{10}_{\ul{24}} &=& {-} \iu \pi_{10}(\iu \t{V}'_{\ul{24}} {+} \iu (b,b)') ~, 
\npb \\
\mc{6}{l}{ W^{10}_{\ul{24}} {+} W^{10}_{\ul{24}}{}^* 
={-} \tfrac{1}{3} \iu \pi_{10}(\iu W^5 {+} \iu W^5{}^* ) ~, } &
\hs*{-6em} W^{10}_{\ul{24}} {-} W^{10}_{\ul{24}}{}^* 
&=& \tfrac{1}{3} \pi_{10}(W^5 {-} W^5{}^* ) ~.
      }
Here, the term $\t{V}_{\ul{24}}' \in \iu \f[a]$ is obtained as follows: We 
decompose $\iu \t{V}_{\ul{24}}' := \tsum_{j=1}^{24} a_j \beta_j\,,$ 
where $\{\beta_j\}_{j=1,\dots,24}$ is an orthonormal basis of $\f[a]\,,$ 
$\tr(\beta_i\beta_j)=-\delta_{ij}\,.$ One finds the equation 
$a_j= -\tfrac{1}{3} \tr(\pi_{10}(\iu \t{V}_{\ul{24}}') \pi_{10}(\beta_j)) 
\equiv -\tfrac{1}{3} \tr(\iu \t{V}^{10} \pi_{10}(\beta_j))\,,$ therefore, 
\eq[tv24]{
\iu \t{V}_{\ul{24}}'= -\tfrac{1}{3} \tsum_{j=1}^{24} 
\tr(\iu \t{V}^{10} \pi_{10}(\beta_j)) \beta_j~.~~
   }
This formula shows the way how to obtain the other formulae of \rf[spla]. One 
can prove
\eqas{rcl}{
\tsum_{j=1}^{24} \tr(\iu A^{10} \pi_{10}(\beta_j)) &\equiv&
\tsum_{j=1}^{24} \tr(\iu A^5 \pi_5(\beta_j)) ~,~~ \yn \npb \\
\tsum_{j=1}^{24} \tr(\iu \t{W}^{10} \pi_{10}(\beta_j)) &\equiv& 
\tsum_{j=1}^{24} \tr(\iu \t{W}^5 \pi_5(\beta_j)) ~.~~
	  }
However, there is no such simple relation between the 10-- and 5--components of 
the $\t{V}$--part. 

Due to \rf[MVa] we can modulo $J_2$ replace $A^{10}_{\ul{75}} \ot 
M_{10}'{\!}^2$ by 
\eq{
A^{10}_{\ul{75}} \ot (2 \Mn' \Mu'{}^* + 2 \Mu' \Mn'{}^* - 4 \Mn' \Mn'{}^* 
- \th M_N' M_N'{}^*)~.~~
   }
Now we add to $\tau^2$ the element $j_0 \in J_0$ given by 
\seq{
\eqa{rcl}{
\alpha &=& \tr(A^5) \alpha_A + B \alpha_B + \tr(U^{10}) \alpha_U 
+ \tr(V^5) \alpha_V~,~~\npb \\
\iu a &=& A^5_{\ul{24}} \beta_A - \iu \pi_{10}^{-1}( \iu U^{10}_{\ul{24}}) 
\beta_U + V^5_{\ul{24}} \check{\beta}_V 
- \iu \pi_{10}^{-1} (\iu V^{10}_{\ul{24}}) \beta_V \npb \\ 
&& + (W^5 {+} W^5{}^*) \beta_W + \iu (W^5 {-} W^5{}^*) \beta_W' ~,  \yn 
\\ 
\iu v &=& (V^{10}_{\ul{75}} - 4 A^{10}_{\ul{75}}) \gamma_V 
+ (U^{10}_{\ul{75}} - \th A^{10}_{\ul{75}}) \gamma_U 
+ (\t{U}^{10}_{\ul{75}} + \t{U}^{10}_{\ul{75}}{}^*) \t{\gamma}_U 
+ \iu (\t{U}^{10}_{\ul{75}} - \t{U}^{10}_{\ul{75}}{}^*) \t{\gamma}_U' \npb \\ 
&& + (W^{10}_{\ul{75}}+W^{10}_{\ul{75}}{}^* + 4 A^{10}_{\ul{75}}) \gamma_W 
+ \iu (W^{10}_{\ul{75}} - W^{10}_{\ul{75}}{}^*) \gamma_W' ~.~~ 
    } 
Moreover, we add the element $j_2 \in J_2'$ given by 
\eqa{rcl}{ 
\iu a &=& A^5_{\ul{24}} \delta_A 
-\iu \pi_{10}^{-1}(\iu U^{10}_{\ul{24}}) \delta_U 
+ V^5_{\ul{24}} \check{\epsilon}_V 
- \iu \pi_{10}^{-1} (\iu V^{10}_{\ul{24}}) \delta_V \npb \\ && 
+ (W^5 {+} W^5{}^*) \delta_W +	\iu (W^5 {-} W^5{}^*) \delta_W'~, \npb \yn
\\ 
\iu v &=& (V^{10}_{\ul{75}} - 4 A^{10}_{\ul{75}}) \epsilon_V 
+ (U^{10}_{\ul{75}} - \th A^{10}_{\ul{75}}) \epsilon_U 
+ (U^{10}_{\ul{75}} + U^{10}_{\ul{75}}{}^*) \t{\epsilon}_U 
+ \iu (U^{10}_{\ul{75}} - U^{10}_{\ul{75}}{}^*) \t{\epsilon}_U'  \npb \\ && 
+  (W^{10}_{\ul{75}}+W^{10}_{\ul{75}}{}^* + 4 A^{10}_{\ul{75}}) \epsilon_W 
+  \iu (W^{10}_{\ul{75}} - W^{10}_{\ul{75}}{}^*) \epsilon_W' ~,~~ 
    } 
and the element $j_3 \in J_3$ determined by 
\eq{ 
\nu= \tr(A^5) \zeta_A + B \zeta_B + \tr(U^{10}) \zeta_U + \tr(V^5) \zeta_V~.~~ 
   } 
   } 
As result, the matrix elements $\h{\tau}_{10},\h{\tau}_5,\h{\tau}_1$ of 
$\h{\tau}^2=\tau^2+j_0+j_2'+j_3$ take the form 
\eqa{rcl}{ 
\h{\tau}_{10} &=& \tr(A^5) \one_{10} \ot \h{M}^{10}_{aa} 
+ \tr(U^{10}) \one_{10} \ot \h{M}^{10}_{cc} + \tr(V^5) \one_{10} 
\ot \h{M}^{10}_{nn} + B \one_{10} \ot \h{M}^{10}_{bb} \npb \\ 
&-& \tfrac{1}{3} \iu \pi_{10}(\iu A^5_{\ul{24}}) \ot M^{10}_{aa} 
+ U^{10}_{\ul{24}} \ot M^{10}_{cc} + V^{10}_{\ul{24}} \ot M^{10}_{nn} 
-\iu \pi_{10}(\iu V^5_{\ul{24}}) \ot \check{M}^{10}_{nn} \npb \\ && 
- \tfrac{1}{3} \iu \pi_{10}(\iu W^5 + \iu W^5{}^*) \ot M^{10}_{\{un\}} 
+ \tfrac{1}{3} \iu \pi_{10} (W^5 - W^5{}^*) \ot M^{10}_{[un]} \npb \\ 
&+& (V^{10}_{\ul{75}} - 4 A^{10}_{\ul{75}}) \ot \t{M}^{10}_{nn} 
+ (U^{10}_{\ul{75}} - \th A^{10}_{\ul{75}}) \ot \t{M}^{10}_{cc} 
+ (\t{U}^{10}_{\ul{75}} + \t{U}^{10}_{\ul{75}}) \ot \t{M}^{10}_{\{cd\}} \npb \\ 
&& + \iu (\t{U}^{10}_{\ul{75}} - \t{U}^{10}_{\ul{75}}) \ot \t{M}^{10}_{[cd]} + 
(W^{10}_{\ul{75}} + W^{10}_{\ul{75}}{}^* + 4 A^{10}_{\ul{75}}) \ot 
\t{M}^{10}_{\{un\}} \npb \\ && 
+ \iu (W^{10}_{\ul{75}} - W^{10}_{\ul{75}}{}^*) \ot \t{M}_{[un]}^{10} 
\yn \label{ht1051} 
\\ 
\h{\tau}_5 &=& \tr(A^5) \one_5 \ot \h{M}^5_{aa} 
+ \tr(U^{10}) \one_5 \ot \h{M}^5_{cc} + \tr(V^5) \one_5 \ot \h{M}^5_{nn} 
+ B \one_5 \ot \h{M}_{bb}^5 \npb \\ 
&+& A^5_{\ul{24}} \ot M^5_{aa} + V^5_{\ul{24}} \ot \check{M}_{nn}^5 
- \iu \pi_{10}^{-1}(\iu V^{10}_{\ul{24}}) \ot M_{nn}^5 
- \iu \pi_{10}^{-1}(\iu U^{10}_{\ul{24}}) \ot M_{cc}^5 \npb \\ && 
+ (W^5 + W^5{}^*) \ot M_{\{un\}}^5 + \iu (W^5 - W^5{}^*) \ot M^5_{[un]} \npb 
\\ 
\h{\tau}_1 &=& \tr(A^5) \ot \h{M}_{aa}^1 + \tr(U^{10}) \ot \h{M}_{cc}^1 + B \ot 
\h{M}_{bb}^1 + \tr(V^5) \ot \h{M}_{nn}^1 ~,~~ 
    } 
where the matrices $M_{ij}^k,\t{M}_{ij}^k,\h{M}_{ij}^k$ and $\check{M}_{ij}^k$ 
are given in Appendix~\ref{appa}.

The last step before including the function algebra is to apply the map 
$\h{\sigma} \circ \h{\pi}^{-1}$ defined in \rf[hsig] to elements 
$\tau^1 \in \p{1}\,.$ Calculating $\h{\sigma} \circ \h{\pi}^{-1} (\tau^1)$ 
means to calculate $j_2$ in \rf[j2], however with the r.h.s.\ of \rf[j3] equal 
to the given element $\tau^1$ and not equal to zero. We have listed the matrix 
elements of $\M^2$ in \rf[Msq]. Again, terms in $\M^2$ proportional to the 
identities $\one_{10},\one_{5},1$ do not contribute to $j_2\,.$ Next, the terms 
proportional to $- \iu v_0\,,$ $- \iu \pi_{10;5}(\frac{m}{2}+I_3)$ and 
$-\iu n''$ contribute to the ideal $\jj{2}\,,$ as explained above. Since we 
regard $\h{\sigma} \circ \h{\pi}^{-1} (\tau^1)$ modulo $\jj{2}\,,$ it is not 
necessary to consider these terms. Therefore, there remain only the terms 
\seq{
\label{dme}
\eqa{l}{
-\iu \pi_{10}(m) \ot \tfrac{1}{3} (\tfrac{1}{5} M_{10}'{\!}^2 - 4 \Mu' \Mn'{}^* 
- 4 \Mn' \Mu'{}^* + 8 \Mn' \Mn'{}^*  + M_N' M_N'{}^*)~, \qquad{} 
\npb \yn \label{dmea} \\ 
-\iu \pi_5(m) \ot (\tfrac{1}{5} M_5'{}^2 - 4 \Mn'{}^* \Mu' - 4 \Mu'{}^* \Mn' 
+ 8 \Mn'{}^* \Mn' )~ \yn \label{dmeb} 
      }}
in the diagonal blocks $(\M^2)_{10}$ and $(\M^2)_{5}$ as 
well as the off--diagonal blocks $(\M^2)_{i,j}\,,$ which give a contribution 
to $\h{\sigma} \circ \h{\pi}^{-1} (\tau^1)\,.$ As we have already noticed, the 
contribution of $(\M^2)_{5,1}$ is $\tfrac{3\iu}{5} \ot M_5'{}^T$ times 
the contribution of $\pi_{5,1}(n) \ot M_e'$ to \rf[j3]. We get analogous 
contributions from the other terms $(\M^2)_{i,j}$ and $(\M^2)_i\,.$ 
Thus, we obtain in the same notations as in \rf[impb] the formula
\eqa{rcl}{
\mc{3}{l}{ \h{\sigma} \circ \h{\pi}^{-1} (\tau^1) = \mb{D{13}|D{13}}{
\sigma_{10} & 0 & 0 & \sigma_{10,10} & \sigma_{10,5} & 0 \\  
0 & \sigma_5^T & 0 & \sigma_{10,5}^T & 0 & \sigma_{5,1} \\  
0 & 0 & 0 & 0 & \sigma_{5,1}^T & 0 \\  \hline 
\sigma_{10,10}^* & \overline{\sigma_{10,5}} & 0 & \sigma_{10}^T & 0 & 0 \\
\sigma_{10,5}^* & 0 & \overline{\sigma_{5,1}} & 0 & \sigma_5 & 0  \\  
0 & \sigma_{5,1}^* & 0 & 0 & 0 & 0 }  ,~~  \mbox{where} ~~ } \yn
\label{ims} \\
\sigma_{10} &=& -\iu \pi_{10}(a) \ot \tfrac{1}{3} (\tfrac{1}{5} M_{10}'{\!}^2 
- 4 \Mu' \Mn'{}^* - 4 \Mn' \Mu'{}^* + 8 \Mn' \Mn'{}^* + M_N' M_N'{}^*)~,~~ 
\npb \\
\sigma_5 &=& -\iu \pi_5(a) \ot (\tfrac{1}{5} M_5'{}^2 - 4 \Mn'{}^* \Mu' 
- 4 \Mu'{}^* \Mn' + 8 \Mn'{}^* \Mn' )~,~~ \\
\sigma_{10,10} &=& \tfrac{3\iu}{5} \pi_{10,10}(b) \ot \th (M_{10}' M_d' 
{+} M_d' M_{10}'{\!}^T) \npb \\ && 
+ \tfrac{\iu}{2} \pi_{10,10}(w) \ot \th (M_{10}' M_d' {-} M_d' M_{10}'{\!}^T) 
- \tfrac{12\iu}{5} c \ot \th (M_{10}' M_N' + M_N' \overline{M_{10}'}) ~,~~ \\
\sigma_{5,1} &=& \tfrac{3\iu}{5} \pi_{5,1}(b) \ot M_5'{}^T M_e' ~,~~ \\
\sigma_{10,5} &=& \iu \pi_{10,5} (b) \ot (-\tfrac{9}{20} M_{10}' \Mu' 
- \tfrac{3}{20} \Mu' M_5' + \tfrac{3}{4} M_{10}' \Mn' 
- \tfrac{3}{4} \Mn' M_5' ) \npb \\ 
&+& \iu w \ot ( \tfrac{1}{4} M_{10}' \Mu' - \tfrac{1}{4} \Mu' M_5' 
- \tfrac{19}{20} M_{10}' \Mn' + \tfrac{7}{20} \Mn' M_5' ) ~.~~
      }
Now, it remains to perform the factorization in the diagonal blocks \rf[dmea] 
and \rf[dmeb]. The same method as before yields that the representatives 
orthogonal to $J_2' \op (J_0+J_3)$ are
\seq{
\eqa{rcl}{
\rf[dmea] &\mapsto& -\tfrac{1}{3} \iu \pi_{10}(a) \ot 
(\tfrac{1}{5} M^{10}_{aa} - 8 M^{10}_{\{un\}} + 8 M^{10}_{nn} 
+ 24 \check{M}^{10}_{nn} + M^{10}_{cc} )~, \qquad{} \yn \npb \\
\rf[dmeb] &\mapsto& - \iu \pi_5(a) \ot (\tfrac{1}{5} M^5_{aa} 
- 8 M^5_{\{un\}} + \tfrac{8}{3} M^5_{nn} + 8 \check{M}^5_{nn} + 
\tfrac{1}{3} M^5_{cc} )~.~~ \yn
          }}

\section{The Action of the Unification Model}
\label{aum}

\subsection{The Curvature}
\label{cum}

Now we are able to construct the bosonic action of the flipped $\SU5 \times 
\U1$--grand unification model. We choose $X$ to be a four dimensional 
Riemannian spin manifold. When using a local basis $\{\g[\mu]\}_{\mu=1,2,3,4}$ 
of $\Lambda^1$ then the basis elements $\g[\mu]$ are selfadjoint as complex 
sections of the Clifford bundle. Elements of $\Lambda^1$ are locally 
represented by real linear combinations of $\{\g[\mu]\}_{\mu=1,2,3,4}\,.$ The 
grading operator is $\g=\g[1]\g[2]\g[3]\g[4]\,.$ 

The first step is to write down the connection form $\rho\,,$ which has 
according to \rf[gc] the structure 
\eqas[rabw]{l}{
\rho=\pi(A)+\pi(A'')+ \g \pi(H)~,~~  \\ 
A \in \Lambda^1 \ot \su5~, \quad A'' \in \Lambda^1 \ot \u1 ~, \quad
H \in \Lambda^0 \ot \W[a]{1}~.
   }
Here, $\g$ acts componentwise and $\pi=\id \ot \h{\pi}\,,$ where the matrix 
parts of $\pi(A)$ and $\pi(A'')$ are given by \rf[piofa] and \rf[u1], 
respectively. Elements of $\p{1}$ are specified by elements of 
$\f[a],\f[b],\f[c]$ and $\f[w]\,,$ see \rf[tau1]. Thus, we consider $H$ as a 
sum 
\eqas[hjfw]{rcl}{
H &=& \itJ + \itF + \itX + \itY~, \\ 
\itJ &\in& \Lambda^0 \ot \f[a]~, \quad \itF \in \Lambda^0 \ot \f[b]~, \quad 
\itX  \in  \Lambda^0 \ot \f[c]~, \quad \itY \in \Lambda^0 \ot \f[w]~. 
	 }
Inserting \rf[rabw] and \rf[hjfw] into formula \rf[th] for the curvature, we 
find with \rf[ims]
\seq{
\eqas[rmmr]{rcl}{
\theta &=& \bfd \pi(A) + \bfd \pi(A'') + \th \{\pi(A),\pi(A)\}  \\ 
&-& \g \big(\bfd \pi(\itJ) +\bfd \pi(\itF) + \bfd \pi(\itX) + \bfd \pi(\itY) 
\\ && 
+ [\pi(A){+}\pi(A''), \pi(\itJ) {+} \pi(\itF) {+} \pi(\itX) {+} \pi(\itY) {-} 
\iu \M ] \big)	\\ 
&+& \big( \th \{\pi(\itJ) {+} \pi(\itF) {+} \pi(\itX) {+} \pi(\itY) ,
\pi(\itJ) {+} \pi(\itF) {+} \pi(\itX) {+} \pi(\itY) \} \\ && 
+ \{\pi(\itJ) {+} \pi(\itF) {+} \pi(\itX) {+} \pi(\itY) ,-\iu \M\} 
+ \hsg (\rho) \mod  \Lambda^0 \ot \cj{2} \big) \,, 
    }
where
\eqa{rcl}{
\hsg (\rho) &:=& -\tfrac{12\iu}{5} \pi(\itX \!\ot\! \th (M_{10}' M_N' {+} M_N' 
M_{10}'{\!}^T )) + \tfrac{\iu}{2} \pi(\pi_{10,10}(\itY) \!\ot\! \th 
(M_{10}' M_d' {-} M_d' M_{10}'{\!}^T)) \npb \\ 
&+& \tfrac{3\iu}{5} \pi(\pi_{10,10}(\itF) \ot \th ( M_{10}' M_d' 
{+} M_d' M_{10}'{\!}^T) 
+ \tfrac{3\iu}{5} \pi(\pi_{5,1}(\itF) \ot M_5'{}^T M_e') \npb \\ 
&-& \iu \pi( \pi_{10,5}(\itF) \ot (\tfrac{9}{20} M_{10}' \Mu' 
+ \tfrac{3}{20} \Mu' M_5' - \tfrac{3}{4} M_{10}' \Mn' 
+ \tfrac{3}{4} \Mn' M_5')) \npb \\ 
&-& \iu \pi(\pi_{10,5}(\itY) \ot (-\tfrac{1}{4} M_{10}' \Mu' 
+ \tfrac{1}{4} \Mu' M_5' + \tfrac{19}{20} M_{10}' \Mn' 
- \tfrac{7}{20} \Mn' M_5')) \yn \label{srho} \npb \\ 
&-& \tfrac{1}{3} \iu \pi( \pi_{10}(\itJ) \ot ( \tfrac{1}{5} M^{10}_{aa} 
- 8 M^{10}_{\{un\}} + 8 M^{10}_{nn} + 24 \check{M}^{10}_{nn} 
+ M^{10}_{cc} )) \npb \\
&-& \iu \pi(\pi_5(\itJ) \ot ( \tfrac{1}{5} M^5_{aa} - 8 M^5_{\{un\}} 
+ \tfrac{8}{3} M^5_{nn} + 8 \check{M}^5_{nn} + \tfrac{1}{3} M^5_{cc} )) ~.
    }}
Here we have denoted by $\pi$ the embedding of the selected matrix elements of 
\rf[ims] into the matrix \rf[ims]. We have
\seq{
\eqa{rl}{
\th \{ & \pi(\itJ){+}\pi(\itF) {+} \pi(\itX) {+} \pi(\itY) , 
\pi(\itJ){+}\pi(\itF) {+} \pi(\itX) {+} \pi(\itY) \} \\ &
{+} \{\pi(\itJ) {+}\pi(\itF) {+} \pi(\itX) {+}\pi(\itY) ,{-}\iu \M\} \\ &
=\th \{\pi(\ittJ){+}\pi(\ittF) {+} \pi(\ittX) {+} \pi(\ittY) , 
\pi(\ittJ){+}\pi(\ittF){+} \pi(\ittX) {+}\pi(\ittY)\} + \M^2~, \quad{} \yn
      }
where
\al{
\label{ittj}
\ittJ &:=\itJ+m ~,& \ittF &:=\itF+n ~,& \ittX &:=\itX+m'~,& \ittY &:=\itY+n' ~. 
   }
   }
Let 
\eq{
\hsg (\t{\rho}):= \mbox{ formula \rf[srho] with $~\itJ \mapsto \ittJ~,~~
\itF \mapsto \ittF~,~~\itX \mapsto \ittX~,~~\itY \mapsto \ittY$}~.
   }
Then we obtain from \rf[rmmr] and \rf[Msq]
\eqa{rcl}{
\theta &=& \bfd \pi(A) + \th \{\pi(A),\pi(A)\} \yn \label{rmr} \npb \\ 
&-& \g (\bfd \pi(\ittJ) +\bfd \pi(\ittF) + \bfd \pi(\ittY) + \bfd \pi(\ittX) \\
&+& [\pi(A)+\pi(A''), \pi(\ittJ) +\pi(\ittF)+\pi(\ittX)+\pi(\ittY)]) \npb \\ 
&+& \big( \th \{\pi(\ittJ)+\pi(\ittF) + \pi(\ittX) + \pi(\ittY) , 
\pi(\ittJ)+\pi(\ittF) + \pi(\ittX) + \pi(\ittY) \} + \hsg (\t{\rho}) \\ 
&+& \diag \big( \begin{array}[t]{l} 
	\one_{10} \ot ( \tfrac{6}{5}  \h{M}^{10}_{aa} {+} \h{M}^{10}_{bb}
	{+} 12 \h{M}^{10}_{nn} {+} \h{M}^{10}_{cc} )\,,~ 
	\one_{5} \ot ( \tfrac{6}{5} \h{M}^5_{aa} {+} \h{M}^5_{bb}
	{+} 12 \h{M}^5_{nn} {+} \h{M}^5_{cc} )^T \,, \npb \\
	\tfrac{6}{5} \h{M}^1_{aa} {+} \h{M}^1_{bb} {+} 12 \h{M}^1_{nn} 
	{+} \h{M}^1_{cc} \,,~
	\one_{10} \ot ( \tfrac{6}{5}  \h{M}^{10}_{aa} {+} \h{M}^{10}_{bb}
	{+} 12 \h{M}^{10}_{nn} {+} \h{M}^{10}_{cc} )^T\,, \npb \\
	\one_{5} \ot ( \tfrac{6}{5} \h{M}^5_{aa} {+} \h{M}^5_{bb}
	{+} 12 \h{M}^5_{nn} {+} \h{M}^5_{cc} ) \,,~ 
	(\tfrac{6}{5} \h{M}^1_{aa} {+} \h{M}^1_{bb} {+} 12 \h{M}^1_{nn} 
	{+} \h{M}^1_{cc} )^T~
	\big) \end{array}  \npb \\
&& \hs*{5em}  \mod \Lambda^0 \ot \cj{2} \big) \,. 
      }
We define 
\eqa{rclrcl}{
\cittF &:=& \pi_{10,5}(\ittF)\;, \qquad{} & 
\hittF &:=& \pi_{10,10}(\ittF)\;, \npb \\
(\ittF, \ittF)' &:=& \ittF \ittF^* - \tfrac{1}{5} \tr(\ittF \ittF^*) \one_5~, 
\qquad{} & (\ittX \ittX^*)' &:=& 
- \iu \pi_{10}^{-1}(\iu (\ittX \ittX^*)_{\ul{24}}) 
{}~. 
        }
Using \rf[ht1051] and \rf[tau2] we obtain the following matrix representation 
of $\f[e](\theta)$:
\seq{
\label{thta1}
\eqa{rcl}{
\mc{3}{c}{
\f[e](\theta)= \mb{D{12}|D{12}}{
\theta_{10} & \theta_{\widetilde{10,5}} & \theta_{10,1} & \theta_{10,10} & 
\theta_{10,5} & 0 \\  
\theta_{\widetilde{10,5}}^*  & \theta_5^T & 0 & \theta_{10,5}^T & 0 & 
\theta_{5,1} \\  
\theta_{10,1}^* & 0 & \theta_1 & 0 & \theta_{5,1}^T & 0 \\  \hline 
\theta_{10,10}^* & \overline{\theta_{10,5}} & 0 & \theta_{10}^T & 
\overline{\theta_{\widetilde{10,5}}} & \overline{\theta_{10,1}} \\  
\theta_{10,5}^* & 0 & \overline{\theta_{5,1}} & \theta_{\widetilde{10,5}}^T & 
\theta_5 & 0  \\  
0 & \theta_{5,1}^* & 0 & \theta_{10,1}^T & 0 & \theta_1^T } ~,~~ 
\mbox{where}~~  \yn 
\label{thta} }	  
\eqnskip \eqnskip 
\\
\theta_{10} &=& \pi_{10}(\bfd A + \th \{A,A\}) \ot \one_6 
- \th \bfd A'' \one_{10} \ot \one_6 - \g \pi_{10}(\bfd \ittJ +[A,\ittJ]) 
\ot M_{10}' \npb \\ 
&+& (\tfrac{6}{5} + \tr(\ittJ^2) ) \one_{10} \ot \h{M}^{10}_{aa} 
+ (1-\ittF^* \ittF) \one_{10} \ot \h{M}^{10}_{bb}  \npb \yn \\
&+& (1-\tr( \ittX \ittX^* )) \one_{10} \ot \h{M}^{10}_{cc} 
+ (12-\tr( \ittY \ittY^* )) \one_{10} \ot \h{M}^{10}_{nn} \npb \\
&-& \tfrac{1}{3} \iu \pi_{10}(\iu (\ittJ^2 - \tfrac{1}{5} \tr(\ittJ^2) 
\one_5 - \tfrac{1}{5} \iu \ittJ )) \ot M^{10}_{aa} 
+ \iu \pi_{10}(\iu (\ittX \ittX^*)' - \tfrac{1}{3} \ittJ) \ot M^{10}_{cc} 
\npb \\
&+& \iu \pi_{10}(\iu ((\ittY \ittY^* )' + \tfrac{8}{3}\iu \ittJ 
- (\ittF, \ittF)')) \ot M^{10}_{nn} \npb \\ 
&+& \iu \pi_{10}(\iu (\ittY^* \ittY - \tfrac{1}{5} \tr (\ittY^* \ittY) \one_5
+ 8 \iu \ittJ + 9 (\ittF, \ittF)')) \ot \check{M}^{10}_{nn} \\ 
&+& \tfrac{1}{3} \iu \pi_{10}(\iu (\ittY^* \cittF 
\!+\! \cittF{}^* \ittY \!-\! 8 \iu \ittJ \!-\! 6 (\ittF, \ittF)')) 
\ot M^{10}_{\{un\}} \!-\! \tfrac{1}{3} \iu \pi_{10}(\ittY^* \cittF 
\!-\! \cittF{}^* \ittY) \ot M^{10}_{[un]} \\ 
&-& \big( \ittX \ittX^* + \th (\pi_{10}(\ittJ))^2-\tfrac{1}{10} 
( \tr(\ittX \ittX^*) + \tfrac{3}{2} \tr(\ittJ^2)) \one_{10} \npb \\ && 
+ \iu \pi_{10}( \iu (\ittX \ittX^* )' + \tfrac{1}{6} \iu( \ittJ^2 
- \tfrac{1}{5} \tr(\ittJ^2) \one_5 ) ) \big) \ot \t{M}^{10}_{cc} \\ 
&-& \big( \ittX \hittF^* + \hittF \ittX^* \big) \ot \t{M}^{10}_{\{cd\}} 
- \iu \big( \ittX \hittF^* - \hittF \ittX^* \big) \ot \t{M}^{10}_{[cd]} \\ 
&-& \big( \ittY \ittY^* + 4 (\pi_{10}(\ittJ))^2 -\tfrac{1}{10} 
\tr(\ittY^* \ittY +12 \ittJ^2) \one_{10} \npb \\ && 
+\iu \pi_{10}(\iu ((\ittY \ittY^* )'+ \tfrac{4}{3} \ittJ^2 
- \tfrac{4}{15} \tr(\ittJ^2) \one_5 ) ) \big) \ot \t{M}^{10}_{nn} \npb \\ 
&-& \big( \cittF \ittY^* + \ittY \cittF{}^* 
- 4 (\pi_{10}(\ittJ))^2 + \tfrac{6}{5} \tr(\ittJ^2) \one_{10} \npb \\ && 
+\tfrac{1}{3} \iu  \pi_{10}(\iu (\ittY^* \cittF + \cittF{}^* \ittY 
- 4 \ittJ^2 + \tfrac{4}{5} \tr(\ittJ^2) \one_5 )) \big) 
\ot \t{M}^{10}_{\{un\}} \npb \\ 
&-& \iu \big( \cittF \ittY^* - \ittY \cittF{}^* 
-\tfrac{1}{3} \pi_{10}(\ittY^* \cittF - \cittF{}^* \ittY) \big) 
\ot \t{M}^{10}_{[un]} ~,
\eqnskip \eqnskip \\
\theta_5 &=& \pi_5(\bfd A + \th \{A,A\}) \ot \one_6 
- \tfrac{3}{2} \bfd A'' \one_{5} \ot \one_6 + \g \pi_5 (\bfd \ittJ 
+[A,\ittJ]) \ot M_5' \yn \npb \\ 
&+& (\tfrac{6}{5} + \tr(\ittJ^2) ) \one_5 \ot \h{M}^5_{aa}
+ (1- \ittF^* \ittF) \one_{5} \ot \h{M}^5_{bb} \npb \\
&+& (1-\tr( \ittX \ittX^* )) \one_5 \ot \h{M}^5_{cc} 
+ (12-\tr( \ittY \ittY^* )) \one_{10} \ot \h{M}^{10}_{nn} \\
&-& \iu \pi_5(\iu (\ittJ^2 \!-\! \tfrac{1}{5} \tr(\ittJ^2) \one_5 
- \tfrac{1}{5} \iu \ittJ )) \ot M^5_{aa} 
+ \iu \pi_5( \iu (\ittX \ittX^*)' - \tfrac{1}{3} \ittJ) \ot M^5_{cc} \npb \\
&+& \iu \pi_5(\iu ( (\ittY \ittY^* )' \!+\! \tfrac{8}{3} \iu \ittJ 
- (\ittF, \ittF)')) \!\ot\! M^5_{nn} \npb \\ 
&+& \iu \pi_5(\iu (\ittY^* \ittY - \tfrac{1}{5} \tr (\ittY^* \ittY) \one_5
+8 \iu \ittJ + 9 (\ittF, \ittF)')) \ot \check{M}^5_{nn} \npb \\ 
&+& \iu \pi_5(\iu ( \ittY^* \cittF + \cittF{}^* \ittY -8\iu \ittJ 
- 6 (\ittF, \ittF)')) \ot M^5_{\{un\}} 
- \iu \pi_5( \ittY^* \cittF - \cittF{}^* \ittY ) \ot M^5_{[un]}~,
\eqnskip \eqnskip \\
\theta_1 &=& -\tfrac{5}{2} \bfd A'' \one_6 
+ (\tfrac{6}{5} + \tr(\ittJ^2) ) \ot \h{M}^1_{aa} 
+ (1-\ittF^* \ittF) \ot \h{M}^1_{bb} \yn \npb \\ 
&+& (1-\tr( \ittX \ittX^* )) \ot \h{M}^1_{cc} 
+ \tr(12 -\tr(\itY \itY^* )) \ot M^1_{nn}  \;, 
\eqnskip \eqnskip \\
\theta_{10,10} &=& -\g \pi_{10,10}(\bfd \ittF 
+ (A {+} A'' \one_5) \ittF) \ot M_d' \yn \npb \\ 
&-& \g (\bfd \ittX {+} \pi_{10}(A) \ittX {+} \ittX \pi_{10}(A)^T {-} A'' \ittX) 
\ot M_N' \\
&+& \pi_{10,10} ( \ittJ \ittF - \tfrac{3\iu}{5} \ittF) 
\ot \th (M_{10}' M_d' + M_d' M_{10}'{\!}^T) \npb \\
&+& (\pi_{10}(\ittJ) \pi_{10,10}(\ittF) - \pi_{10,10}(\ittF) \pi_{10}(\ittJ)^T 
+ \tfrac{\iu}{2} \pi_{10,10}(\ittY)) \ot 
\th (M_{10}' M_d' {-} M_d' M_{10}'{\!}^T) 
\\
&+& ( \pi_{10}(\ittJ) \ittX + \ittX \pi_{10}(\ittJ)^T - \tfrac{12\iu}{5} \ittX)
 \ot \th (M_{10}' M_N' + M_N' M_{10}'{\!}^T) \npb \\
&+& ( \pi_{10}(\ittJ) \ittX - \ittX \pi_{10}(\ittJ)^T )
 \ot \th (M_{10}' M_N' - M_N' M_{10}'{\!}^T ) \;, 
\eqnskip 
\\ 
\theta_{10,5} &=& -\g \pi_{10,5}(\bfd \ittF + (A{+}A''\one_5) \ittF ) \ot \Mu' 
\npb \yn \\
&-& \g (\bfd \ittY + \pi_{10}(A) \ittY - \ittY \pi_5(A) + A''\ittY )) \ot \Mn' 
\npb \\ 
&+& (\pi_{10}(\ittJ) \cittF - \tfrac{9\iu}{20} \cittF 
+ \tfrac{\iu}{4} \ittY) \ot M_{10}' \Mu' 
- (\cittF \pi_5(\ittJ) + \tfrac{3\iu}{20} \cittF 
+ \tfrac{\iu}{4} \ittY) \ot \Mu' M_5' \npb \\ 
&+& (\pi_{10}(\ittJ) \ittY + \tfrac{3\iu}{4} \cittF 
- \tfrac{19\iu}{20} \ittY) \ot M_{10}' \Mn' 
- ( \ittY \pi_5(\ittJ) + \tfrac{3\iu}{4} \cittF 
- \tfrac{7\iu}{20} \ittY) \ot \Mn' M_5' ~,
\eqnskip
\\ 
\theta_{5,1} &=& - \g \pi_{5,1}(\bfd \ittF + (A{+}A''\one_5) \ittF ) \ot M_e' 
+ \pi_{5,1}(\ittJ \ittF  -\tfrac{3\iu}{5} \ittF) \ot M_5'{}^T M_e' ~,  \yn 
\\
\theta_{10,1} &=& -\ittY \ittF \ot \Mn' \bar{M}_e'~, \yn
\\
\theta_{\widetilde{10,5}} &=& -(\ittY^* \hittF)_{\ul{10}}^T 
\ot M_d' \bar{M}_{\t{n}}' - (\ittY^* \ittX)_{\ul{10}}^T 
\ot M_N' \bar{M}_{\t{n}}'  \yn \npb \\ 
&-& (\ittY ^*\hittF)_{\ul{40}}^T \ot M_{d\t{n}}' 
- (\tfrac{1}{4} (\ittY^* \ittX)_{\ul{40}} + \tfrac{3}{4} 
(\cittF \ittX) )^T \ot M_{Nu}' ~. 
}}

\subsection{The Bosonic Action}
\label{baum}

It is convenient to put 
\eqas[itbf]{rclrclrcl}{
\bftJ &:=& -\iu \ittJ\;, \qquad{} & \cbftJ &:=& -\iu \pi_{10}(\ittJ)\;, 
\qquad{} & \bftY &:=& -\iu \ittY\;, \\
\hbftY &:=& -\iu \pi_{10,10}(\ittY)\;, \qquad{} &
\bftF &:=& -\iu \ittF\;, \qquad{} & \cbftF &:=& -\iu \pi_{10,5}(\ittF) \;, \\
\hbftF &:=& -\iu \pi_{10,10}(\ittF)\;, \qquad{} & 
\bftX &:=& -\iu \ittX\;, \qquad{} & \check{A} &:=& \pi_{10}(A)\;.
              }
It turns out that the computation of the bosonic action is not difficult now. 
The only problem is the length. All what one needs are the orthogonality 
of different irreducible representations and the relations 
\eqas[AtA]{rcl}{
\tr(\pi_{10}(a) \pi_{10}(\t{a})) &=& 3 \tr(\pi_5(a) \pi_5(\t{a})) 
=3 \tr(a\t{a})~,~~ \\
\mc{3}{l}{\tr \big(( A-\tfrac{1}{10} \tr(A) \one_{10}- A_{\ul{24}})
( \t{A}-\tfrac{1}{10} \tr(\t{A}) \one_{10}- \t{A}_{\ul{24}})\big) } \\
&=& \tr(A\t{A})-\tfrac{1}{10} \tr(A) \,\tr(\t{A}) 
- \tr(A_{\ul{24}} \t{A}_{\ul{24}})~,~~ 
       }
for $a,\t{a} \in \f[a]$ and skew--adjoint $A,\t{A} \in \mat{10}\,.$ 
We compute the Lagrangian $\Lgr = \tfrac{1}{192\,g_0^2}\,
\tr_c ((\f[e](\theta))^2)\,,$ where $g_0$ is a coupling constant 
and $\tr_c$ the combination of the trace over the matrix structure with the 
trace in the Clifford algebra. For functions $f \in C^\infty(X)$ we have 
$\tr_c(f)=4f\,.$ We find: 
\seq{
\label{lagw}
\eqa{rcl}{
\mc{3}{l}{
\tfrac{1}{192\,g_0^2}\,\tr_c ((\f[e](\theta))^2) 
= \Lgr[2] + \Lgr[1] + \Lgr[0] ~,~~ \yn }
\\
\Lgr[2] &=& \tfrac{1}{4\, g_0^2} \, \tr_c ((\bfd A+ \th \{A,A\} )^2) 
+ \tfrac{5}{4\, g_0^2} \, \tr_c ((\bfd A'')^2) ~,~~ \yn \label{L2w} 
\\
\Lgr[1] &=& \tfrac{1}{g_0^2} \mu_0\, \tr_c ((\bfd \bftJ+ [A,\bftJ])^2)
	  \yn \label{L1w} \npb \\ 
&+& \tfrac{1}{g_0^2} \mu_1\, \tr_c ((\bfd \bftF + (A + A'' \one_5)\bftF)^* 
(\bfd \bftF + (A + A'' \one_5)\bftF)) \npb \\
&+& \tfrac{1}{g_0^2} \mu_2\, \tr_c ((\bfd \bftY + \check{A} \bftY - \bftY A 
+ A''\bftY)^* (\bfd \bftY + \check{A} \bftY - \bftY A + A'' \bftY)) \npb \\
&+& \tfrac{1}{g_0^2} \mu_3\, \tr_c ((\bfd \bftX + \check{A} \bftX 
+ \bftX \check{A}^T - A''\bftX)^* (\bfd \bftX + \check{A} \bftX 
+ \bftX \check{A}^T - A''\bftX)) \,,
\\
\Lgr[0] &=&  \tfrac{1}{24\, g_0^2} \big\{ \la[a] (\tr(\bftJ^2) -\tfrac{6}{5})^2 
+ \la[b] (\bftF^*\bftF -1)^2 + \la[c] (\tr(\bftY^* \bftY) -12)^2 
\yn \label{L0w} \npb \\
&+& \la[d] (\tr(\bftJ^2) -\tfrac{6}{5})(\bftF^*\bftF -1) 
+ \la[e] (\tr(\bftJ^2) -\tfrac{6}{5})(\tr(\bftY^* \bftY) -12) \npb \\
&+& \la[f] (\bftF^*\bftF -1)(\tr(\bftY^* \bftY) -12) \eqnskip \\
&+& \lac[a] (\tr(\bftX \bftX^* ) -1)^2
+ \lac[b] (\tr(\bftX \bftX^* ) -1)(\tr(\bftJ^2) -\tfrac{6}{5}) \\
&+& \lac[c] (\tr(\bftX \bftX^* ) -1)(\bftF^*\bftF -1)
+ \lac[d] (\tr(\bftX \bftX^* ) -1)(\tr(\bftY^* \bftY) -12) \eqnskip \\
&+& \lac[e]\, \tr (( \cbftJ \bftX + \bftX \cbftJ^T - \tfrac{12}{5} \bftX) 
( \cbftJ \bftX + \bftX \cbftJ^T - \tfrac{12}{5} \bftX)^*) \eqnskip \\
&+& \lac[f]\, \tr (( \cbftJ \bftX - \bftX \cbftJ^T ) 
( \cbftJ \bftX - \bftX \cbftJ^T)^*) \eqnskip \\
&+&  \lac[g]\, \mathrm{Re} (\tr(( \cbftJ \hbftF - \hbftF \cbftJ^T 
+ \tfrac{1}{2} \hbftY) ( \cbftJ \bftX - \bftX \cbftJ^T )^*)) \eqnskip \\
&+&  \lac[h]\, \mathrm{Im} (\tr(( \cbftJ \hbftF - \hbftF \cbftJ^T 
+ \tfrac{1}{2} \hbftY) ( \cbftJ \bftX - \bftX \cbftJ^T )^*)) \eqnskip 
\\
&+& \la[g]\, \bftF^* (\bftJ - \tfrac{3}{5} \one_5)^2 \bftF 
+  \la[h]\, \tr(( \cbftJ \hbftF - \hbftF \cbftJ^T + \tfrac{1}{2} \hbftY) 
( \cbftJ \hbftF - \hbftF \cbftJ^T + \tfrac{1}{2} \hbftY)^*) \eqnskip \\
&+& \la[i]\, \tr( (\cbftJ \cbftF - \tfrac{9}{20} \cbftF 
+\tfrac{1}{4} \bftY)^* (\cbftJ \cbftF - \tfrac{9}{20} 
\cbftF +\tfrac{1}{4} \bftY) ) \eqnskip \\
&+& \la[j]\, \tr( (\cbftF \bftJ + \tfrac{3}{20} \cbftF +\tfrac{1}{4} \bftY)^* 
(\cbftF \bftJ + \tfrac{3}{20} \cbftF +\tfrac{1}{4} \bftY)) \eqnskip \\
&+& \la[k]\, \tr( (\cbftJ \bftY + \tfrac{3}{4} \cbftF - \tfrac{19}{20} \bftY)^* 
(\cbftJ \bftY + \tfrac{3}{4} \cbftF - \tfrac{19}{20} \bftY) ) \eqnskip \\
&+& \la[l]\, \tr( (\bftY \bftJ + \tfrac{3}{4} \cbftF 
- \tfrac{7}{20} \bftY)^* (\bftY \bftJ + \tfrac{3}{4} 
\cbftF - \tfrac{7}{20} \bftY) ) \eqnskip \\
&-& \la[m]\, \mathrm{Re}( \tr( (\cbftJ \cbftF - \tfrac{9}{20} 
\cbftF +\tfrac{1}{4} \bftY)^* (\cbftF \bftJ + \tfrac{3}{20} 
\cbftF +\tfrac{1}{4} \bftY) )) \eqnskip \\ 
&+& \la[n]\, \mathrm{Re}( \tr( (\cbftJ \cbftF - \tfrac{9}{20} 
\cbftF +\tfrac{1}{4} \bftY)^* (\cbftJ \bftY 
+ \tfrac{3}{4} \cbftF -\tfrac{19}{20} \bftY) )) \eqnskip  \\  
&+& \la[o]\, \mathrm{Im} ( \tr( (\cbftJ \cbftF - \tfrac{9}{20} 
\cbftF +\tfrac{1}{4} \bftY)^* (\cbftJ \bftY 
+ \tfrac{3}{4} \cbftF -\tfrac{19}{20} \bftY))) \eqnskip  \\
&-& \la[p]\, \mathrm{Re} ( \tr( (\cbftJ \cbftF - \tfrac{9}{20} 
\cbftF +\tfrac{1}{4} \bftY)^* ( \bftY \bftJ + \tfrac{3}{4} 
\cbftF -\tfrac{7}{20} \bftY) \npb \\ &&  \hs*{3em} 
+ (\cbftF \bftJ + \tfrac{3}{20} \cbftF 
+\tfrac{1}{4} \bftY)^* (\cbftJ \bftY + \tfrac{3}{4} \cbftF 
-\tfrac{19}{20} \bftY) )) \eqnskip  \\
&-& \la[q]\, \mathrm{Im} ( \tr( (\cbftJ \cbftF - \tfrac{9}{20} 
\cbftF +\tfrac{1}{4} \bftY)^* ( \bftY \bftJ + \tfrac{3}{4} \cbftF 
-\tfrac{7}{20} \bftY) \npb \\ && \hs*{3em}
+ (\cbftF \bftJ + \tfrac{3}{20} \cbftF 
+\tfrac{1}{4} \bftY)^* (\cbftJ \bftY + \tfrac{3}{4} \cbftF 
-\tfrac{19}{20} \bftY) )) \eqnskip  \\
&+& \la[r]\, \mathrm{Re}(\tr( (\cbftF \bftJ +\tfrac{3}{20} \cbftF 
+\tfrac{1}{4} \bftY)^* (\bftY \bftJ + \tfrac{3}{4} \cbftF 
-\tfrac{7}{20} \bftY) )) \eqnskip  \\
&+& \la[s]\, \mathrm{Im}(\tr( (\cbftF \bftJ +\tfrac{3}{20} \cbftF 
+\tfrac{1}{4} \bftY)^* (\bftY \bftJ + \tfrac{3}{4} \cbftF 
-\tfrac{7}{20} \bftY) )) \eqnskip  \\
&-& \la[t]\, \mathrm{Re}( \tr( (\cbftJ \bftY +\tfrac{3}{4} \cbftF 
-\tfrac{19}{20} \bftY)^* (\bftY \bftJ + \tfrac{3}{4} \cbftF 
-\tfrac{7}{20} \bftY) )) \eqnskip \\
&+& \la[u]\, \bftF^* \bftY^* \bftY \bftF 
+ \la[v]\, \tr((\bftY^* \hbftF)_{\ul{10}} (\bftY^* \hbftF)_{\ul{10}}^* ) 
+ \la[w]\, \tr((\bftY^* \hbftF)_{\ul{40}} (\bftY^* \hbftF)_{\ul{40}}^* ) 
\eqnskip \\
&+& \lac[i]\, \tr((\bftY^* \bftX)_{\ul{10}} (\bftY^* \bftX)_{\ul{10}}^*)
+ \lac[l]\, \tr ((\tfrac{1}{4} (\bftY^* \bftX)_{\ul{40}} + \tfrac{3}{4} 
\cbftF \bftX )(\tfrac{1}{4} (\bftY^* \bftX)_{\ul{40}} + \tfrac{3}{4} 
\cbftF \bftX )^*) \eqnskip \\
&+& \lac[j]\, \mathrm{Re}(\tr((\bftY^* \hbftF)_{\ul{10}} 
(\bftY^* \bftX)_{\ul{10}}^*)) 
+ \lac[m]\, \mathrm{Re}(\tr ((\tfrac{1}{4} (\bftY^* \bftX)_{\ul{40}} 
+ \tfrac{3}{4} \cbftF \bftX )(\bftY^* \hbftF)_{\ul{40}}^* )) 
\eqnskip \\
&+& \lac[k]\, \mathrm{Im}(\tr((\bftY^* \hbftF)_{\ul{10}} 
(\bftY^* \bftX)_{\ul{10}}^*)) 
+ \lac[n]\, \mathrm{Im}(\tr ((\tfrac{1}{4} (\bftY^* \bftX)_{\ul{40}} 
+ \tfrac{3}{4} \cbftF \bftX )(\bftY^* \hbftF)_{\ul{40}}^* ))
\smallskip
\\
&+& \lat[a]\, \tr(( \bftJ^2- \tfrac{1}{5} \tr(\bftJ^2) \one_5 
- \tfrac{1}{5} \bftJ )^2) \\
&+& \lat[b]\, \tr((\bftY^* \bftY - \tfrac{1}{5} \tr(\bftY^* \bftY) 
\one_5 - 8 \bftJ + 9 \bftF \bftF^* - \tfrac{9}{5} \bftF^* \bftF \one_5)^2) \\
&+& \lat[c]\, \tr(((\bftY \bftY^* )' - \tfrac{8}{3} \bftJ 
- \bftF \bftF^* + \tfrac{1}{5} \bftF^* \bftF \one_5)^2) \\
&+& \lat[d]\, \tr(( \bftY^* \cbftF {+} \cbftF{}^* \bftY 
{+} 8 \bftJ {-} 6 \bftF \bftF^* {+} \tfrac{6}{5} \bftF^* \bftF \one_5)^2) 
+ \lat[e]\, \tr(- ( \bftY^* \cbftF {-} \cbftF{}^* \bftY )^2) \\
&+& \lat[f]\, \tr(( \bftJ^2 {-} \tfrac{1}{5} \tr(\bftJ^2) \one_5 
{-} \tfrac{1}{5} \bftJ )(\bftY^* \bftY {-} \tfrac{1}{5} \tr(\bftY^* \bftY) 
\one_5 {-} 8 \bftJ {+} 9 \bftF \bftF^* {-} \tfrac{9}{5} \bftF^* \bftF \one_5)) 
\\
&+& \lat[g]\, \tr(( \bftJ^2- \tfrac{1}{5} \tr(\bftJ^2) \one_5 
- \tfrac{1}{5} \bftJ )((\bftY \bftY^* )' - \tfrac{8}{3} \bftJ - \bftF \bftF^* 
+ \tfrac{1}{5} \bftF^* \bftF \one_5)) \\
&+& \lat[h]\, \tr(( \bftJ^2- \tfrac{1}{5} \tr(\bftJ^2) \one_5 
- \tfrac{1}{5} \bftJ )( \bftY^* \cbftF + \cbftF{}^* \bftY 
+ 8 \bftJ - 6 \bftF \bftF^* + \tfrac{6}{5} \bftF^* \bftF \one_5)) \\
&+& \lat[i]\, \iu \,\tr(( \bftJ^2- \tfrac{1}{5} \tr(\bftJ^2) \one_5 
- \tfrac{1}{5} \bftJ )( \bftY^* \cbftF - \cbftF{}^* \bftY )) 
\eqnskip \\
&+& \lat[j]\, \tr((\bftY^* \bftY - \tfrac{1}{5} \tr(\bftY^* \bftY) 
\one_5 - 8 \bftJ + 9 \bftF \bftF^* - \tfrac{9}{5} \bftF^* \bftF \one_5) \times 
\npb \\ && \hs*{2em} 
\times ((\bftY \bftY^* )' - \tfrac{8}{3} \bftJ - \bftF \bftF^* 
+ \tfrac{1}{5} \bftF^* \bftF \one_5)) \eqnskip \\
&+& \lat[k]\, \tr((\bftY^* \bftY - \tfrac{1}{5} \tr(\bftY^* \bftY) 
\one_5 - 8 \bftJ + 9 \bftF \bftF^* - \tfrac{9}{5} \bftF^* \bftF \one_5) \times
\npb \\ && \hs*{2em} 
\times ( \bftY^* \cbftF + \cbftF{}^* \bftY + 8 \bftJ 
- 6 \bftF \bftF^* + \tfrac{6}{5} \bftF^* \bftF \one_5)) \eqnskip \\
&+& \lat[l]\, \iu \, \tr((\bftY^* \bftY - \tfrac{1}{5} \tr(\bftY^* \bftY) 
\one_5 - 8 \bftJ + 9 \bftF \bftF^* - \tfrac{9}{5} \bftF^* \bftF \one_5) 
( \bftY^* \cbftF - \cbftF{}^* \bftY )) \eqnskip \\
&+& \lat[m]\, \tr(((\bftY \bftY^* )' - \tfrac{8}{3} \bftJ - \bftF \bftF^* 
+ \tfrac{1}{5} \bftF^* \bftF \one_5) \times \npb \\ && \hs*{3em}
\times ( \bftY^* \cbftF + \cbftF{}^* \bftY + 8 \bftJ 
- 6 \bftF \bftF^* + \tfrac{6}{5} \bftF^* \bftF \one_5)) \eqnskip \\
&+& \lat[n] \, \iu \, \tr(((\bftY \bftY^* )' - \tfrac{8}{3} \bftJ 
- \bftF \bftF^* + \tfrac{1}{5} \bftF^* \bftF \one_5) 
( \bftY^* \cbftF - \cbftF{}^* \bftY )) \\
&+& \lat[o] \, \iu \, \tr(( \bftY^* \cbftF + \cbftF{}^* \bftY 
+ 8 \bftJ - 6 \bftF \bftF^* + \tfrac{6}{5} \bftF^* \bftF \one_5)
( \bftY^* \cbftF - \cbftF{}^* \bftY )) \eqnskip \\
&+& \lah[a]\, \tr (((\bftX \bftX^*)' - \tfrac{1}{3} \bftJ)^2) 
+ \lah[b]\, \tr (((\bftX \bftX^*)' - \tfrac{1}{3} \bftJ)
( \bftJ^2- \tfrac{1}{5} \tr(\bftJ^2) \one_5 - \tfrac{1}{5} \bftJ )) \\
&+& \lah[c]\, \tr (((\bftX \bftX^*)' - \tfrac{1}{3} \bftJ)
(\bftY^* \bftY - \tfrac{1}{5} \tr(\bftY^* \bftY) 
\one_5 - 8 \bftJ + 9 \bftF \bftF^* - \tfrac{9}{5} \bftF^* \bftF \one_5)) \\
&+& \lah[d]\, \tr (((\bftX \bftX^*)' - \tfrac{1}{3} \bftJ)
((\bftY \bftY^* )' - \tfrac{8}{3} \bftJ - \bftF \bftF^* 
+ \tfrac{1}{5} \bftF^* \bftF \one_5)) \\
&+& \lah[e]\, \tr (((\bftX \bftX^*)' - \tfrac{1}{3} \bftJ)
( \bftY^* \cbftF + \cbftF{}^* \bftY + 8 \bftJ - 6 \bftF \bftF^* 
+ \tfrac{6}{5} \bftF^* \bftF \one_5)) \\
&+& \lah[f]\, \iu \, \tr (((\bftX \bftX^*)' - \tfrac{1}{3} \bftJ)
( \bftY^* \cbftF - \cbftF{}^* \bftY )) \eqnskip \\
&+& \lat[p] \big( \tr((\bftY \bftY^*  - 4 \cbftJ{}^2)^2) - \tfrac{1}{10} 
(\tr(\bftY^* \bftY -12 \bftJ^2))^2 - \npb \\ && \hs*{4em} 
- 3 \tr( ((\bftY \bftY^* )' - \tfrac{4}{3} \bftJ^2 + \tfrac{4}{15} \tr(\bftJ^2) 
\one_5 )^2) \big) \\ 
&+& \lat[q] \big( \tr(( \cbftF \bftY^* + \bftY \cbftF{}^* 
+ 4 \cbftJ{}^2 )^2) -  \tfrac{72}{5} (\tr(\bftJ^2))^2 - 
\npb \\ && \hs*{4em} 
- \tfrac{1}{3} \tr((\bftY^* \cbftF + \cbftF{}^* \bftY + 4 \bftJ^2 
- \tfrac{4}{5} \tr(\bftJ^2) \one_5 )^2) \big) \\
&+& \lat[r] \big( \tr(-( \cbftF \bftY^* - \bftY \cbftF{}^* )^2)
+ \tfrac{1}{3} \tr((\bftY^* \cbftF - \cbftF{}^* \bftY)^2) \big)\\
&+& \lat[s] \big( \tr((\bftY \bftY^* -4 \cbftJ{}^2) 
(\cbftF \bftY^* + \bftY \cbftF{}^* + 4 \cbftJ{}^2)) 
- \tfrac{6}{5} \tr( \bftJ^2) \tr(\bftY^* \bftY - 12 \bftJ^2) - 
\npb \\ && \hs*{1em} 
- \tr(((\bftY \bftY^* )'- \tfrac{4}{3} \bftJ^2 + \tfrac{4}{15} 
\tr(\bftJ^2) \one_5 ) (\bftY^* \cbftF + \cbftF{}^* \bftY + 
4 \bftJ^2 - \tfrac{4}{5} \tr(\bftJ^2) \one_5 )) \big) \\
&+& \lat[t] \,\iu \big( \tr( (\bftY \bftY^* \!- 4 \cbftJ{}^2) 
(\cbftF \bftY^*  \!- \bftY \cbftF{}^* )) 
-  \tr( ((\bftY \bftY^* )'- \tfrac{4}{3} \bftJ^2 )
(\bftY^* \cbftF - \cbftF{}^* \bftY)) \big) 
\\ 
&+& \lat[u] \,\iu \big( \tr( ( \cbftF \bftY^* + \bftY \cbftF{}^* 
+ 4 \cbftJ{}^2 )(\cbftF \bftY^* - \bftY \cbftF{}^* )) -
\npb \\ && \hs*{4em} 
- \tfrac{1}{3} \tr((\bftY^* \cbftF + \cbftF{}^* \bftY + 4 \bftJ^2 )
(\bftY^* \cbftF - \cbftF{}^* \bftY) ) \big) \eqnskip \\
&+& \lah[g] \big( \tr((\bftX \bftX^*  - \th \cbftJ{}^2)^2) - \tfrac{1}{10} 
(\tr(\bftX \bftX^*) - \tfrac{3}{2} \tr(\bftJ^2))^2 - \npb \\ && \hs*{4em} 
- 3 \tr(( (\bftX \bftX^* )' - \tfrac{1}{6} \bftJ^2 + \tfrac{1}{30} 
\tr(\bftJ^2) \one_5 )^2) \big) \eqnskip \\ 
&+& \lah[h] \tr((\bftX \hbftF^* {+} \hbftF \bftX^* )^2) 
- \lah[i] \, \tr((\bftX \hbftF^* {-} \hbftF \bftX^* )^2) \\
&+& \lah[j] \, \tr((\bftX \bftX^*  - \th \cbftJ{}^2)
(\bftX \hbftF^* {+} \hbftF \bftX^* )) \\
&+& \lah[k] \, \iu \, \tr((\bftX \bftX^*  - \th \cbftJ{}^2)
(\bftX \hbftF^* {-} \hbftF \bftX^* )) 
+ \lah[l] \, \iu \, \tr((\bftX \hbftF^* {+} \hbftF \bftX^* )
(\bftX \hbftF^* {-} \hbftF \bftX^* )) \\
&+& \lah[m] \big(\! \tr((\bftX \bftX^* \! {-} \th \cbftJ{}^2)
(\bftY \bftY^* \! {-} 4 \cbftJ{}^2)) 
-\! \tfrac{1}{10} (\tr(\bftX \bftX^*) {-} \tfrac{3}{2} \tr(\bftJ^2)) 
\tr(\bftY^* \bftY {-} 12 \bftJ^2) {-} \npb \\ && \hs*{4em} 
- 3 \tr(((\bftX \bftX^* )' {-} \tfrac{1}{6} \bftJ^2 
{+} \tfrac{1}{30} \tr(\bftJ^2) \one_5 )((\bftY \bftY^* )' 
{-} \tfrac{4}{3} \bftJ^2 + \tfrac{4}{15} \tr(\bftJ^2) \one_5 ) )\big) \eqnskip 
\\
&+& \lah[n] \big( \tr((\bftX \bftX^*  {-} \th \cbftJ{}^2)( \cbftF \bftY^* 
+ \bftY \cbftF{}^* {+} 4 \cbftJ{}^2 )) 
- \tfrac{12}{10} (\tr(\bftX \bftX^*) {-} \tfrac{3}{2} \tr(\bftJ^2)) \, 
\tr(\bftJ^2) - \npb \\ && \hs*{3.5em} 
- \tr(((\bftX \bftX^* )' {-} \tfrac{1}{6} \bftJ^2 
{+} \tfrac{1}{30} \tr(\bftJ^2) \one_5 )(\bftY^* \cbftF {+} \cbftF{}^* \bftY 
{+} 4 \bftJ^2 {-} \tfrac{4}{5} \tr(\bftJ^2) \one_5 ))\big) \eqnskip \\
&+& \lah[o] \big( \iu\, \tr((\bftX \bftX^*  - \th \cbftJ{}^2)
( \cbftF \bftY^* - \bftY \cbftF{}^* )) - \npb \\ && \hs*{4em} 
- \iu \, \tr(((\bftX \bftX^* )' - \tfrac{1}{6} \bftJ^2 
+ \tfrac{1}{30} \tr(\bftJ^2) \one_5 )(\bftY^* \cbftF - \cbftF{}^* \bftY)) \big) 
\\
&+& \lah[p] \, \tr((\bftX \hbftF^* {+} \hbftF \bftX^* )
(\bftY \bftY^*	- 4 \cbftJ{}^2)) \\
&+& \lah[q] \, \tr((\bftX \hbftF^* {+} \hbftF \bftX^* )( \cbftF \bftY^* 
+ \bftY \cbftF{}^* + 4 \cbftJ{}^2 )) \\
&+& \lah[r] \, \iu\, \tr((\bftX \hbftF^* {+} \hbftF \bftX^* )
( \cbftF \bftY^* - \bftY \cbftF{}^* ))
+ \lah[s] \, \iu\,\tr((\bftX \hbftF^* {-} \hbftF \bftX^* )
(\bftY \bftY^*	- 4 \cbftJ{}^2)) \\
&+& \lah[t] \, \iu \tr((\bftX \hbftF^* {-} \hbftF \bftX^* )( \cbftF \bftY^* 
{+} \bftY \cbftF{}^* + 4 \cbftJ{}^2 )) \\
&-& \lah[u] \, \tr((\bftX \hbftF^* {-} \hbftF \bftX^* )
( \cbftF \bftY^* {-} \bftY \cbftF{}^* )) 
\big\} \,,
     }
     }
where the coefficients $\mu^i$ are given in Appendix~\ref{appb}.

The group of local gauge transformations associated to our model is 
\eq{
\mathcal{U}_0 (\f[g]) = \exp(\pi(\CX \!\ot\! (\su5 \!\op\! \u1))) 
\cong \CX \!\ot\! (\SU5 \!\times\! \U1) \;.~~
         }
The Lagrangian \rf[lagw] is invariant under the gauge transformations 
\seq{
\label{gu}
\eqas{rclrcl}{
\gamma_u(A) &=& u_5 \bfd u_5^* + u_5 A u_5^*\,, \qquad{} &
\gamma_u(\check{A}) &=& u_{10} \bfd u_{10}^* + u_{10} \check{A} u_{10}^*\,,~ \\
\gamma_u(A'')   &=& u_1 \bfd u_1^* + A'' \,, \\
\gamma_u(\bftY) &=& u_1 u_{10} \bftY u_5^* ~,~~ &
\gamma_u(\hbftY) &=& u_1^* u_{10} \hbftY u_{10}^T ~,~~ \\
\gamma_u(\bftJ) &=& u_5 \bftJ u_5^*~,~~ &
\gamma_u(\cbftJ) &=& u_{10} \cbftJ u_{10}^*~, \\
\gamma_u(\bftF) &=& u_1 u_5 \bftF ~, &
\gamma_u(\cbftF) &=& u_1 u_{10} \cbftF u_5^* ~,~~ \\
\gamma_u(\hbftF) &=& u_1^* u_{10} \hbftF u_{10}^T ~,~~ &
\gamma_u(\bftX) &=& u_1^* u_{10} \bftX u_{10}^T ~,~~ 
     }
where 
\eqas{ll}{
u_5 = \exp(t_5)~, \qquad u_{10} = \exp(\pi_{10}(t_5)) ~, \qquad{} & 
t_5 \in \CX \ot \su5~,~~ \\ 
u_1 = \exp(t_1)~,~~ & t_1 \in \CX \ot \u1~.~~
       }}

To determine the spontaneous symmetry breaking pattern, we must search for 
a local minimum of the Higgs potential $\Lgr[0]\,.$ This problem is easy to 
solve. We know that, applying the transformation \rf[ittj] in the other 
direction, the $\Lambda^0$--part of the curvature $\f[e](\theta)$ (and hence 
the Higgs potential $\Lgr[0]$) is zero for 
\eqas[minh]{rclrclrclrcl}{
\itJ &=& 0~, \quad{} & \itF &=& 0~, \quad{} & \itX &=& 0~, \quad{} & 
\itY &=& 0~ \quad  \mbox{or} \\
\ittJ &=& m~, & \ittF &=& n~, & \ittX &=& m'~, & \ittY &=& n'~.
         }
Since the Higgs potential $\Lgr[0]$ is not negative as the trace of the square 
of the $\Lambda^0$--part of the selfadjoint matrix $\f[e](\theta)\,,$ the point 
\rf[minh] is a global minimum of $\Lgr[0]\,.$ But \rf[minh] is clearly a local 
minimum as well: In the vicinity of \rf[minh], the $\Lambda^0$--part of 
$\f[e](\theta)$ is linear in the components of $\itJ, \itF, \itX$ and $\itY$ so 
that the Higgs potential $\Lgr[0]$ is in lowest order quadratic in these 
components. 

We underline that, given the fermion masses and the spontaneous symmetry 
breaking pattern as the input, our formalism provides a straightforward 
algorithm to determine the occurring Higgs multiplets and their most general 
gauge invariant Higgs potential. 

\subsection{The Bosonic Lagrangian in Local Coordinates}
\label{bllc}

In this subsection we will write down the Lagrangian \rf[lagw] in terms of 
local coordinates. We must restrict ourselves concerning the Higgs 
potential \rf[L0w] to the terms quadratic in the fields, because the complete 
expansion of $\Lgr[0]$ is too voluminous. Let us introduce in the same way as 
in \rf[itbf] the bold matrices
\eqa{l}{
\bfm:=-\iu \pi_5(m) \equiv \diag(-\tfrac{2}{5},-\tfrac{2}{5},-\tfrac{2}{5}, 
\tfrac{3}{5}, \tfrac{3}{5})~,~~ \npb \\
\check{\bfm}:=-\iu \pi_{10}(m) \equiv  \diag(\tfrac{1}{5},\tfrac{1}{5}, 
\tfrac{1}{5}, \tfrac{1}{5}, \tfrac{1}{5},\tfrac{1}{5},
-\tfrac{4}{5},-\tfrac{4}{5}, -\tfrac{4}{5}, \tfrac{6}{5})~,~~ \yn \label{bfn} 
\eqnskip \\
\bfn':=-\iu n' \equiv \mb{E{12}E{10}E{12}}{
\one_3 & 0_{3 \times 1} & 0_{3 \times 1} \\ 0_{3 \times 3} & 0_{3 \times 1} & 
0_{3 \times 1} \\ 0_{3 \times 3} & 0_{3 \times 1} & 0_{3 \times 1} \\ 
0_{1 \times 3} & 0_{1 \times 1} & 3 } ~,~~  
\check{\bfn}:=-\iu \pi_{10,5}(n) \equiv \mb{E{12}E{10}E{12}}{
\one_3 & 0_{3 \times 1} & 0_{3 \times 1} \\ 0_{3 \times 3} & 0_{3 \times 1} & 
0_{3 \times 1} \\ 0_{3 \times 3} & 0_{3 \times 1} & 0_{3 \times 1} \\ 
0_{1 \times 3} & 0_{1 \times 1} & -1 } , \eqnskip \eqnskip \\ 
\bfn:=\overline{-\iu \pi_{5,1}(n)} \equiv \mb{E{12}}{ 0_{3 \times 1} \\ 1 \\ 
0_{1 \times 1} } ~,~~
\h{\bfn}:=-\iu \pi_{10,10}(n) \equiv \mb{Z{12}Z{12}}{
0_{3 \times 3} & 0_{3 \times 3} & 0_{3 \times 3} & 0_{3 \times 1} \\
0_{3 \times 3} & 0_{3 \times 3} & -\one_3    & 0_{3 \times 1} \\ 
0_{3 \times 3} & -\one_3    & 0_{3 \times 3} & 0_{3 \times 1} \\
0_{1 \times 3} & 0_{1 \times 3} & 0_{1 \times 3} & 0_{1 \times 1} } ,
\eqnskip \\
\bfm':=-\iu m' \equiv \mb{Z{12}Z{12}}{
0_{3 \times 3} & 0_{3 \times 3} & 0_{3 \times 3} & 0_{3 \times 1} \\
0_{3 \times 3} & 0_{3 \times 3} & 0_{3 \times 3} & 0_{3 \times 1} \\ 
0_{3 \times 3} & 0_{3 \times 3} & 0_{3 \times 3} & 0_{3 \times 1} \\
0_{1 \times 3} & 0_{1 \times 3} & 0_{1 \times 3} & -1 }~,~~ 
\eqnskip
\\
\h{\bfn}' :=-\iu \pi_{10,10}(n') \equiv \mb{Z{12}Z{12}}{
0_{3 \times 3} & 0_{3 \times 3} & 0_{3 \times 3} & 0_{3 \times 1} \\
0_{3 \times 3} & 0_{3 \times 3} & 2 \one_3   & 0_{3 \times 1} \\ 
0_{3 \times 3} & - 2 \one_3 & 0_{3 \times 3} & 0_{3 \times 1} \\
0_{1 \times 3} & 0_{1 \times 3} & 0_{1 \times 3} & 0_{1 \times 1} }~,~~
      }
see \rf[mnnp]. We shall write our formulae in terms of the ``physical'' fields 
$\bfJ,\bfF, \bX, \bY$ given by 
\eq{
\bftJ=\bfJ+\bfm~, \qquad \bftF=\bfF+\bfn~, \qquad \bftX=\bX+\bfm'~, \qquad 
\bftY=\bY+\bfn'~.
   }

The subgroup of $\CX \ot (\SU5 \times \U1)\,,$ which leaves \rf[minh] 
invariant, is the group $\CX \ot (\SU3_C \times \U1_{EM})\,.$ 
The Higgs mechanism consists in reducing the symmetry of the whole theory to 
the symmetry of the vacuum. This means that we fix the gauge transformations 
corresponding to 
\[
\CX \ot \big((\SU5 \times \U1)\,/\,(\SU3_C \times \U1_{EM}) \big)
\]
in such a way that the Higgs multiplets $\bfJ\,,$ $\bfF$ and $\bX$ take the 
form
\seq{
\eqa{rcl}{
\bfJ &=& \mb{E{33}E{26}}{ -\sqrt{\tfrac{4}{15}} \itJ_0 \one_3 + \bfJ_g & 0 \\ 
0 & \sqrt{\tfrac{3}{5}} \itJ_0 + \bfJ_w } \,, \yn \label{blj} \\
\bfJ_g &=& \! \mb{Z{22}E{17}}{ \sqrt{\tfrac{1}{3}} \itJ_8 {+} \itJ_3 & 
\itJ_1 {-} \iu \itJ_2 & \itJ_4 {-} \iu \itJ_5 \\ \itJ_1 {+} \iu \itJ_2 & 
\sqrt{\tfrac{1}{3}} \itJ_8 {-} \itJ_3 & 
\itJ_6 {-} \iu \itJ_7 \\ \itJ_4 {+} \iu \itJ_5 & \itJ_6 {+} \iu \itJ_7 & 
{-} \sqrt{\tfrac{4}{3}} \itJ_8 } \!= \sum_{a=1}^8 \itJ_a \lambda^a~,~~ 
\itJ_a \in \CX ~, \hspace*{3em} \yn \label{blg} \\ 
\bfJ_w &=& \mb{Z{20}}{	\itJ_3' & \itJ_1' - \iu \itJ_2' \\ 
\itJ_1' + \iu \itJ_2' & - \itJ_3' ~~ } = \sum_{a=1}^3 \itJ_a' \sigma^a~, \quad
\itJ_a' \in \CX~, \label{blw}  \yn
\\
\bfF &=& \mb{E{8}}{ \bfF_g \\ \bfF_w }~,~~ \bfF_g=\mb{E{20}}{ \itF_1 
+\iu \itF_4 \\ \itF_2+\iu \itF_5 \\ \itF_3 +\iu \itF_6 }~,~~ 
\bfF_w=\mb{E{8}}{ \itF_0 \\ 0 } ~, \quad \itF_a \in \CX~,  \yn \label{blf}
\\
\bX &=& \mb{Z{25}Z{25}}{
\overline{\bX_A} & \overline{\bX_D} - \th \vre(\overline{\bX_c}) & 
(\bX_E^0)^* & \bX_a \\ 
\overline{\bX_D} + \th \vre(\overline{\bX_c}) & \overline{\bX_B} & 
(\bX_F^0)^* & \bX_b \\ 
\overline{\bX_E^0} & \overline{\bX_F^0} & \bX_C & \overline{\bX_c} \\ 
\bX_a^T & \bX_b^T & \bX_c^* & - \bX_0 }~,  \yn \label{blx}
    }}
where $\bX_0 \in \CX$ is a \emph{real} function. The explicit form of 
$\bX$ is presented in \rf[bxt], where $\itX_i \in \CX\,,$ $i=0,\dots,98\,.$	   
\begin{table}[p]
\rotatebox{90}{
\vbox{
\eqa{l}{
\hspace*{-8cm}\bX= \npb \yn \label{bxt} \\
\hspace*{-8cm}\includegraphics{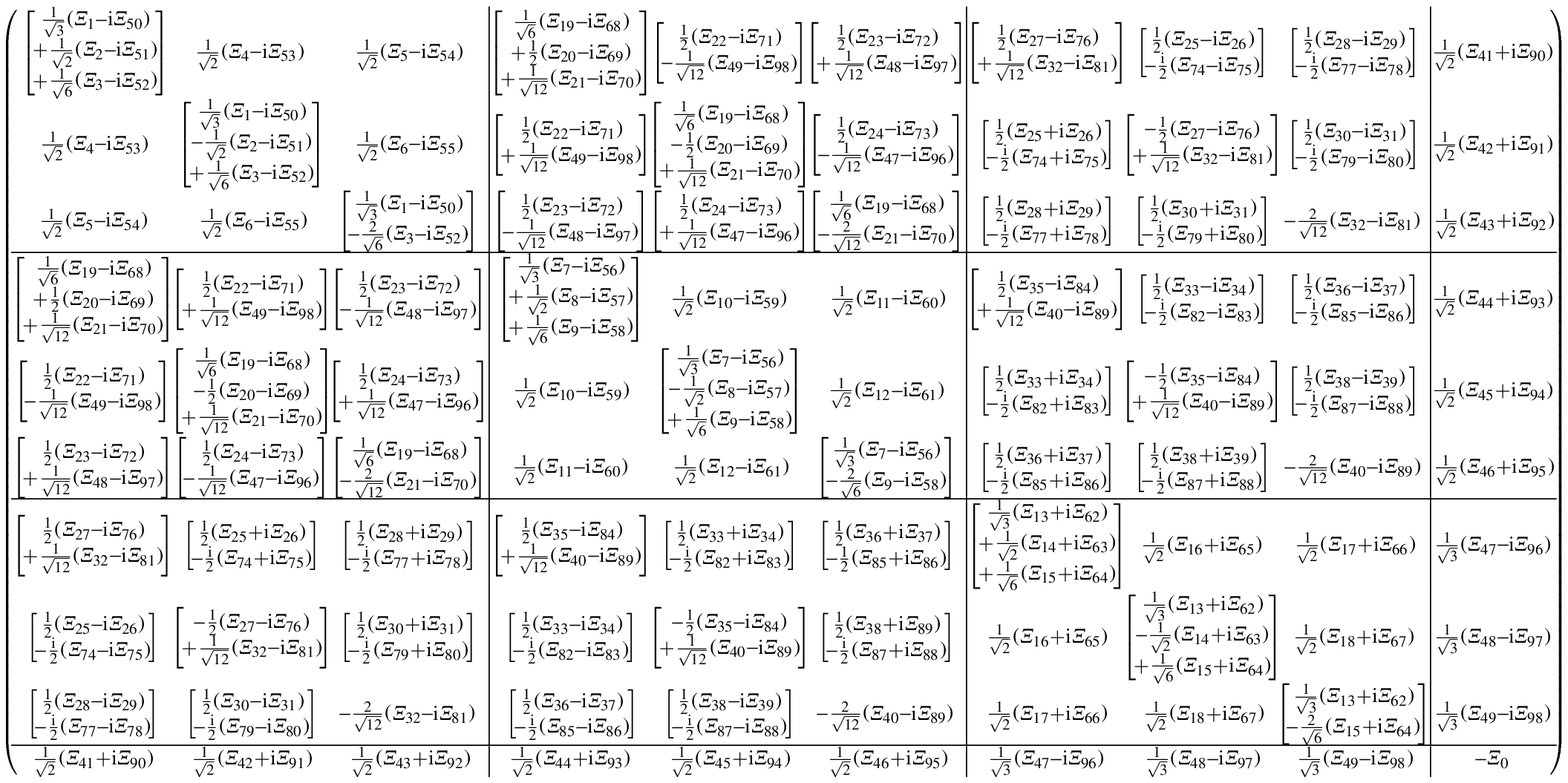}
}}}
\end{table}
Here, $\lambda^a$ are the Gell--Mann matrices and $\sigma^a$ the Pauli 
matrices. The matrix $\bY$ is an arbitrary element of $\iu \f[w]$ as displayed 
in \rf[bly], where $\itY_i \in \CX\,.$ 
\begin{table}[p]
\vs*{-2ex}
\eqa{l}{
\bY= \npb \yn \label{bly} \\
\mbox{\footnotesize{$\mb{ccc|c|c}{
\ru{8}{7}
\mpa{c}{ \tfrac{1}{\sqrt{6}} (\itY_0 {+} \iu \itY_{45}) \\ {+} \itY_3 
{+} \iu \itY_{48} \\ {+} \tfrac{1}{\sqrt{3}} (\itY_{8}{+} \iu \itY_{53}) } & 
\mpa{c}{ (\itY_1 {-} \iu \itY_2) \\ {+} \iu (\itY_{46} {-} \iu \itY_{47}) } & 
\mpa{c}{ (\itY_4 {-} \iu \itY_5) \\ {+} \iu (\itY_{49} {-} \iu \itY_{50}) } & 
\mpa{c}{ \tfrac{1}{\sqrt{2}} (\itY_{9} {+} \iu \itY_{54}) \\ 
{+} \itY_{12} {+} \iu \itY_{57} } & \sqrt{2} (\itY_{15} {+} \iu \itY_{60}) 
\\ 
\mpa{c}{ (\itY_1 {+} \iu \itY_2) \\ {+} \iu (\itY_{46} {+} \iu \itY_{47}) } & 
\mpa{c}{ \tfrac{1}{\sqrt{6}} (\itY_0 {+} \iu \itY_{45}) \\ 
{-} \itY_3 {-} \iu \itY_{48} \\ {+} \tfrac{1}{\sqrt{3}} (\itY_{8} {+} \iu 
\itY_{53}) } 
& \mpa{c}{ (\itY_6 {-} \iu \itY_7) \\ {+} \iu (\itY_{51} {-} \iu \itY_{52}) } & 
\mpa{c}{ \tfrac{1}{\sqrt{2}} (\itY_{10} {+} \iu \itY_{55}) \\ 
{+} \itY_{13} {+} \iu \itY_{58} } & \sqrt{2} (\itY_{16} {+} \iu \itY_{61}) 
\\
\mpa{c}{ (\itY_4 {+} \iu \itY_5) \\ {+} \iu (\itY_{49} {+} \iu \itY_{50}) } & 
\mpa{c}{ (\itY_6 {+} \iu \itY_7) \\ {+} \iu (\itY_{51} {+} \iu \itY_{52}) } & 
\mpa{c}{ \tfrac{1}{\sqrt{6}} (\itY_0 {+} \iu \itY_{45}) \\ 
{-} \tfrac{2}{\sqrt{3}} (\itY_{8} {+} \iu \itY_{53}) } & 
\mpa{c}{ \tfrac{1}{\sqrt{2}} (\itY_{11} {+} \iu \itY_{56}) \\ 
{+} \itY_{14} {+} \iu \itY_{59} } & \sqrt{2} (\itY_{17} {+} \iu \itY_{62}) 
\ru{6}{5} 
\\ \hline
\ru{8}{7}
\mpa{c}{ \tfrac{1}{\sqrt{6}} (\itY_{18} {+} \iu \itY_{63}) \\ {+} \itY_{21} 
{+} \iu \itY_{66} \\ {+} \tfrac{1}{\sqrt{3}} (\itY_{26}{+} \iu \itY_{71}) }\! & 
\mpa{c}{ (\itY_{19} {-} \iu \itY_{20}) \\ {+} \iu (\itY_{64} {-} \iu \itY_{65}) 
} & 
\mpa{c}{ (\itY_{22} {-} \iu \itY_{23}) \\ {+} \iu (\itY_{67} {-} \iu \itY_{68}) 
} & 
\sqrt{2} (\itY_{27} {+} \iu \itY_{72}) & \mpa{c}{ \tfrac{1}{\sqrt{2}} (\itY_{9} 
{+} \iu \itY_{54}) \\ {-} \itY_{12} {-} \iu \itY_{57} } 
\\ 
\mpa{c}{ (\itY_{19} {+} \iu \itY_{20}) \\ {+} \iu (\itY_{64} 
{+} \iu \itY_{65}) } & 
\! \mpa{c}{ \tfrac{1}{\sqrt{6}} (\itY_{18} {+} \iu \itY_{63}) \\ 
{-} \itY_{21} {-} \iu \itY_{66} \\ {+} \tfrac{1}{\sqrt{3}} (\itY_{26} 
{+} \iu \itY_{71}) } & \mpa{c}{ (\itY_{24} {-} \iu \itY_{25}) \\ 
{+} \iu (\itY_{69} {-} \iu \itY_{70}) } & 
\sqrt{2} (\itY_{28} {+} \iu \itY_{73}) & \mpa{c}{ \tfrac{1}{\sqrt{2}} 
(\itY_{10} {+} \iu \itY_{55}) \\ {-} \itY_{13} {-} \iu \itY_{58} } 
\\
\mpa{c}{ (\itY_{22} {+} \iu \itY_{23}) \\ {+} \iu (\itY_{67} {+} \iu \itY_{68}) 
} & \mpa{c}{ (\itY_{24} {+} \iu \itY_{25}) \\ {+} \iu (\itY_{69} 
{+} \iu \itY_{70}) } & \mpa{c}{ \tfrac{1}{\sqrt{6}} (\itY_{18} 
{+} \iu \itY_{63}) \\ {-} \tfrac{2}{\sqrt{3}} (\itY_{26} {+} \iu \itY_{71}) } & 
\sqrt{2} (\itY_{29} {+} \iu \itY_{74}) & 
\mpa{c}{ \tfrac{1}{\sqrt{2}} (\itY_{11} {+} \iu \itY_{56}) \\ 
{-} \itY_{14} {-} \iu \itY_{59} } 
\ru{6}{5} 
\\ \hline
\ru{8}{7}
\mpa{c}{ \sqrt{\tfrac{2}{3}} (\itY_{30} {+} \iu \itY_{75}) \\ {+} \itY_{31} 
{+} \iu \itY_{76} \\ {+} \tfrac{1}{\sqrt{3}} (\itY_{32}{+} \iu \itY_{77}) }\! & 
\mpa{c}{ \itY_{33} {+} \iu \itY_{78} \\ {-} \tfrac{1}{\sqrt{2}} (\itY_{11} 
{+} \iu \itY_{56}) } & \mpa{c}{ \itY_{34} {+} \iu \itY_{79} \\ 
{+} \tfrac{1}{\sqrt{2}} (\itY_{10} {+} \iu \itY_{55}) } & 
\sqrt{2} (\itY_{36} {+} \iu \itY_{81}) & \sqrt{2} (\itY_{39} {+} \iu \itY_{84}) 
\\ 
\mpa{c}{ \itY_{33} {+} \iu \itY_{78} \\ {+} \tfrac{1}{\sqrt{2}} (\itY_{11} 
{+} \iu \itY_{56}) } & \mpa{c}{ \sqrt{\tfrac{2}{3}} (\itY_{30} 
{+} \iu \itY_{75}) \\ {-} \itY_{31} {-} \iu \itY_{76} \\ 
{+} \tfrac{1}{\sqrt{3}} (\itY_{32}{+} \iu \itY_{77}) } & 
\! \mpa{c}{ \itY_{35} {+} \iu \itY_{80} \\ 
{-} \tfrac{1}{\sqrt{2}} (\itY_{9} {+} \iu \itY_{54}) } & 
\sqrt{2} (\itY_{37} {+} \iu \itY_{82}) & \sqrt{2} (\itY_{40} {+} \iu \itY_{85}) 
\\ 
\mpa{c}{ \itY_{34} {+} \iu \itY_{79} \\ {-} \tfrac{1}{\sqrt{2}} (\itY_{10} 
{+} \iu \itY_{55}) }\! & \mpa{c}{ \itY_{35} {+} \iu \itY_{80} \\ 
{+} \tfrac{1}{\sqrt{2}} (\itY_{9} {+} \iu \itY_{54}) } & \mpa{c}{ 
\sqrt{\tfrac{2}{3}} (\itY_{30} {+} \iu \itY_{75}) \\ {-} \tfrac{2}{\sqrt{3}} 
(\itY_{32}{+} \iu \itY_{77}) } & \sqrt{2} (\itY_{38} {+} \iu \itY_{83}) & 
\sqrt{2} (\itY_{41} {+} \iu \itY_{88}) 
\ru{6}{5} 
\\ \hline
\ru{3.5}{0}
\sqrt{2} (\itY_{42} {-} \iu \itY_{87}) & \sqrt{2} (\itY_{43} {-} \iu \itY_{88}) 
& \sqrt{2} (\itY_{44} {-} \iu \itY_{89}) & {-}\tfrac{3}{\sqrt{6}} (\itY_{18} 
{+} \iu \itY_{63}) & \tfrac{3}{\sqrt{6}} (\itY_{0} {+} \iu \itY_{45})  
}\! $}} 
\\
\ru{8}{6}
= \mb{D{30}}{
\bY_A & \bY_a+\bY_b & \bY_c \\ \bY_B & \bY_d & \bY_a -\bY_b \\
\bY_C - \vre(\bY_a) & \bY_e & \bY_f \\ \bY_g^* & -\tr(\bY_B) 
& \tr(\bY_A) }~,~~ 
\\
(\vre(A))_{\alpha\beta}=\tsum_{\gamma=1}^3 \vre_{\alpha\beta\gamma} A_{\gamma} 
{}~.~~	      
       }
\end{table}

For $A$ and $A''$ we make the ansatz
\seq{
\label{bfield}
\eqa{rcl}{
A &=& \frac{\iu g_0}{2} \mb{Z{35}}{
\sqrt{\tfrac{4}{15}} A' \one_3 + \bfG & \bfX \\ \bfX^* & 
-\sqrt{\tfrac{3}{5}} A' \one_2+ \bfW }\;, \quad A' \in \Lambda^1~, 
\hs*{3em} \yn \label{bfa} \npb \\
A''&=& \tfrac{\iu g_0}{2} \sqrt{\tfrac{2}{5}} \t{A} ~, \quad
\t{A} \in \Lambda^1~, 
\yn \label{bfap} \\
\bfG &=& \mb{D{22}}{ \sqrt{\tfrac{1}{3}} G^8 + G^3 & G^1-\iu G^2 &  
G^4 - \iu G^5 \\ G^1+\iu G^2 & \sqrt{\tfrac{1}{3}} G^8 - G^3 & G^6 - \iu G^7 \\
G^4 + \iu G^5 & G^6+\iu G^7 & -\sqrt{\tfrac{4}{3}} G^8 }= \sum_{a=1}^8 G^a 
\lambda^a~, \quad
G^a \in \Lambda^1~, \hs*{3em} \yn \label{bfg} \\ 
\bfW &=& \mb{Z{23}}{ W^3 & W^1-\iu W^2 \\ W^1 + \iu W^2 & - W^3 }
= \sum_{a=1}^3 W^a \sigma^a~, \quad W^a \in \Lambda^1~,
\label{bfw} \yn \\
\bfX &=& \Big( X\,,\,Y \Big)\,, \quad X =\! \mb{E{18}}{ X^1 {-} \iu X^2 \\  
X^3 {-} \iu X^4 \\ X^5 {-} \iu X^6 } , \quad 
Y =\! \mb{E{18}}{ Y^1{-}\iu Y^2 \\ Y^3 {-} \iu Y^4 \\ Y^5 {-} \iu Y^6 } , 
\quad X^a,Y^a \in \Lambda^1\,. \hspace*{3em} \label{bfx} \yn
     }}
In terms of the local basis $\{\g[\mu]\}_{\mu=1,2,3,4}$ of $\Lambda^1$ we put
\eqa{rclrclrclrclrcl}{
\bfG &=& \bfG_{\mu} \g[\mu]\;,\quad{} & G^a &=& G^a_{\mu} \g[\mu]\;,\quad{} & 
W^a &=& W^a_{\mu} \g[\mu]\;,\quad{} & A' &=& A'_{\mu} \g[\mu]\;,\quad{} & 
\t{A} &=& \t{A}_{\mu} \g[\mu]\;, \npb \\
\bfX &=& \bfX_{\mu} \g[\mu]\;,\quad{} & X &=&X_{\mu} \g[\mu]\;,\quad{} & 
X^a &=& X^a_{\mu} \g[\mu]\;,\quad{} & Y &=& Y_{\mu} \g[\mu]\;,\quad{} & 
Y^a &=& Y^a_{\mu} \g[\mu]\;. 
       }
Moreover, we introduce the abbreviation 
\[
S_{[\mu} T_{\nu]}:=S_\mu T_\nu-S_\nu T_\mu~.~~
\]

Now we start to write down the explicit form of the Lagrangian $\Lgr[2]\,,$
where we restrict ourselves to the interesting part and denote the rest by 
$I.T$ (interaction terms). We obtain in terms of the local basis 
$\g[\mu] \wedge \g[\nu] \equiv \th(\g[\mu] \g[\nu] - \g[\nu] \g[\mu])$ of 
$\Lambda^2$
\seq{
\eq{
\bfd A + \th \{A,A\} = \dfrac{\iu g_0}{4} \mbox{\small{$ 
\mb{cc}{ \mpa{c}{ \tfrac{2}{3} (\sqrt{\tfrac{3}{5}} A'_{\mu\nu} 
- X_{\mu\nu}^0) \one_3 \\ 
+ \sum_{a=1}^8 (G^a_{\mu\nu} - X^a_{\mu\nu} ) \lambda^a } &  (D \bfX)_{\mu\nu} 
\\ 
(D\bfX)^*_{\mu\nu} & \mpa{c}{ (-\sqrt{\tfrac{3}{5}} A'_{\mu\nu} 
+ X_{\mu\nu}^0) \one_2 \\ + \sum_{a=1}^3 (W_{\mu\nu}^a - \t{X}_{\mu\nu}^a) 
\sigma^a } } $}} \g[\mu] \wedge \g[\nu] \,,  \raisetag{3ex}
     }
where
\eqa{rcl}{
G_{\mu\nu}^a &=& \partial_{[\mu} G^a_{\nu]} - g_0 \tsum_{b,c=1}^8 f_{abc} 
G^b_{\mu} G^c_\nu\;, \qquad W_{\mu\nu}^a = \partial_{[\mu} W^a_{\nu]} 
- g_0 \tsum_{b,c=1}^3 \vre_{abc} W^b_{\mu} W^c_\nu~,~~ \npb \\ 
A'_{\mu\nu} &=& \partial_{[\mu} A'_{\nu]}\;, \qquad (D\bfX)_{\mu\nu} = \big(~ 
\partial_{[\mu} X_{\nu]} + I.T ~,~ \partial_{[\mu} Y_{\nu]} + I.T~ \big)~,~~
\yn
  }
  }
and $X^a_{\mu\nu}, \t{X}^{\t{a}}_{\mu\nu}$ are interaction terms. Moreover, 
\eq{
\bfd A''=\iu \tfrac{g_0}{4} \sqrt{\tfrac{2}{5}} \t{A}_{\mu\nu} ~, \qquad
\t{A}_{\mu\nu}:=\partial_{[\mu} \t{A}_{\nu]} ~.~~
   }
Then, using 
\eqa{l}{
\tr(\sigma^a \sigma^b)=2 \,\delta^{ab}~, \qquad{} 
\tr(\lambda^a \lambda^b)=2 \,\delta^{ab}~,~~ \npb \label{sl} \yn \\
\tr_c( (\g[\kappa] \wedge \g[\lambda])(\g[\mu] \wedge \g[\nu])) 
= 4(\delta^{\lambda\mu}\delta^{\kappa\nu} 
- \delta^{\kappa\mu}\delta^{\lambda\nu}) \;,~ 
\tr_c( \g[\mu] \g[\nu])=4 \delta^{\mu\nu}\;,~ \tr_c(1)=4 \;,
       }
we obtain for \rf[L2w]
\eqa{rl}{
\Lgr[2] = \tfrac{1}{4} \delta^{\kappa\mu}\delta^{\lambda\nu} \big( &
\tsum_{a=1}^8 G_{\kappa\lambda}^a G^a_{\mu\nu} 
+ \tsum_{a=1}^3 W_{\kappa\lambda}^a W^a_{\mu\nu} 
+ A'_{\kappa\lambda} A'_{\mu\nu} + \t{A}_{\kappa\lambda} \t{A}_{\mu\nu} \npb \\
&+ \tsum_{a=1}^6 \partial_{[\kappa} X^a_{\lambda]} \, 
\partial_{[\mu} X^a_{\nu]} +
\tsum_{a=1}^6 \partial_{[\kappa} Y^a_{\lambda]} \, 
\partial_{[\mu} Y^a_{\nu]} \big) + I.T ~.~~  \yn \label{C2}
  }

We proceed with the calculation of $\Lgr[1]\,,$ where we restrict ourselves 
again to the interesting part. Using \rf[blj] and \rf[bfa] we get
\eqa{rl}{
\bfd \bfJ + & [A,\bfJ+\bfm]= \bfd \bfJ +[A,\bfm] + I.T = \yn \label{L11} \\ &
\mb{cc}{~
-\sqrt{\tfrac{4}{15}} \partial_{\mu} \itJ_0 \one_3 
+ \tsum_{a=1}^8 \partial_{\mu} 
\bfJ_a \, \lambda^a & \iu \tfrac{g_0}{2} \bfX_{\mu} \\ 
(\iu \tfrac{g_0}{2} \bfX_{\mu})^* & 
\sqrt{\tfrac{3}{5}} \partial_{\mu} \itJ_0 \one_2 + \tsum_{a=1}^3 
\partial_{\mu} \bfJ_a'\, \sigma^a~{} } \g[\mu] + I.T\,. 
  }
Now, using \rf[bfa] and \rf[blf] we calculate
\eqa{l}{
\bfd \bfF+(A {+} A''\one_5)(\bfF {+} \bfn) = \bfd \bfF + (A {+} A''\one_5) \bfn 
+ I.T =\mb{E{15}}{ D_{\mu} \bfF_g  \\ D_{\mu} \bfF_w } \g[\mu]~,~~{}
\yn \label{L12} \\
D_{\mu} \bfF_g = \partial_{\mu} \bfF_g + \iu \tfrac{g_0}{2} X_{\mu} + I.T~,~~
\\
D_{\mu} \bfF_w = \mb{c}{ ~\partial_{\mu} \itF_0 + \iu \tfrac{g_0}{2} 
( W^3_{\mu} {-} (\sqrt{\tfrac{3}{5}} A'_{\mu} {-} 
\sqrt{\tfrac{2}{5}} \t{A}_{\mu})) + I.T~{} \\ 
\iu \tfrac{g_0}{2} (W_{\mu}^1+\iu W_{\mu}^2) + I.T } ~.~~
       }
Next, using \rf[bfa] and \rf[bly] we calculate
\eqa{rcl}{
\mc{3}{l}{
\bfd \bY + \check{A} (\bY {+} \bfn')- (\bY {+} \bfn') A + A'' (\bY {+} \bfn') 
=\bfd \bY + \check{A} \bfn' {-} \bfn' A {+} A'' \bfn' + I.T ~~} 
\yn \label{L13} \npb \\ 
&=& \!\! \mbox{\small{$ \mb{ccc}{\,
\mpa{c}{ \prt[\mu] \bY_A + \iu \tfrac{g_0}{2} ( \sqrt{\tfrac{2}{5}} 
 \t{A}_{\mu} \\ {-} \tfrac{3}{\sqrt{15}} 
A_{\mu}' {+} W_{\mu}^3) \one_3 } \!\! & 
\mpa{c}{ \prt[\mu] \bY_a + \iu \tfrac{g_0}{2} X_{\mu} + \\ 
\prt[\mu] \bY_b + \iu  \tfrac{g_0}{2} ({-}2 X_{\mu}) } \hs*{-0.9em} & 
\prt[\mu] \bY_c + \iu \tfrac{g_0}{2} ( {-}4 Y_{\mu}) 
\\ 
\prt[\mu] \bY_B + \iu  \tfrac{g_0}{2} (W_{\mu}^1 {+} \iu W_{\mu}^2) \one_3 
\!\!\! & 
\partial_{\mu}\bY_d & \mpa{c}{ \prt[\mu] \bY_a + \iu  \tfrac{g_0}{2} X_{\mu} \\
- \prt[\mu] \bY_b - \iu  \tfrac{g_0}{2} ({-}2 X_{\mu}) } 
\\
\partial_{\mu} \bY_C - \vre(\prt[\mu] \bY_a {+} \iu  \tfrac{g_0}{2} X_{\mu} ) 
\!\!\! & \partial_{\mu}\bY_e & \partial_{\mu}\bY_f 
\\ 
\!\!\!\! (\prt[\mu] \bY_g + \iu  \tfrac{g_0}{2} ( 4 Y_{\mu} ))^* & \hs*{-1.5em}
\mpa{c}{ {-} \tr( \prt[\mu] \bY_B) - \\ 3 \iu \tfrac{g_0}{2} 
(W_{\mu}^1 {+} \iu W_{\mu}^2) } & \hs*{-0.9em}
\mpa{c}{ \tr(\prt[\mu] \bY_A) + 3 \iu \tfrac{g_0}{2} ({-}\tfrac{3}{\sqrt{15}} 
A_{\mu}' \\ {+} \sqrt{\tfrac{2}{5}}  \t{A}_{\mu} 
{+} W_{\mu}^3) }\,
}$}} \! \g[\mu] + I.T {}.
	 }
Finally, using \rf[bfa] and \rf[blx] we calculate
\eqa{rcl}{
\mc{3}{l}{
\bfd \bX + \check{A} (\bX {+} \bfm')+ (\bX {+} \bfm') \check{A}^T 
- A'' (\bX {+} \bfm') 
= \bfd \bX {+} \check{A} \bfm' {+} \bfm' \check{A}^T {-} A'' \bfm' 
{+} I.T } \npb \\ 
&=& \!\! \mbox{\small{$ \mb{cccc}{\,	 
\overline{\prt[\mu] \bX_A} & \overline{\prt[\mu] \bX_D} 
- \th \vre(\prt[\mu] \overline{\bX_c}) & 
(\prt[\mu] \bX_E^0)^* & \prt[\mu] \bX_a + \iu \tfrac{g_0}{2} Y_{\mu} 
\eqnskip \\ 
\overline{\prt[\mu] \bX_D} + \th \vre(\prt[\mu] \overline{\bX_c}) & 
\overline{\prt[\mu] \bX_B} & (\prt[\mu] \bX_F^0)^* & 
\prt[\mu] \bX_b - \iu \tfrac{g_0}{2} X_{\mu} 
\eqnskip \\ 
\overline{\prt[\mu] \bX_E^0} & \overline{\bX_F^0} & \prt[\mu] \bX_C & 
\overline{\prt[\mu] \bX_c} 
\eqnskip \\ 
(\prt[\mu] \bX_a + \iu \tfrac{g_0}{2} Y_{\mu})^T & 
(\prt[\mu] \bX_b - \iu \tfrac{g_0}{2} X_{\mu})^T & (\prt[\mu] \bX_c)^* & 
- (\prt[\mu] \bX_0 - \iu \tfrac{g_0}{2} (\tfrac{12}{\sqrt{15}} A_{\mu}' 
+ \sqrt{\tfrac{2}{5}} \t{A}_{\mu} ))  }$}} \! \g[\mu] \npb \\ 
&+& I.T~.  \yn	\label{L14}
         }
The Lagrangian $\Lgr[1]$ is obtained from formulae \rf[L11], \rf[L12], \rf[L13] 
and \rf[L14], where one has to use \rf[sl]:
\eqa{rcl}{
\Lgr[1] &=& \tfrac{8 \mu_0}{g_0^2} \delta^{\mu\nu} \big( 
\tsum_{a=0}^8 \prt[\mu] \bfJ_a \, \prt[\nu] \bfJ_a 
+ \tsum_{a=1}^3 \prt[\mu] \bfJ_a' \, \prt[\nu] \bfJ_a' \big) 
+ \tfrac{4\mu_1}{g_0^2} \delta^{\mu\nu} \; 
\tsum_{a=0}^6 \prt[\mu] \itF_a \, \prt[\nu] \itF_a 
\npb \eqnskip \\
&+& \tfrac{8\mu_2}{g_0^2} \delta^{\mu\nu} \; \tsum_{i=0}^{89} 
\prt[\mu] \itY_i \, \prt[\nu] \itY_i + \tfrac{4\mu_3}{g_0^2} \delta^{\mu\nu} \; 
\tsum_{i=0}^{98} \prt[\mu] \itX_i \, \prt[\nu] \itX_i 
\eqnskip \\ 
&+& (\mu_1  + 12 \mu_2) \delta^{\mu\nu} \big( \begin{array}[t]{l}
W^1_{\mu} W^1_{\nu} + W^2_{\mu} W^2_{\nu} + \\ 
+ (W^3_{\mu} - \sqrt{\tfrac{3}{5}} A'_{\mu} +\sqrt{\tfrac{2}{5}} \t{A}_{\mu})
(W^3_{\nu} - \sqrt{\tfrac{3}{5}} A'_{\nu} + \sqrt{\tfrac{2}{5}} \t{A}_{\nu}) 
\big) \end{array} 
\yn \label{L1fin} \eqnskip \\ 
&+& \mu_3 \delta^{\mu\nu} (4 \sqrt{\tfrac{3}{5}}  A_{\mu}' 
+ \sqrt{\tfrac{2}{5}} \t{A}_{\mu} ) (4 \sqrt{\tfrac{3}{5}}  A_{\nu}' 
+ \sqrt{\tfrac{2}{5}} \t{A}_{\nu} ) 
\npb \eqnskip \\
&+& \delta^{\mu\nu} \tsum_{a=1}^6 \big( (2 \mu_0 {+} \mu_1 {+} 12 \mu_2 
{+} 2 \mu_3) X^a_{\mu} X^a_{\nu} 
+ (2 \mu_0 {+} 32 \mu_2 {+} 2 \mu_3) Y^a_{\mu} Y^a_{\nu} \big) + I.T~.
     }

We perform the orthogonal transformation by Euler angles 
\seq{
\label{ortr}
\eq[ort3]{
\mb[\!\!]{c}{ Z_{\mu} \\ Z_{\mu}' \\ P_{\mu} } \!=\!
\mb[\!]{ccc}{
\cos \phi_E & {}~~{-} \sin \phi_E~~{} & 0 \\ \sin \phi_E & \cos \phi_E & 0 \\ 
0 & 0 & 1} \!\!\! \mb[\!]{ccc}{
1 & 0 & 0 \\ 0 & {}~~\cos \theta_E~~{} & {-} \sin \theta_E \\ 
0 & \sin \theta_E & \cos \theta_E } \!\!\! \mb[\!]{rcr}{
\cos \psi_E & {}~~{-} \sin \psi_E~~{} & 0 \\ \sin \psi_E & \cos \psi_E & 0 \\ 
0 & 0 & 1 } \!\!\!
\mb[\!]{c}{ W^3_{\mu} \\  A'_{\mu} \\  \t{A}_{\mu} } \!.
  }
The photon $P_{\mu}$ is the massless linear combination, which is perpendicular 
to the plane spanned by $(W^3_{\mu} {-} \sqrt{\tfrac{3}{5}} A'_{\mu} 
{+} \sqrt{\tfrac{2}{5}} \t{A}_{\mu})$ and $(4 \sqrt{\tfrac{3}{5}}  A_{\mu}' 
{+} \sqrt{\tfrac{2}{5}} \t{A}_{\mu} )\,,$ see \rf[L1fin]. Calculating the 
vector product yields immediately
\eq{
P_{\mu}=\sqrt{\tfrac{3}{8}} W_{\mu}^3 + \sqrt{\tfrac{1}{40}} A'_{\mu} 
-\sqrt{\tfrac{3}{5}} \t{A}_{\mu}~,
   }
which implies
\eq{
\cos \theta_E= -\sqrt{\tfrac{3}{5}}~,\quad 
\sin \theta_E = \sqrt{\tfrac{2}{5}}~, \qquad 
\cos \psi_E= \tfrac{1}{4}~,\quad 
\sin \psi_E = \sqrt{\tfrac{15}{16}}~. 
   }
The Euler angle $\phi_E$ is determined by the diagonalization of the mass 
matrix. The result is 
\al{
\tan 2 \phi_E &= -\tfrac{3}{4} + \tfrac{25}{4} \lambda_4~, &
\lambda_4 &:= \tfrac{(\mu_1  + 12 \mu_2)}{25 \mu_3}~.
   }
We choose $\cos \phi_E < 0$ and $\sin \phi_E > 0\,.$ Then, 
the inverse transformation is for $\lambda_4 \ll 1$ given by
\eqas[invtr]{rcl}{
W^3_{\mu} &=& \sqrt{\tfrac{5}{8}} Z_{\mu} + \sqrt{\tfrac{3}{8}} P_{\mu} 
- \sqrt{\tfrac{5}{2}} \lambda_4 Z_{\mu}' ~, \\
A'_{\mu} &=&  -\sqrt{\tfrac{3}{200}} (1-16 \lambda_4) Z_{\mu} 
+ \sqrt{\tfrac{1}{40}} P_{\mu} 
+ \sqrt{\tfrac{24}{25}} (1+\tfrac{1}{4} \lambda_4) Z_{\mu}' ~,\\
\t{A}_{\mu} &=& \tfrac{3}{5} (1+\tfrac{2}{3} \lambda_4) Z_{\mu} 
- \sqrt{\tfrac{3}{5}} P_{\mu} + \tfrac{1}{5} (1-6 \lambda_4) Z_{\mu}' ~.
   }
   }

The Lagrangian \rf[L1fin] requires to perform the reparametrizations
\eqas[rpa]{rclrclrclrcl}{
\itJ_i &=& \tfrac{g_0}{ \sqrt{16\,\mu_0}} \psi_i\;, \quad{}& 
i &=& 0, \dots ,8\;, \qquad{}& 
\itJ_i' &=& \tfrac{g_0}{\sqrt{16\,\mu_0}} \psi_i'\;, \quad{}& 
i &=& 1, \dots ,3\;, \\
\itF_i &=&  \tfrac{g_0}{\sqrt{8 \,\mu_1}} \phi_i\;, \quad{}& 
i &=& 0, \dots ,6\;, \\
\itY_i &=&  \tfrac{g_0}{\sqrt{16 \, \mu_2}} \upsilon_i\;, \quad{}& 
i &=& 0, \dots ,89\;, \quad{} &
\itX_i &=&  \tfrac{g_0}{\sqrt{8 \, \mu_3}} \xi_i\;, \quad{}& 
i &=& 0, \dots ,98\;.
     }
It remains to compute the quadratic terms of the Higgs potential \rf[L0w]. 
Due to the extremely rich Higgs structure we need computer algebra for that 
calculation. It turns out that it is advantageous to perform an orthogonal 
transformation in the $\phi_0{-}\upsilon_0$--sector: 
\eq[fy]{
\mb[\!]{c}{ \phi_0 \\ \upsilon_0 } \!=\!
\mb[\!]{Z{15}}{
\cos \alpha & {-} \sin \alpha \\ \sin \alpha & \cos \alpha } \!\!
\mb[\!]{c}{ \phi_0' \\  \upsilon_0' } ,\qquad 
\tan \alpha =\sqrt{\tfrac{12 \mu_2}{\mu_1}}~.
  }
The motivation for this transformation is that the linear combination $\phi_0'$ 
receives a much smaller mass than all other Higgs fields, see below. We present 
the quadratic terms of the Higgs potential in Appendix \ref{appc}.

We perform a Wick rotation from the Riemannian manifold $X$ to the 
Minkowskian manifold $X_M$ by introduction of a global minus sign in the action 
and by replacing\footnote{The minus sign in $\delta^{\mu\nu} \mapsto 
-g^{\mu\nu}$ is due to $(\gh)^*=-\gh$ on the Minkowski space.}
\eq{
\delta^{\mu\nu} \mapsto -g^{\mu\nu}~, \qquad g^{\mu\nu}=\diag(1,-1,-1,-1)~.
   }
We define $P_{\mu\nu} := \partial_{\lbrack \mu} P_{\nu \rbrack}$ and 
\eqas[mass]{rrlrrl}{
m_W^2 &=& (2 \,\mu_1 + 24 \,\mu_2) \;, & 
m_Z^2 &=& \tfrac{1}{\cos^2 (\theta_W {-} \theta_W')} m_W^2\;,  \npb \\
m_{Z'}^2 &=& 32 \mu_3 \cos^2 (\theta_W {-} \theta_W')\;, 
\npb  \\
m_X^2 &=& (4 \,\mu_0 + 2\, \mu_1 + 24 \,\mu_2 + 4 \mu_3)\;, \qquad{} & 
m_Y^2 &=& (4 \,\mu_0 + 64 \,\mu_2 + 4 \mu_3)\;, \\
\sin \theta_W &=& \sqrt{\tfrac{3}{8}}~, & 
\theta_W' &=& \th \sqrt{\tfrac{5}{3}} \lambda_4~. 
   }
      
Now we can write down the final formula for the bosonic Lagrangian: 
\seq{
\eqas[final]{rcl}{
\Lgr &=& -\tfrac{1}{4} g^{\kappa\mu} g^{\lambda\nu} 
(\tsum_{a=1}^8 (G_{\kappa\lambda}^a G^a_{\mu\nu}) 
+ P_{\kappa\lambda} P_{\mu\nu} ) \npb \\
&+& \tsum_{a=1}^2 (- \tfrac{1}{4} g^{\kappa\mu} g^{\lambda\nu} 
\partial_{\lbrack \kappa} W^a_{\lambda \rbrack} \, 
\partial_{\lbrack \mu} W^a_{\nu \rbrack} 
+ \tfrac{1}{2} g^{\mu\nu} m_W^2 W^a_{\mu} W^a_{\nu} ) \npb \\
&+& (- \tfrac{1}{4} g^{\kappa\mu} g^{\lambda\nu} 
\partial_{\lbrack \kappa}Z_{\lambda \rbrack} \,
\partial_{\lbrack \mu} Z_{\nu \rbrack} + \tfrac{1}{2} g^{\mu\nu} m_Z^2 
Z_{\mu} Z_{\nu} ) \npb \\
&+& (- \tfrac{1}{4} g^{\kappa\mu} g^{\lambda\nu} 
\partial_{\lbrack \kappa} Z'_{\lambda \rbrack} \,
\partial_{\lbrack \mu} Z'_{\nu \rbrack} + \tfrac{1}{2} g^{\mu\nu} m_{Z'}^2 
Z'_{\mu} Z'_{\nu} ) +\Lgr[ew](P,W,Z,Z') \\
&+& \tsum_{a=1}^6 (- \tfrac{1}{4} g^{\kappa\mu} g^{\lambda\nu} 
\partial_{\lbrack \kappa} X^a_{\lambda \rbrack} \,
\partial_{\lbrack \mu} X^a_{\nu \rbrack} 
+ \tfrac{1}{2} g^{\mu\nu} m_X^2 X^a_{\mu} X^a_{\nu} )\\
&+& \tsum_{a=1}^6 (- \tfrac{1}{4} g^{\kappa\mu} g^{\lambda\nu} 
\partial_{\lbrack \kappa} Y^a_{\lambda \rbrack} \,
\partial_{\lbrack \mu} Y^a_{\nu \rbrack} 
+ \tfrac{1}{2} g^{\mu\nu} m_Y^2 Y^a_{\mu} Y^a_{\nu} ) + \Lgr[H] + I.T~,~~
    }
where
\eqa{rcl}{
\mc{3}{l}{ \Lgr[ew] (P,W,Z,Z') } \yn \\
&=& g_0 \,g^{\kappa\mu} g^{\lambda\nu} 
( \partial_{\lbrack \kappa} W^1_{\lambda \rbrack}\, W^2_{\mu} W^3_{\nu} 
{+} \partial_{\lbrack \kappa} W^2_{\lambda \rbrack}\, W^3_{\mu} W^1_{\nu} 
{+} \partial_{\lbrack \kappa} W^3_{\lambda \rbrack}\, W^1_{\mu} W^2_{\nu} ) \\
&-& \th \,g_0^2 (g^{\kappa\mu} g^{\lambda\nu} {-} g^{\kappa\nu} g^{\lambda\mu})
( W^1_{\kappa} W^1_{\mu} W^2_{\lambda} W^2_{\nu} 
{+} W^1_{\kappa} W^1_{\mu} W^3_{\lambda} W^3_{\nu} 
{+} W^2_{\kappa} W^2_{\mu} W^3_{\lambda} W^3_{\nu} ) ~, 
\\[1ex]
\Lgr[H] &=& \th g^{\mu\nu} \big( \begin{array}[t]{l} 
\tsum_{i=0}^8  \prt[\mu] \psi_i\, \prt[\nu] \psi_i 
+ \tsum_{i=1}^3  \prt[\mu] \psi_i'\, \prt[\nu] \psi_i' 
+ \prt[\mu] \phi_0'\, \prt[\nu] \phi_0' 
+ \prt[\mu] \upsilon_0'\, \prt[\nu] \upsilon_0' \\
+ \tsum_{i=1}^6 \prt[\mu] \phi_i\, \prt[\nu] \phi_i 
+ \tsum_{i=0}^{98}  \prt[\mu] \xi_i\, \prt[\nu] \xi_i 
+ \tsum_{i=1}^{89}  \prt[\mu] \upsilon_i\, \prt[\nu] \upsilon_i \big) 
- \Lgr[0]~.~~ 
\end{array} \yn  \label{LH0}	 
    }
    }
This is precisely the bosonic Lagrangian of the flipped 
$\SU5 \times \U1$--model, where the masses of the gauge bosons are given in 
\rf[mass]. The parameters $\mu_1,\mu_2,\mu_3$ and the Weinberg angle 
$\theta_W$ will be determined in Section~\ref{fa} when discussing the fermionic 
action. Within our framework there is no possibility to determine $\mu_0\,.$ 
However, we will find in Section~\ref{fa} that the $X$ and $Y$ bosons lead to 
proton decay. In order to suppress the proton decay sufficiently we need 
$\mu_0 \gg \max(\mu_1,\mu_2)\,.$ Then, it remains to derive the masses of gauge 
and Higgs bosons in Section~\ref{thp}. 

\subsection{The Fermionic Action}
\label{fa}

Now we write down the fermionic action $S_F$ defined in \rf[sb]. However, we 
pass immediately to the Minkowski space $X_M\,.$ We denote the gamma matrices 
in Minkowski space by $\{\gh[\mu]\}$ and use the convention
\eq[gam]{
\gh[0]=\mb{Z{8}}{ 0 & \one_2 \\ \one_2 & 0 },~~
\gh[a]=\mb{E{8}E{10}}{ 0 & {-}\sigma^a \\ \sigma^a & 0 },~~
\gh[5]=\iu \gh[0] \gh[1] \gh[2] \gh[3] 
= \mb{E{8}E{10}}{ \one_2 & 0 \\ 0 & {-}\one_2 } .
  }
Then, the invariant fermionic action is 
\eq[sfee]{
S_F = \tfrac{1}{4} \int_{X_M} \!\! dx \: \bsj^* \gh[0] (D + \iu \rho_M) 
\bsj~.~~ 
         }
The factor $\tfrac{1}{4}$ additional to \rf[sb] occurs because 
we are going to impose constraints on $\bsj\,,$ which require precisely the 
form \rf[sfee] for the action, see below. More explicitly, inserting \rf[rabw] 
and \rf[hjfw] and using \rf[AMM] we obtain
\seq{
\eqa{l}{
D + \iu \rho_M = \yn \label{dpr1} \\ 
\mb[\!]{cccc}{
\sfD+\iu\t{\pi}(A {+} A'') & -\gh \t{\pi}(\bfJ+\bfm) & 
-\gh \t{\pi}(\bftF {+} \bftX {+} \bftY) & 0 \\
\gh \t{\pi}(\bftJ)^* & \sfD+\iu\t{\pi}(A {+} A'') & 0 & 
-\gh \t{\pi}(\bftF {+} \bftX {+} \bftY) \\
\gh \t{\pi}(\bftF {+} \bftX {+} \bftY)^* & 
0 & \!\! \sfD - \gh[2] \overline{ (\iu \t{\pi}(A {+} A''))} \gh[2] \!\! & 
- \gh \overline{ \t{\pi}(\bftJ)} \\
0 & \gh \t{\pi}(\bftF {+} \bftX {+} \bftY)^* & \gh \t{\pi}(\bftJ)^T & 
\!\! \sfD - \gh[2] \overline{ (\iu \t{\pi}(A {+} A''))} \gh[2] } ,
       }
where
\eqa{rcl}{
\t{\pi}(A {+} A'') &:=& \diag((\pi_{10}(A) {-} \th A''\one_{10}) \!\ot\! \one_3 
\,,\, \gh[2] \overline{(\pi_5(A) {-} \tfrac{3}{2} A'' \one_5 )} \gh[2] 
\!\ot\! \one_3\,,\, {-} \tfrac{5}{2} A'' \!\ot\! \one_3)\,,  \\  
\t{\pi}(\bftJ) &:=& \diag\big( (\check{\bfJ}+ \check{\bfm}) \ot M_{10}~,~
- \overline{(\bfJ + \bfm) \ot M_5} ~,~	0_{3 \times 3}\big)~,~~  
\yn  \label{tpA}
\\
\mc{3}{l}{
\t{\pi}(\bftF {+} \bftX {+} \bftY) := \mb[\!]{ccc}{
\mpa{c}{ (\h{\bfF}+\h{\bfn}) \ot M_d \\ + (\bX+\bfm') \ot M_N } & 
\mpa{c}{ (\check{\bfF}+\check{\bfn}) \ot \Mu \\
+ (\bY+\bfn') \ot \Mn } & 0 \\
\mpa{c}{ (\check{\bfF}+\check{\bfn})^T \ot \Mu^T \\
+ (\bY+\bfn')^T \ot \Mn^T } & 0 & \overline{(\bfF + \bfn)} \ot M_e \\ 
0 & (\bfF + \bfn)^* \ot M_e^T & 0 } .  }
    }
    }
We have used that within our convention \rf[gam] we have $\gh=-(\gh)^*$ and 
$[\sfD,\bar{f}]=-\gh[2] \ol{[\sfD,f]} \gh[2]\,.$ Recall \cite{rw2} that 
$\t{\pi}(A {+} A'')$ is given by commutators of $\sfD \ot \one_{192}$ 
with an arbitrary number of elements of the form $f \ot \h{\pi}(a)\,,$ where 
$a \in \f[a]$ and $f \in \CX\,.$ This fact and the complex conjugation in 
\rf[piofa] are the reasons why terms of the form $[\sfD,\bar{f}]$ occur in 
$\t{\pi}(A {+} A'')\,.$ 

Minkowskian fermions $\bsj$ live in the space $h_M=L^2(X_M,S) \ot \C^{192}$ 
and have in terms of the decomposition \rf[dpr1] the form 
\eq{
\bsj=(\bsj_1,\bsj_2,\bsj_3,\bsj_4)^T~,\quad \bsj_i \in L^2(X_M,S) \ot \C^{48}~.
   }
However, we shall restrict ourselves to the subspace of $h_M$ invariant under 
the charge conjugation $\mathcal{C}\,,$ the chirality operator 
$\t{\Gamma}$ and a symmetry transformation $\mathcal{S}$ defined in terms of 
$48 \times 48$--blocks by 
\eqa{rcl}{ 
\renewcommand{\arraystretch}{1}
\mathcal{C} &:=& \! \mb{cccc}{
0 & 0 & \!\!\!\!{-}\gh[2] \!\ot\! \one_{48}\!\! & 0 \\ 
0 & 0 & 0 & \!\!{-}\gh[2] \!\ot\! \one_{48} \;{} \\ 
\;{-}\gh[2] \!\ot\! \one_{48}\!\! & 0 & 0 & 0 \\ 
0 & \!\!{-}\gh[2] \!\ot\! \one_{48}\!\!\!\! & 0 & 0 }  \circ \mbox{ c.c}~,~~ 
\mathcal{S}:=\mb{Z{8.5}Z{8.5}}{ 0 & \one_{48} & 0 & 0 \\ 
\one_{48} & 0 & 0 & 0 \\ 0 & 0 & 0 & \one_{48}~{} \\ 
0 & 0 & \one_{48} & 0 } , \npb \\
\t{\Gamma} &:=& \diag ( {-} \gh \ot \one_{48}\;,~
{-} \gh \ot \one_{48}\;,~\gh \ot \one_{48}\;,~\gh \ot \one_{48})\;, \yn 
\label{CSG}
   }
where $\mathrm{c.c}$ means complex conjugation. Thus, we consider elements 
$\bsj \in h_M$ of the form
\eq[csg]{
\bsj= \mathcal{C} \bsj = \h{\Gamma} \bsj = \mathcal{S} \bsj = \! \mb[]{c}{
\th(1-\gh)\bsj_1 \\ \th(1-\gh)\bsj_1 \\ -\th(1+\gh) \gh[2] \bar{\bsj}_1 \\
-\th(1+\gh) \gh[2] \bar{\bsj}_1 } \! \equiv \! \mb[]{c}{
\th(1-\gh)\bsj_1 \\ \th(1-\gh)\bsj_1 \\ -\gh[2] \ol{\th(1-\gh) 
\bsj_1} \\ -\gh[2] \ol{\th(1-\gh) \bsj_1} }.
       }
Observe that the choice \rf[CSG] for the chirality operator breaks the 
structure of the model, which is precisely our intention. Since $\t{\Gamma}$ 
commutes with $\pf\,,$ the gauge invariance is not destroyed. But $\t{\Gamma}$ 
no longer anticommutes with the whole $D\,.$ We see that $D$~-- applied on 
chiral fermions \rf[csg]~--
\[
\th(\id_h - \t{\Gamma}) D \th(\id_h + \t{\Gamma})~,
\]
differs from the matrix \rf[dpr1] by the absence of $\gh \t{\pi}(\bftJ)\,.$ In 
other words, the choice \rf[CSG] for the chirality condition eliminates the 
disturbing terms $\gh \t{\pi}(\bftJ)$ in the fermionic action. 

Within our conventions one has the block structure 
\eq{
\th (1-\gh) \bsj_1=\binom{ 0}{\bsj_0 } ~, \quad \bsj_0 \in L^2(X_M) \ot 
\C^2 \ot \C^{48}~,~~
   }
where $L^2(X_M)$ denotes the space of square integrable functions on the 
Minkowski space. In local bases we have 
\eq[bas]{
\sfD=\iu \gh[\mu] \partial_{\mu}~, \quad  A=A_{\mu} \gh[\mu]~, \quad 
A''=A''_{\mu} \gh[\mu]\;.
	}
We define $\sigma^0=\t{\sigma}^0=\one_2$ and 
$\t{\sigma}^a=-\sigma^a\,,\ a=1,2,3\,,$ or in a symbolic notation
\eq{
\sigma^{\mu}=\{\one_2, \sigma^a\}~, \quad \t{\sigma}^{\mu} 
=\{\one_2, -\sigma^a\}~,~ \quad \mu=0,1,2,3~,~~a=1,2,3~.
   }
Then, from \rf[sfee], \rf[dpr1], \rf[csg] and \rf[gam] we get
\eq[Sf2]{
S_F = \tfrac{1}{2} \int_{X_M} \!\!\!\!\! \mathrm{v}_M \:
\big( \bsj_0^* \,,\, \bsj_0^T \sigma^2 \big) \!\ \mbox{\small{$ \mb[\!]{c}{
\iu \t{\sigma}^{\mu} (\prt[\mu] + \t{\pi}(A_{\mu} {+} A''_{\mu}))~;~~{}
- \t{\pi}(\bftF {+} \bftX {+} \bftY) \\ 
- \t{\pi}(\bftF {+} \bftX {+} \bftY)^*	~;~~{}
\iu \sigma^{\mu} (\prt[\mu] + \overline{ \t{\pi}(A_{\mu} {+} A''_{\mu})}) 
}$}} \!\!\! \mb[\!]{c}{ \bsj_0 \\ \sigma^2 \overline{\bsj_0} } .  
    }
This formula can be further simplified, because we have 
\eqa{rl}{
{\ds \int_{X_M} } \!\!\!\! \mathrm{v}_M & \: \bsj_0^T \sigma^2 \iu \sigma^{\mu} 
(\prt[\mu] + \overline{ \t{\pi}(A_{\mu} {+} A''_{\mu})} ) 
\sigma^2 \overline{\bsj_0} 
={\ds \int_{X_M}} \!\!\!\! \mathrm{v}_M \: \bsj_0^T \iu (\t{\sigma}^{\mu})^T
(\prt[\mu] + \overline{ \t{\pi}(A_{\mu} {+} A''_{\mu})} ) \overline{\bsj_0} 
\npb \\ &
= {\ds \int_{X_M}} \!\!\!\! \mathrm{v}_M \: \big((-\iu \prt[\mu] \bsj_0^T) 
(\t{\sigma}^{\mu})^T \overline{\bsj_0} + \bsj_0^T (\t{\sigma}^{\mu})^T 
(-\iu \t{\pi}(A_{\mu} {+} A''_{\mu}))^T \overline{\bsj_0} \big) 
\yn \label{rl1} \npb \\ &
={\ds \int_{X_M}} \!\!\!\! \mathrm{v}_M \: 
\bsj_0^* \iu \t{\sigma}^{\mu} (\prt[\mu] 
+ \t{\pi}(A_{\mu} {+} A''_{\mu})) \bsj_0 ~.~~
     }
Here, we have partially integrated and made use of 
$\t{\pi}(A_{\mu} {+} A''_{\mu}) = -\t{\pi}(A_{\mu} {+} A''_{\mu})^*\,.$ In 
the last step we took into account that in quantum mechanics the fields 
$\bsj_0$ are annihilation operators and the 
fields $\overline{\bsj_0}$ creation operators. In normal ordered products, 
all creation operators must stand on the left of all annihilation operators. 
This means that in \rf[rl1] we have to exchange $\bsj_0$ and 
$\overline{\bsj_0}\,.$ But since they represent fermions, which anticommute, 
this change of order gives a minus sign. Now, \rf[Sf2] takes the form 
\eq[Sf3]{
S_F=\int_{X_M} \!\!\!\! \mathrm{v}_M \: \big( \bsj_0^* \iu  \t{\sigma}^{\mu} 
(\prt[\mu] + \t{\pi}(A_{\mu} {+} A''_{\mu})) \bsj_0 - \th ( \bsj_0^* 
\t{\pi}(\bftF {+} \bftX {+} \bftY) \sigma^2 \overline{\bsj_0} + \mathrm{h.c} ) 
\big) ~,  \raisetag{1.5ex}
   }
where $\mathrm{h.c}$ denotes the Hermitian conjugate of the preceding term, 
without change of signs when exchanging fermion fields. 
For $\bsj_0 \in L^2(X_M) \ot \C^2 \ot \C^{48}$ we choose the following 
parametrization: 
\eqas{rcl}{
\bsj_0 &=& \big( u_L^r \,,\, u_L^b \,,\, u_L^g \,,\, d_L^r \,,\, 
d_L^b \,,\, d_L^g \,, \sigma^2 \bar{d}_R^r \,, \sigma^2 \bar{d}_R^b \,,
\sigma^2 \bar{d}_R^g \,,\, \sigma^2 \bar{\nu}_R \,,\, \hs*{4em} \\
\mc{3}{r}{
{-}\sigma^2 \bar{u}_R^r \,,\, {-}\sigma^2 \bar{u}_R^b \,,\, 
{-}\sigma^2 \bar{u}_R^g \,,\, {-}e_L \,,\, \nu_L \,,\, 
\sigma^2 \bar{e}_R \big)^t , } \npb \\
\sigma^2 \bar{\bsj}_0 &=& \big( \sigma^2 \bar{u}_L^r \,, 
\sigma^2 \bar{u}_L^b \,,\sigma^2 \bar{u}_L^g \,,\, \sigma^2 \bar{d}_L^r \,,\, 
\sigma^2 \bar{d}_L^b \,,\, \sigma^2 \bar{d}_L^g \,, {-}d_R^r \,, 
{-}d_R^b \,, {-}d_R^g \,, {-} \nu_R \,,\, \hs*{4em} \\
\mc{3}{r}{u_R^r \,,\, u_R^b \,,\, u_R^g \,, {-}\sigma^2 \bar{e}_L 
\,,\, \sigma^2 \bar{\nu}_L \,, {-} e_R \big)^t , }
    }
where $u_L^r, \dots , e_R \in L^2(X_M) \ot \C^2 \ot \C^3$ and $~^t$ means 
transposition only of the row, without transposing the matrix elements. 

Inserting the matrix structures of \rf[blf], \rf[blx], \rf[bly] and \rf[bfield]
into formulae \rf[tpA], it is straightforward to 
write down the explicit formula for the fermionic action \rf[Sf3]. Here, one 
must insert the explicit form \cite{phd} of the embeddings $\pi_{10}, 
\pi_{10,10}, \pi_{10,5}$ and $\pi_{5,1}\,.$ The transformation \rf[invtr] 
requires some care. Let us derive the coefficients of $P,Z,Z'$ corresponding 
to the left electron. From \rf[tpA], \rf[bfa] and \rf[mass] we find 
for $\lambda_4 \ll 1$ in good approximation
\eqas[ec]{rcl}{
\pi_{e_L}(A_{\mu}+ A''_{\mu}) &\to& -\iu\tfrac{g_0}{2} (W^3_{\mu}
- \sqrt{\tfrac{3}{5}} A' - \tfrac{3}{2} \sqrt{\tfrac{2}{5}}  \t{A}_{\mu} ) \npb 
\\
&=& -\tfrac{1}{2 \cos (\theta_W {-} 2 \theta_W')} \iu g_0 \t{Z}_{\mu} 
- \iu \t{e} \t{Z}_{\mu}' + \iu e \t{P}_{\mu} ~, 
\qquad \mbox{where}
\npb \\
\t{P}_{\mu} &:=& P_{\mu} - \tan (\theta_W {-} 2 \theta_W') Z_{\mu} 
+ (\tfrac{4}{\sqrt{15}} + \tfrac{12}{5} \theta_W' ) Z_{\mu}'~, \\
\t{Z}_{\mu} &:=& Z_{\mu} - \tfrac{1}{2} (1 + 2 \sqrt{15} \theta_W') Z_{\mu}' ~, 
\\
\t{Z}_{\mu}' &:=&  Z_{\mu}' + 4 \theta_W' \tan \theta_W Z_{\mu} ~, \\
e &:=& \sin \theta_W g_0 ~, \qquad  \t{e}:=\cos \theta_W g_0 ~. 
  }
Moreover, we express $\itF_0,\itF_g,\itX_A, \dots, \itX_c, \itY_A, \dots , 
\itY_g$ in terms of the physical Higgs bosons $\phi_0,\phi_g,\xi_A, \dots, 
\xi_c, \upsilon_A, \dots, \upsilon_g\,,$ see \rf[blf], \rf[bly] and \rf[rpa]. 
Then we arrive at the following formula for the fermionic Lagrangian: 
\seq{
\label{SFfin}
\eqa{rcl}{
S_F &=& \int_{X_M} \!\!\! \mathrm{v}_M \: ( \Lgr[q]+\Lgr[\ell] + \Lgr[m] 
+ \Lgr[x] + \Lgr[h] + \Lgr[h]' + \Lgr[h]'') ~, \qquad \mbox{where} \quad{} \yn
\eqnskip \eqnskip 
\\
\Lgr[q] &=& \mbox{\small{$ \Big( \bsu_L^*\,,\, \bsd_L^* \Big) \big( 
\t{\sigma}^{\mu} \!\mb{cc}{ \mpa[\!\!]{c}{ \iu \prt[\mu] - \tfrac{g_0}{2} 
\bfG_{\mu} \\ 
{-} ( \tfrac{g_0}{2 \cos (\theta_W {-} 2\theta_W')} \t{Z}_{\mu}  \\
{+} \tfrac{2}{3} e \t{P}_{\mu} {-} \tfrac{1}{3} \t{e} \t{Z}_{\mu}' ) \one_3 } 
&
-\tfrac{g_0}{2} (W^1_{\mu}-\iu W^2_{\mu}) \one_3
\\[-2ex]
-\tfrac{g_0}{2} (W^1_{\mu}+\iu W^2_{\mu}) \one_3 
&
\mpa[\!\!]{c}{ \iu \prt[\mu] - \tfrac{g_0}{2} \bfG_{\mu} \\
{-} ( {-} \tfrac{g_0}{2 \cos (\theta_W {-} 2\theta_W')} \t{Z}_{\mu} \\
{-} \tfrac{1}{3} e \t{P}_{\mu} {-} \tfrac{1}{3} \t{e} \t{Z}_{\mu}' ) \one_3 }} 
\!\! \ot \one_3 \big)
\!\! \mb[\!]{c}{ \bsu_L \\ \bsd_L } $}}
\npb \\
&+& \bsu_R^* \big( \sigma^{\mu}(\iu \prt[\mu]  -  \tfrac{g_0}{2}
\bfG_{\mu}  -  ( \tfrac{2}{3} e \t{P}_{\mu} - \tfrac{1}{3} \t{e} \t{Z}_{\mu}' )
\one_3 ) \ot \one_3 \big) \bsu_R 
\npb \\
&+& \bsd_R^* \big( \sigma^{\mu} (\iu \prt[\mu]	-  \tfrac{g_0}{2}
\bfG_{\mu}  - ( -\tfrac{1}{3} e \t{P}_{\mu} - \tfrac{1}{3} \t{e} \t{Z}_{\mu}' )
\one_3 ) \ot \one_3 \big) \bsd_R \,,  \yn 
\eqnskip \eqnskip
\smallskip
\\
\Lgr[\ell] &=& \mbox{\small{$ \Big( \nu_L^*\,,\, e_L^* \Big) \big( 
\t{\sigma}^{\mu} \! \mb[\!\!]{cc}{ {}\,\iu \prt[\mu] 
- ( \tfrac{g_0}{2 \cos (\theta_W {-} 2\theta_W')} 
\t{Z}_{\mu} {+} \t{e} \t{Z}_{\mu}' ) \hs*{-2ex}
&
\hs*{1.5ex} - \tfrac{g_0}{2} (W^1_{\mu}-\iu W^2_{\mu})
\\
-\tfrac{g_0}{2} (W^1_{\mu}+\iu W^2_{\mu}) \hs*{1.5ex}
&
\hs*{-2ex} \mpa{c}{\iu \prt[\mu] 
- ({-} \tfrac{g_0}{2 \cos (\theta_W {-} 2\theta_W')} \t{Z}_{\mu} \\ 
{-} e \t{P}_{\mu} {+} \t{e} \t{Z}_{\mu}' ) }} 
\!\! \ot \one_3 \big) \!\! \mb[\!]{c}{ \nu_L \\ e_L } $}}
\npb \\
&+& \nu_R^* \big( \sigma^{\mu} ( \iu \prt[\mu] - \t{e} \t{Z}_{\mu}' ) 
\ot \one_3 \big) \nu_R \npb \\
&+& e_R^* \big( \sigma^{\mu} ( \iu \prt[\mu] - ( - e \t{P}_{\mu}
+ \t{e} \t{Z}_{\mu}' )) \ot \one_3 \big) e_R \,,  \yn
\smallskip
\eqnskip \eqnskip
\\
\Lgr[m] &=& \big( -\bsd_L^* ( \one_3 \ot ( M_d + \tfrac{g_0}{\sqrt{8 \mu_1}} 
\phi_0 M_d ) - \tfrac{g_0}{\sqrt{8 \mu_3}} (\xi_F^0)^* \ot M_N ) \bsd_R \npb \\
&-& e_L^* ( M_e+ \tfrac{g_0} {\sqrt{8 \mu_1}} \phi_0 M_e ) e_R \npb\\
&-& \bsu_L^* (\one_3 \ot (M_u + \tfrac{g_0}{\sqrt{8 \mu_1}} \phi_0 \Mu )
+ \tfrac{g_0}{4 \sqrt{\mu_2}} \upsilon_A \ot \Mn ) \bsu_R \npb \\
&-& \nu_L^* ( M_n^T + \tfrac{g_0} {\sqrt{8 \mu_1}} \phi_0 \Mu^T
- \tfrac{3 g_0}{4 \sqrt{6 \mu_2}} 
(\upsilon_0 + \iu \upsilon_{45}) \Mn^T ) \nu_R \npb \\
&-& \bsd_L^* ( \tfrac{g_0}{4 \sqrt{\mu_2}} \upsilon_B \ot \Mn ) \bsu_R
- e_L^* ( \tfrac{3 g_0}{4 \sqrt{6 \mu_2}} 
(\upsilon_{18} + \iu \upsilon_{63}) \ot \Mn^T ) \nu_R \npb \\
&+& \bsu_L^* ( (\xi_E^0)^* \ot M_N ) \bsd_R 
- \th \nu_R^T \sigma_2 (M_N + \tfrac{g_0}{\sqrt{8 \mu_3}} \xi_0 M_N) \nu_R 
\big) + \mathrm{h.c} \,,\yn
\smallskip
\eqnskip \eqnskip
\\
\Lgr[x] &=& \tfrac{g_0}{2} \big( -\bsu_L^* ( \t{\sigma}^{\mu} 
\sigma^2 \epsilon(\bar{X}_{\mu}) \ot \one_3) \bar{\bsd}_R 
- \bsd_L^* ( \t{\sigma}^{\mu} \sigma^2 \epsilon(\bar{Y}_{\mu}) \ot \one_3) 
\bar{\bsd}_R \npb \\
&+& \bsu_L^* ( \t{\sigma}^{\mu} \sigma^2 Y_{\mu} \ot \one_3 ) \bar{\nu}_R
- \bsd_L^* ( \t{\sigma}^{\mu} \sigma^2	X_{\mu} \ot \one_3 ) \bar{\nu}_R 
\npb \\
&-& \nu_L^* ( \t{\sigma}^{\mu} \sigma^2  Y^T_{\mu} \ot \one_3 ) \bar{\bsu}_R
+ e_L^* ( \t{\sigma}^{\mu} \sigma^2  X^T_{\mu} \ot \one_3 ) \bar{\bsu}_R \big) 
+ \mathrm{h.c} \,, \yn
\smallskip
\eqnskip \eqnskip
\\
\Lgr[h] &=& \tfrac{g_0}{\sqrt{8 \mu_1}} \big( 
-\bsu_L^* \{ \sigma^2 \epsilon(\bar{\phi}_g) \ot M_d \} \bar{\bsd}_L 
+ \nu_R^T \{ \sigma^2 \phi_g^* \ot M_d \} \bsd_R \npb \\
&-& \bsu_L^* \{ \sigma^2 \phi_g \ot \Mu \} \bar{e}_L 
+ \bsd_L^* \{ \sigma^2 \phi_g \ot \Mu \} \bar{\nu}_L \npb \\
&+& \bsd_R^T \{ \sigma^2 \epsilon(\phi_g) \ot \Mu \} \bsu_R
- \bsu_R^T \{ \sigma^2 \bar{\phi}_g \ot M_e \} e_R \big) + \mathrm{h.c} \,,
\quad{} \yn
\smallskip
\eqnskip \eqnskip
\\
\Lgr[h]' &=& \tfrac{g_0}{4 \sqrt{\mu_2}} \big( 
\bsu_L^* ( \sigma^2 (\upsilon_a + \upsilon_b) \ot \Mn) \bar{e}_L 
- \bsd_L^* ( \sigma^2 (\upsilon_a- \upsilon_b) \ot \Mn ) \bar{\nu}_L
\npb \\
&-& \bsd_R^T ( \sigma^2 (\upsilon_C -\epsilon(\upsilon_a)) \ot \Mn) \bsu_R
- \bsu_L^* ( \sigma^2 \upsilon_c \ot \Mn) \bar{\nu}_L 
+ \bsd_L^* ( \sigma^2 \upsilon_d \ot \Mn) \bar{e}_L \npb \\
&-& e_L^* ( \upsilon_e^* \ot \Mn^T ) \bsd_R + \nu_L^* (\upsilon_f^* \ot \Mn^T)
\bsd_R - \nu_R^T ( \sigma^2 \upsilon_g^* \ot \Mn ) \bsu_R \big) 
+ \mathrm{h.c}\,,  \yn
\smallskip
\eqnskip \eqnskip
\\
\Lgr[h]'' &=& \tfrac{g_0}{\sqrt{8 \mu_3}} \big( 
-\th \bsu_L^* ( \sigma^2 \overline{\xi_A} \ot M_N) \bar{\bsu}_L 
- \bsu_L^* ( \sigma^2 (\overline{\xi_D} - \th \vre(\overline{\xi_c})) \ot M_N ) 
\bar{\bsd}_L \npb \\
&+& \bsu_L^* ( \xi_a \ot M_N ) \nu_R 
- \th \bsd_L^* ( \sigma^2 \overline{\xi_B} \ot M_N) \bar{\bsd}_L 
+ \bsd_L^* ( \xi_b \ot M_N) \nu_R \npb \\
&+& \th \bsd_R^T (\sigma_2 \xi_C \ot M_N ) \bsd_R 
+ \bsd_R^T (\sigma_2 \overline{\xi_c} \ot M_N) \nu_R \big) + \mathrm{h.c}\,.  
\yn
      }
      }
The Lagrangian $\Lgr[q]$ contains the kinetic terms and the strong and 
electroweak interactions of quarks. The Lagrangian $\Lgr[\ell]$ contains the 
kinetic terms and electroweak interactions of leptons. The Lagrangian $\Lgr[m]$ 
contains the mass terms of the fundamental fermions and their interactions with 
the Higgs fields $\phi_0,\xi_E^0,\xi_F^0,\xi_0,\upsilon_A$ and $\upsilon_B\,.$ 
The masses of the $u,c,t$--quarks, the $d,s,b$--quarks and the 
$e,\mu,\tau$--leptons are the eigenvalues of $M_u, M_d$ and $M_e\,.$ The mass 
Lagrangian of the neutrino sector is given by 
\eq[neut]{
- \th \big( -\nu_L^*\,,\, \nu_R^T \sigma_2 \big) \mb[\!]{cc}{ 0 & - M_n \\ 
- M_n & M_N } \!\! \mb[\!]{c}{ \sigma_2 \bar{\nu}_L \\ \nu_R } + \mathrm{h.c}~.
   }
The diagonalization of the mass matrix occurring in \rf[neut] yields the 
masses of the neutrinos. The mixing angles are small for $\|M_N\| \gg 
\|M_n\|\,.$ In this case, the left--handed neutrinos receive a mass of the 
order $\tfrac{\|M_n\|^2}{2 \|M_N\|}$ and the right--handed neutrinos a mass of 
the order $\th \|M_N\|\,.$ Thus, for $\|M_n\|$ being of the order of the mass 
of the top quark and $\|M_N\|$ being of the order of the unification scale, we 
obtain very low masses for the left--handed neutrinos, which is compatible 
with experiments (seesaw mechanisms). Moreover, the matrices $M_u,M_d,M_e, M_n$ 
and $M_N$ contain mixing angles between the fermions, which constitute 
generalized Kobayashi--Maskawa matrices. Finally, the Lagrangians 
$\Lgr[x],\Lgr[h], \Lgr[h]'$ and $\Lgr[h]''$ describe the coupling of the 
fundamental fermions to the $X$ and $Y$ leptoquarks, the Higgs bosons $\phi_g$ 
and the remaining Higgs bosons $\upsilon_i$ and $\xi_i\,,$ respectively. All 
terms of these Lagrangians contribute to the proton decay. 

Observe that the Lagrangians $\Lgr[q]$ and $\Lgr[\ell]$ differ from the 
corresponding Lagrangians of the standard model in two aspects: First, there 
occurs the massive gauge field $Z'\,,$ which of course is not a terrible 
problem if its mass is sufficiently large. Second, the universal Weinberg 
angle $\theta_W$ of the standard model is modified by angles of the order 
$\theta_W'\,.$ However, this angle $\theta_W'$ is extremely small if $m_{Z'}$ 
is very large against $m_Z\,.$ This means that experiments will certainly not 
detect $\theta_W'\,.$

\section{The Masses of Yang--Mills and Higgs Fields}
\label{thp}

The final step is to compute the boson masses. For that purpose we must compute 
the parameters $\mu^i,\t{\mu}^i,\check{\mu}^i,\h{\mu}^i$ of the Higgs 
potential \rf[lah], which depend according to Appendix~\ref{appb} 
on the mass matrices occurring in the generalized Dirac operator 
$\M\,.$ We have found in Section \ref{fa} that the eigenvalues 
\eqas[eva]{lclclcllclclcl}{
\mbox{of} &~~& M_u M_u^*  &~~& \mbox{are} &~~& m_u^2,m_c^2,m_t^2~, 
\hs*{3em} 
& \mbox{of} &~~& M_d M_d^*  &~~& \mbox{are} &~~& m_d^2,m_s^2,m_b^2~,~~ 
\npb \\
\mbox{of} &~~& M_e M_e^*  &~~& \mbox{are} &~~& m_e^2,m_{\mu}^2,m_{\tau}^2 ~, 
    }
referring to the usual names of the fermions. By unitary transformations we 
can achieve that $M_u$ is diagonal, 
\seq{
\label{udn}
\eq[uma]{
M_u = \diag ( m_u,m_c,m_t )~.
        }
It is necessary to make several assumptions to simplify the calculation: Since 
the Kobayashi--Maskawa matrix between $M_u$ and $M_d$ is approximately 
the identity matrix, let us assume
\eq[dma]{
M_d = \diag (m_d, m_s, m_b)~. 
        }
The experimental data show that 
$m_t$ is much bigger than all other eigenvalues. Among the remaining 
eigenvalues we neglect all but $m_b^2$ and $m_{\tau}^2\,.$ 
For simplicity we also neglect $m_{\tau}^2$ against $m_b^2\,,$ although this is 
not completely justified. Unfortunately, there are no experimental values 
for the matrix $M_n\,.$ Therefore, we can only estimate its contribution: 
We assume that in the case \rf[uma] we have
\eq[MNU]{
M_n=\diag(0,0,\mathrm{e}^{\iu \chi} m_n)~.~~  
   }
   }
Quantum corrections suggest that $m_n$ is of the order $m_t\,.$ Using 
\rf[mtu] we find for \rf[nu012] approximately 
\eqas{rcl}{
\mu_1 &=& \tfrac{1}{8} m_b^2 + \tfrac{1}{96} (9 m_t^2 + 6 m_t m_n \cos \chi 
+ m_n^2) + \tfrac{1}{24} m_{\tau}^2~, \\
\mu_2 &=& \tfrac{1}{384} (m_t^2 - 2 m_t m_n \cos \chi + m_n^2)~,
   }
which yields according to \rf[mass] for the mass $m_W$ of the $W$ boson 
\eq{
m_W^2 = \tfrac{1}{4}( m_t^2 + m_b^2 + \tfrac{1}{3} m_n^2 
+ \tfrac{1}{3} m_{\tau}^2)~.
   }
The comparison with the experimental values for $m_t$ and $m_W$ requires that 
$m_n$ is small against $m_t\,.$ Thus, we shall neglect $m_n$ against $m_t$ 
whenever this is possible. 

Since (at energies accessible at present) the standard model is in excellent 
agreement with experiments, the parameter $\mu_3 \sim \tr(M_N M_N^*)$ must be 
very large, see Sections~\ref{bllc} and~\ref{fa}. We choose the 
parametrization
\eq[NUdU]{
M_N = m_N \, U \diag( \sin \theta_N \cos \phi_N \,,\, 
\sin \theta_N \sin \phi_N \,,\, \cos \theta_N ) U^T~,
   }
for $U \in \U3\,,$ where the parameter $m_N \gg m_t$ determines the mass scale. 

The mass of the $X$ and $Y$ bosons must be very large in order to suppress 
the proton decay. This could be achieved by a sufficiently large $\mu_3\,,$ 
however, there are also Higgs bosons which induce an insufficient lifetime for 
the proton if $\mu_0$ is too small. Therefore, we must demand 
\eq[maxt]{
\max(\tr (M_{10} M_{10}^*)\,,\, \tr(M_5 M_5^*)) \gg \tr (M_u M_u^*)~.~~  
	 }
We put\footnote{The choice $M_{10} = (M \one_3 + m_{10})~,~~
M_5 = \mathrm{e}^{\iu \chi_0} (M \one_3 + m_5)$ yields the same results.}
\eq[splMn]{
M_{10} = M \one_3 + m_{10}~, \quad M_5 = M \one_3 + m_5~, \quad M \in \R~,~~ 
	  }
where $m_{10},m_5 \in \mat{3}$ are perturbations, which we consider for the 
time being as small against $M \one_3\,.$ Thus, we obtain for \rf[nu012] 
approximately
\al{
\label{nufin}
\mu_0 &= \tfrac{1}{4} M^2~, & \mu_1 &= \tfrac{3}{32} m_t^2  ~, &
\mu_2 &= \tfrac{1}{384} m_t^2 ~, & \mu_3 &= \tfrac{1}{48} m_N^2~.
   }

Inserting the leading approximation \rf[splMn] into the quadratic terms 
\rf[lah] of the Higgs potential, we can distinguish linear combinations of 
$\la[i]$ to $\la[t]$ that do not depend on $M\,.$ It turns out that the 
following combinations are essential:
\seq{
\label{lagt}
\eqa{l}{
\tfrac{1}{4} \la[i] {+} \tfrac{1}{4} \la[j] {+} \tfrac{1}{4} \la[k] 
{+} \tfrac{1}{4} \la[l] {-} \tfrac{1}{4} \la[m] {+} \tfrac{1}{4} \la[n] 
{-} \tfrac{1}{2} \la[p] {+} \tfrac{1}{4} \la[r] {-} \tfrac{1}{4} \la[t] 
= \th \tr(\t{M}_u \t{M}_u^* {+} \h{M}_u \h{M}_u^*) =: \t{\lambda}_1^2 m_t^4~, 
\npb \\
\tfrac{1}{4} \la[i] {{+}} \tfrac{1}{4} \la[j] {{+}} \tfrac{9}{4} \la[k] 
{{+}} \tfrac{9}{4} \la[l] {-} \tfrac{1}{4} \la[m] {-} \tfrac{3}{4} \la[n] 
{{+}} \tfrac{3}{2} \la[p] {-} \tfrac{3}{4} \la[r] {-} \tfrac{9}{4} \la[t] 
\npb \\
\hs*{7em} = \th \tr(\t{M}_{n} \t{M}_n^* {{+}} \h{M}_{n} \h{M}_n^* ) 
=: \t{\lambda}_2^2 m_t^2 m_n^2 \;, \\
\tfrac{1}{2} \la[i] {+} \tfrac{1}{2} \la[j] {-} \tfrac{3}{2} \la[k] 
{-} \tfrac{3}{2} \la[l] {-} \tfrac{1}{2} \la[m] {-} \tfrac{1}{2} \la[n] 
{+} \la[p] {-} \tfrac{1}{2} \la[r] {+} \tfrac{3}{2} \la[t] 
\yn  \npb \label{lagta} \\
\hs*{7em} = \th \tr(\t{M}_u \t{M}_n^* {+} \h{M}_u \h{M}_n^* 
{+} \t{M}_n \t{M}_u^* {+} \h{M}_n \h{M}_u^* ) 
=: 2 \t{\lambda}_3^2 m_t^3 m_n \cos \chi ~,~~ 
\npb \\
\la[o] {-} 2 \la[q] {+} \la[s] 
= \th \iu \, \tr(\t{M}_u \t{M}_{n}^* {+} \h{M}_u \h{M}_{n}^* 
{-} \t{M}_{n} \t{M}_u^* {-} \h{M}_{n} \h{M}_u^* )
=: 2 \t{\lambda}_4^2 m_t^3 m_n \sin \chi ~,~~
    }
where
\eqas{rclrcl}{
\t{M}_u &=& m_{10} M_u - M_u m_5 ~, \qquad \quad{} & 
\h{M}_u &=& m_{10}^* M_u - M_u m_5^* ~, \\ 
\t{M}_n &=& m_{10} M_{n} - M_{n} m_5 ~, \qquad \quad {} &
\h{M}_n &=& m_{10}^* M_{n} - M_{n} m_5^* ~,
	}
see \rf[mtu] and \rf[la]. Within our assumptions \rf[udn] we have 
\eq{
\t{\lambda}_1=\t{\lambda}_2=\t{\lambda}_3=\t{\lambda}_4 \equiv \lambda < 
\tfrac{M}{m_t} ~.~~
   }
   }

The matrices $M_{10}'$ and $M_5'$ enter the matrices \rf[Mht] only 
quadratically. Neglecting quadratic terms in $m_{10}$ and $m_5$ we have 
\[
M_i^2 = \diag( M^2 \one_3 + M (m_i + m_i^*)\,,\, 
M^2 \one_3 + M (m_i + m_i^*))~, \quad i=10,\,5~.
\]
Thus, we may assume $m_{10}=m_{10}^*$ and $m_5=m_5^*\,.$ Moreover, we may 
assume $\tr(m_{10})=0\,,$ because the transformation $m_5 \mapsto m_5 
+ \nu \one_3$ and $m_{10} \mapsto m_{10} + \nu \one_3\,,$ for $\nu \in \R\,,$ 
leaves the matrices $M_{aa}^i$ and $\h{M}_{aa}^i$ invariant. Therefore, we make 
the ansatz ($\nu_i^j \in \R\,,\ j \in \{10,5\}$)
\eq[mat10]{
m_j =\! \mbox{\small{$ \mb[\!]{ccc}{ \sqrt{\tfrac{1}{3}} \nu^j_0 {+} \nu^j_3 
{+} \sqrt{\tfrac{1}{3}} \nu^j_8 & \nu^j_1 {-} \iu \nu^j_2 & \nu^j_4 
{-} \iu \nu^j_5 \\ 
\nu^j_1{+}\iu \nu^j_2 & \sqrt{\tfrac{1}{3}} \nu^j_0 {-} \nu^j_3 
{+} \sqrt{\tfrac{1}{3}} \nu^j_8 & \nu^j_6 {-} \iu \nu^j_7 \\ 
\nu^j_4 {+} \iu \nu^j_5 & \nu^j_6{+}\iu \nu^j_7 & 
\sqrt{\tfrac{1}{3}} \nu^j_0 {-}\sqrt{\tfrac{4}{3}} \nu^j_8~ } $}}\,,\quad 
j=10,5~,~~ \nu^{10}_0 \equiv 0~.  \raisetag{3ex}
    }
We introduce the abbreviations
\eqas{l}{
\cos^4 \theta_N + \sin^4 \theta_N ( \cos^4 \phi_N + \sin^4 \phi_N ) 
\equiv \tfrac{1}{3} (1 + 2 \cos^2 \chi_N)~,  \\
\nu_{10}^2 = 2 \tsum_{i=1}^8 (\nu^{10}_i)^2~, \qquad 
(\nu^{10}_1)^2 + (\nu^{10}_2)^2 = \tfrac{1}{\sqrt{2}} \nu_{10} 
\sin \t{\chi} \sin \t{\chi}' \cos \t{\chi}'' ~. 
   }
For physical reasons we assume 
\eq{
M,m_N \gg \lambda m_t, \t{\lambda} m_t \gg m_t \gg m_b,m_n,m_{\tau}~,
   }
where $\t{\lambda}^2 m_t^2:= \tfrac{8}{5} (\tfrac{9}{2} \nu_{10}^2 
+ \tfrac{9}{16} (\nu^5_0)^2 + \nu_5^2 )\,.$ Inserting \rf[udn], \rf[mat10], 
\rf[MNU] and \rf[NUdU] into \rf[la], \rf[lac], \rf[lat] and \rf[lad] and 
this result into the Higgs potential \rf[lah], we find that~-- apart from the 
combinations \rf[lagta]~-- only the following parameters are relevant in 
leading approximation:
\eqa{rclrcl}{
\la[b]	&=& \tfrac{351}{160} m_t^4 \;, & 
\la[c]	&=& \tfrac{13}{7680} m_t^4\;,  \\
\la[f]	&=& \tfrac{39}{320} m_t^4 \;,  &
\la[g]	&=& 12 M^2 m_b^2 \;, \\
\la[h]	&=& \th m_b^2 \nu_{10}^2 \sin^2 \t{\chi} \sin^2 \t{\chi}' 
	    \sin^2 \t{\chi}'' &
\la[i]	&=& \la[j]=\th\la[m] = \tfrac{9}{4} M^2 m_t^2 \;, \\
\la[k]	&=& \la[l]=\th\la[t] = \tfrac{1}{4} M^2 m_t^2\;, &
\la[n]	&=& \la[p]=\la[r] = \tfrac{3}{2} M^2 m_t^2\;,\;\; \\
\lac[a] &=& \tfrac{1}{120} m_N^4 ( 1+16 \cos^2 \chi_N) \;, &
\lac[c] &=& \tfrac{3}{10} m_N^2 m_t^2 (|1+2 \cos \chi_N \cos \h{\chi}| 
	    {-} \tfrac{5}{4} ) \;, \\
\lac[d] &=& \tfrac{1}{120} m_N^2 m_t^2 ( |1+2 \cos \chi_N \cos \h{\chi}| 
	    {-} \tfrac{5}{4} )\;,~~{} &
\lac[e] &=& 2 M^2 m_N^2\;, \\
\lat[b] &=& \tfrac{1}{176} m_t^4 \;, & 
\lat[d] &=& \tfrac{9}{176} m_t^4 \;, \yn \label{laf} \\
\lat[k] &=& \tfrac{3}{88} m_t^4 \;, &
\lat[p] &=& \tfrac{1}{192} m_t^4 \sin^2 \h{\chi}\;, \\
\lat[q] &=& \tfrac{3}{64} m_t^4 \sin^2 \h{\chi}\;, &
\lat[s] &=& \tfrac{1}{32} m_t^4 \sin^2 \h{\chi}\;, \\
\lah[a] &=& 2 m_N^4 ((1+2 \cos^2 \chi_N) - \tfrac{1}{4} ) \cos^2 \h{\chi}_a  
\;, &
\lah[c] &=& \tfrac{1}{88} m_N^2 m_t^2 (|1+2 \cos \chi_N \cos \h{\chi}| {-} 1) 
\;, \\
\lah[e] &=& \tfrac{3}{88} m_N^2 m_t^2 (|1+2 \cos \chi_N \cos \h{\chi}| {-} 1) 
\;.
     }
The parameters $\h{\chi}$ and $\h{\chi}_a$ are complicated functions of the 
mass matrices. Now we find for \rf[LH0] in tree--level approximation
\eqa{rcl}{
\Lgr[H] &=& \th g^{\mu\nu} (\prt[\mu] \phi_0'\, \prt[\nu] \phi_0'
+ \prt[\mu] \upsilon_0'\, \prt[\nu] \upsilon_0' 
+ \prt[\mu] \upsilon_{45}\, \prt[\nu] \upsilon_{45} 
\yn \label{LH} \npb \\ &&
+ \prt[\mu] \psi_0\, \prt[\nu] \psi_0 + \prt[\mu] \psi_3'\, \prt[\nu] \psi_3' 
+ \prt[\mu] \xi_0\, \prt[\nu] \xi_0 ) \\ &&
- \tfrac{1}{2} \big( \lambda^2 m_t^2 \upsilon_0^2 
+ \tfrac{3}{4} \lambda^2 m_t^2 \upsilon_{45}^2 
+ \tfrac{m_N^4}{M^2} (\tfrac{1}{144} \cos^2 \h{\chi}_a 
+ \tfrac{1}{54} \cos^2 \chi_N \cos^2 \h{\chi}_a ) \psi_3'{}^2 \npb \\ &&
+ \tfrac{1}{48 M^2} ( \tfrac{48}{5} \lac[e] + \tfrac{2}{9} \lah[a] ) \psi_0^2 
+ \tfrac{1}{2 m_N^2} ( 4 \lac[a] + \tfrac{8}{15} \lah[a] ) \xi_0^2 \npb \\ &&
+ (\tfrac{207}{110} + \tfrac{2}{9} \sin^2 \h{\chi}) m_t^2 {\phi_0'}^2 
+ \tfrac{1}{8 \sqrt{15} M m_t} ( - \tfrac{32}{3} \lah[c] 
+ \tfrac{32}{3} \lah[e] ) \psi_0 \phi_0' \npb \\ &&
+ \tfrac{1}{\sqrt{24} m_N m_t} ( 4 \lac[c] + 48 \lac[d] 
+ \tfrac{64}{5} \lah[c] - \tfrac{64}{5} \lah[e] ) \phi_0' \xi_0 
- \tfrac{2}{ 9 \sqrt{10} M m_N} \lah[a]  \psi_0 \xi_0  \big) 
\\ 
&+& \th g^{\mu\nu} (\prt[\mu] \psi_1'\, \prt[\nu] \psi_1' +
\prt[\mu] \psi_2'\, \prt[\nu] \psi_2' 
+ \prt[\mu] \upsilon_{18}\, \prt[\nu] \upsilon_{18} 
+ \prt[\mu] \upsilon_{63} \, \prt[\nu] \upsilon_{63} ) \npb \\ &&
- \th \big( \tfrac{m_N^4}{M^2} (\tfrac{1}{144} \cos^2 \h{\chi}_a 
+ \tfrac{1}{54} \cos^2 \chi_N \cos^2 \h{\chi}_a )  ({\psi_1'}^2 +{\psi_2'}^2 ) 
+ \tfrac{3}{4} \lambda^2 m_t^2 (\upsilon_{18}^2 + \upsilon_{63}^2) \big)
\\
&+& \th g^{\mu\nu} ( \tsum_{i=1}^8 \prt[\mu] \psi_i\, \prt[\nu] \psi_i +
\tsum_{i=1}^8 \prt[\mu] \upsilon_i\, \prt[\nu] \upsilon_i + \tsum_{i=46}^{53}
\prt[\mu] \upsilon_i\, \prt[\nu] \upsilon_i \npb \\ && 
+ \tsum_{i=33}^{40} \prt[\mu] \xi_i\, \prt[\nu] \xi_i 
+ \tsum_{i=82}^{89} \prt[\mu] \xi_i\, \prt[\nu] \xi_i ) \\ &&
- \th \big( \tfrac{m_N^4}{M^2} (\tfrac{1}{144} \cos^2 \h{\chi}_a 
+ \tfrac{1}{54} \cos^2 \chi_N \cos^2 \h{\chi}_a )  
\tsum_{i=1}^8 \psi_i^2 \npb \\ &&
+ (\lambda^2 m_n^2 + \tfrac{m_b^2 \nu_{10}^2}{m_t^2} \sin^2 \t{\chi}\sin^2 
\t{\chi}'\sin^2 \t{\chi}'' ) ( \tsum_{i=1}^8 \upsilon_i^2 
+ \tsum_{i=46}^{53} \upsilon_i^2) \npb \\ &&
+ 9 M^2 (\tsum_{i=33}^{40} \xi_i^2 + \tsum_{i=82}^{89} \xi_i^2) \big) 
\\
&+& \th g^{\mu\nu} ( \tsum_{i=19}^{26} 
\prt[\mu] \upsilon_i\, \prt[\nu] \upsilon_i 
+ \tsum_{i=64}^{71} \prt[\mu] \upsilon_i\, \prt[\nu] \upsilon_i  \npb \\ &&
+ \tsum_{i=25}^{32} \prt[\mu] \xi_i\, \prt[\nu] \xi_i +
\tsum_{i=74}^{81} \prt[\mu] \xi_i\, \prt[\nu] \xi_i ) \npb \\ &&
- \th \big( (\lambda^2 m_n^2 + \tfrac{m_b^2 \nu_{10}^2}{m_t^2} 
\sin^2 \t{\chi}\sin^2 \t{\chi}'\sin^2 \t{\chi}'' ) 
( \tsum_{i=19}^{26} \upsilon_i^2 + \tsum_{i=64}^{71} \upsilon_i^2) \npb \\ &&
+ 9 M^2 ( \tsum_{i=25}^{32} \xi_i^2 + \tsum_{i=74}^{81} \xi_i^2 ) \big) 
\\
&+& \th g^{\mu\nu} ( \tsum_{i=1}^{6} \prt[\mu] \phi_i\, \prt[\nu] \phi_i 
+ \tsum_{i=9}^{14} \prt[\mu] \upsilon_i\, \prt[\nu] \upsilon_i 
+ \tsum_{i=54}^{59} \prt[\mu] \upsilon_i\, \prt[\nu] \upsilon_i \npb \\ &&
+ \tsum_{i=30}^{35} \prt[\mu] \upsilon_i\, \prt[\nu] \upsilon_i 
+ \tsum_{i=75}^{80} \prt[\mu] \upsilon_i\, \prt[\nu] \upsilon_i 
+ \tsum_{i=39}^{41} \prt[\mu] \upsilon_i\, \prt[\nu] \upsilon_i 
+ \tsum_{i=84}^{86} \prt[\mu] \upsilon_i\, \prt[\nu] \upsilon_i \npb \\ && 
+ \tsum_{i=19}^{24} \prt[\mu] \xi_i\, \prt[\nu] \xi_i 
+ \tsum_{i=68}^{73} \prt[\mu] \xi_i\, \prt[\nu] \xi_i 
+ \tsum_{i=44}^{46} \prt[\mu] \xi_i\, \prt[\nu] \xi_i 
+ \tsum_{i=93}^{95} \prt[\mu] \xi_i\, \prt[\nu] \xi_i \npb \\ &&
+ \tsum_{i=47}^{49} \prt[\mu] \xi_i\, \prt[\nu] \xi_i 
+ \tsum_{i=96}^{98} \prt[\mu] \xi_i\, \prt[\nu] \xi_i ) \\ &&
- \tfrac{1}{2} \big( M^2 (\tsum_{i=1}^6 \phi_i^2
+ \tsum_{i=9}^{14} \upsilon_i^2 + \tsum_{i=54}^{59} \upsilon_i^2 
+ \tsum_{i=30}^{35} \upsilon_i^2 + \tsum_{i=75}^{80} \upsilon_i^2 ) \npb \\ &&
+ 4 M^2 (\tsum_{i=39}^{41} \upsilon_i^2 + \tsum_{i=84}^{86} \upsilon_i^2 
+ \tsum_{i=19}^{24} \xi_i^2 + \tsum_{i=68}^{73} \xi_i^2 
+ \tsum_{i=47}^{49} \xi_i^2 + \tsum_{i=96}^{98} \xi_i^2 ) \npb \\ && 
+ (M^2 + m_N^2 (\tfrac{1}{12} \cos^2 \h{\chi}_a + \tfrac{2}{9} \cos^2 \chi_N 
\cos^2 \h{\chi}_a)) (\tsum_{i=44}^{46} \xi_i^2 + \tsum_{i=93}^{95} \xi_i^2 ) 
\big) 
\\
&+& \th g^{\mu\nu} ( \tsum_{i=7}^{18} \prt[\mu] \xi_i\, \prt[\nu] \xi_i
+ \tsum_{i=56}^{67} \prt[\mu] \xi_i\, \prt[\nu] \xi_i
+ \tsum_{i=41}^{43} \prt[\mu] \xi_i\, \prt[\nu] \xi_i
+ \tsum_{i=90}^{92} \prt[\mu] \xi_i\, \prt[\nu] \xi_i \npb \\ &&
+ \tsum_{i=15}^{17} \prt[\mu] \upsilon_i\, \prt[\nu] \upsilon_i
+ \tsum_{i=60}^{62} \prt[\mu] \upsilon_i\, \prt[\nu] \upsilon_i 
+ \tsum_{i=42}^{44} \prt[\mu] \upsilon_i\, \prt[\nu] \upsilon_i 
+ \tsum_{i=87}^{89} \prt[\mu] \upsilon_i\, \prt[\nu] \upsilon_i \npb \\ &&
+ \tsum_{i=36}^{38} \prt[\mu] \upsilon_i\, \prt[\nu] \upsilon_i 
+ \tsum_{i=81}^{83} \prt[\mu] \upsilon_i\, \prt[\nu] \upsilon_i ) \\ && 
- \tfrac{1}{2} \big( M^2 (\tsum_{i=15}^{17} \upsilon_i^2 
+ \tsum_{i=60}^{62} \upsilon_i^2+ \tsum_{i=42}^{44} \upsilon_i^2 
+ \tsum_{i=87}^{89} \upsilon_i^2 ) \npb \\ &&
+ 4 M^2 (\tsum_{i=36}^{38} \upsilon_i^2 {+} \tsum_{i=81}^{83} \upsilon_i^2  
{+} \tsum_{i=7}^{12} \xi_i^2 {+} \tsum_{i=56}^{61} \xi_i^2 ) \npb \\ &&
+ 16 M^2 (\tsum_{i=13}^{18} \xi_i^2 {+} \tsum_{i=62}^{69} \xi_i^2 ) \npb \\ &&
+ (M^2 + m_N^2 (\tfrac{1}{12} \cos^2 \h{\chi}_a + \tfrac{2}{9} \cos^2 \chi_N 
\cos^2 \h{\chi}_a )) (\tsum_{i=41}^{43} \xi_i^2 + \tsum_{i=90}^{92} \xi_i^2 ) 
\big) 
\\ 
&+& \th g^{\mu\nu} ( \tsum_{i=27}^{29} 
\prt[\mu] \upsilon_i\, \prt[\nu] \upsilon_i
+ \tsum_{i=72}^{74} \prt[\mu] \upsilon_i\, \prt[\nu] \upsilon_i 
+ \tsum_{i=1}^{6} \prt[\mu] \xi_i\, \prt[\nu] \xi_i 
+ \tsum_{i=50}^{55} \prt[\mu] \xi_i\, \prt[\nu] \xi_i  ) \npb \\ &&
- \tfrac{1}{2} \big( 4 M^2 (\tsum_{i=1}^{6} \xi_i^2 
+ \tsum_{i=50}^{55} \xi_i^2) 
+ M^2 (\tsum_{i=27}^{29} \upsilon_i^2 + \tsum_{i=72}^{74} \upsilon_i^2) \big)~.
    }
It remains to find the eigenvalues of the quadratic form\footnote{The 
corresponding matrix is positive definite by construction. This is not apparent 
when inserting \rf[laf], because there are complicated relations between 
$\chi_N,\h{\chi}_a,\h{\chi}\,.$}
determined by the $\phi_0'{-}\psi_0{-}\xi_0$ sector in \rf[LH]. We use the 
general result that the smallest (largest) eigenvalue is smaller (larger) than 
the smallest (largest) diagonal matrix element. This means that the mass of the 
$\phi_0'$ Higgs field is smaller than $\sqrt{\tfrac{2083}{990}}\, m_t 
\approx 1.45\, m_t\,.$ We assume $\tfrac{48}{5} \lac[e] \gg 
\tfrac{2}{9} \lah[a] \,,$ or $M^2 \gg \tfrac{55}{864} m_N^2\,.$ Then, the large 
parameter $\lac[e]$ occurring in the coefficient of $\psi_0^2$ stabilizes the 
other two eigenvalues near the diagonal matrix elements 
$\tfrac{1}{5 M^2} \lac[e]$ and $\tfrac{1}{m_N^2}(2 \lac[a] 
+ \tfrac{4}{15} \lah[a])\,,$ respectively. 

For convenience we list in Table~\ref{tab1} our tree--level predictions for the 
masses of the Higgs fields and the masses of the gauge fields derived in 
Section~\ref{bllc}. 
\begin{table}[p]
{}\hfill$
\begin{array}{|E{34.5}|E{34.5}|E{1.8}|E{34.5}|E{34.5}|}
\cline{1-2} \cline{4-5}
\mbox{ Particle } & \mbox{ Mass} && 
\mbox{ Particle } & \mbox{ Mass} 
\\ 
\cline{1-2} \cline{4-5} 
\mc{5}{c}{ \mbox{ \textbf{1.} The completely neutral Higgs fields: \ru{3}{0} }} 
\\ \cline{1-2} \cline{4-5}
\phi_0'   & (0 \dots 1.45) \,m_t  &&
\xi_0	 & (\sqrt{\tfrac{1}{60}} \dots \sqrt{\tfrac{7}{4}}) m_N \\
\upsilon_0'   & \lambda m_t                 && 
\upsilon_{45} & \th \sqrt{3} \lambda m_t    \\
\psi_0	  & \sqrt{\tfrac{2}{5}} m_N &&
\psi_3'   & (0 \dots \tfrac{1}{12} \sqrt{\tfrac{11}{3}}) \tfrac{m_N^2}{M}
\\ 
\cline{1-2} \cline{4-5} 
\mc{5}{c}{ \mbox{ \textbf{2.} The colour--neutral Higgs fields of charge 
$\mp 1\,$: \ru{3}{0} }} 
\\ \cline{1-2} \cline{4-5} 
\tfrac{1}{\sqrt{2}} (\upsilon_{18} \pm \iu \upsilon_{63} ) & 
\th \sqrt{3} \lambda m_t && 
\tfrac{1}{\sqrt{2}} (\psi_1 \pm \iu \psi_2 ) &	    
(0 \dots \tfrac{1}{12} \sqrt{\tfrac{11}{3}}) \tfrac{m_N^2}{M}
\\ 
\cline{1-2} \cline{4-5} 
\mc{5}{c}{ \mbox{ \textbf{3.} The neutral Higgs fields, for $i=0, \dots ,7\,$: 
\ru{3}{0} }} 
\\ \cline{1-2} \cline{4-5} 
\psi_{i+1} & (0 \dots \tfrac{1}{12} \sqrt{\tfrac{11}{3}}) \tfrac{m_N^2}{M} & && 
\\
\upsilon_{i+1} & (\lambda \dots \lambda {+} \check{\lambda}) m_n &&
\upsilon_{i+45} & (\lambda \dots \lambda {+} \check{\lambda}) m_n \\
\xi_{i+32} &  3 M &&
\xi_{i+81} &  3 M 
\\ 
\cline{1-2} \cline{4-5} 
\mc{5}{c}{ \mbox{ \textbf{4.} The Higgs fields of charge $\mp 1\,,$ for 
$i=0 \dots 7\,$: \ru{3}{0} }} 
\\ \cline{1-2} \cline{4-5} 
\tfrac{1}{\sqrt{2}} (\upsilon_{19+i} \pm \iu \upsilon_{64+i} ) & 
       (\lambda \dots \lambda {+} \check{\lambda}) m_n &&
\tfrac{1}{\sqrt{2}} (\xi_{25+i} \pm \iu \xi_{74+i} ) & 3 M 
\\ 
\cline{1-2} \cline{4-5} 
\mc{5}{c}{ \mbox{ \textbf{5.} The Higgs fields of charge $\mp \tfrac{1}{3}\,,$ 
for $i=0,1,2$ and $j=0,\dots,5\,$: \ru{3}{0} }} 
\\ \cline{1-2} \cline{4-5}
\tfrac{1}{\sqrt{2}} (\phi_{1+i} \pm \iu \phi_{3+i} )	& M &&
\tfrac{1}{\sqrt{2}} (\upsilon_{9+i} \pm \iu \upsilon_{54 +i} )	& M \\
\tfrac{1}{\sqrt{2}} (\upsilon_{12+i} \pm \iu \upsilon_{57 +i} ) & M &&
\tfrac{1}{\sqrt{2}} (\upsilon_{39+i} \pm \iu \upsilon_{84 +i} ) & 2 M \\
\tfrac{1}{\sqrt{2}} (\xi_{44+i} \pm \iu \upsilon_{93 +i} ) & M &&
\tfrac{1}{\sqrt{2}} (\xi_{47+i} \pm \iu \upsilon_{96+i} ) & 2 M \\
\tfrac{1}{\sqrt{2}} (\xi_{19+j} \pm \iu \upsilon_{68+j} ) & 2 M &&
\tfrac{1}{\sqrt{2}} (\upsilon_{30+j} \pm \iu \upsilon_{75+j} ) & M 
\\ 
\cline{1-2} \cline{4-5}
\mc{5}{c}{ \mbox{ \textbf{6.} The Higgs fields of charge $\pm \tfrac{2}{3}\,,$ 
for $i=0,1,2$ and $j=0,\dots,5\,$: \ru{3}{0} }} 
\\ \cline{1-2} \cline{4-5}
\tfrac{1}{\sqrt{2}} (\upsilon_{15+i} \pm \iu \upsilon_{60 +i} )  & M &&
\tfrac{1}{\sqrt{2}} (\upsilon_{36+i} \pm \iu \upsilon_{81 +i} ) & 2 M \\
\tfrac{1}{\sqrt{2}} (\upsilon_{42+i} \pm \iu \upsilon_{87 +i} ) & M &&
\tfrac{1}{\sqrt{2}} (\xi_{41+i} \pm \iu \upsilon_{90 +i} ) & M \\
\tfrac{1}{\sqrt{2}} (\xi_{7 +j} \pm \iu \xi_{56 +j} ) & 2 M &&
\tfrac{1}{\sqrt{2}} (\xi_{13+j} \pm \iu \xi_{62 +j} ) & 4 M 
\\ 
\cline{1-2} \cline{4-5}
\mc{5}{c}{ \mbox{ \textbf{7.} The Higgs fields of charge $\mp \tfrac{4}{3}\,,$ 
for $i=0,1,2$ and $j=0,\dots,5\,$: \ru{3}{0} }} 
\\ \cline{1-2} \cline{4-5}
\tfrac{1}{\sqrt{2}} (\upsilon_{27+i} \pm \iu \upsilon_{72+i} )	& M &&
\tfrac{1}{\sqrt{2}} (\xi_{1+j} \pm \iu \upsilon_{50+j} )  & 2 M 
\\ 
\cline{1-2} \cline{4-5}
\mc{5}{c}{ \mbox{ \textbf{8.} The neutral massive gauge fields: \ru{3}{0} }} 
\\ \cline{1-2} \cline{4-5} 
Z & \sqrt{\tfrac{2}{5}} \,m_t && 
Z' & \th \sqrt{\tfrac{5}{3}} m_N 
\\ 
\cline{1-2} \cline{4-5}
\mc{5}{c}{ \mbox{ \textbf{9.} The massive gauge fields of charge $\pm 1\,$: 
\ru{3}{0} }}
\\ \cline{1-2} \cline{4-5} 
\tfrac{1}{\sqrt{2}} (W_1 \mp \iu W_2) & \th m_t &&
\mc{2}{|c|}{ \mbox{Weinberg angle: } \sin^2 \theta_W = \tfrac{3}{8} } 
\\ 
\cline{1-2} \cline{4-5}
\mc{5}{c}{ \mbox{ \textbf{10.} The leptoquarks leading to proton decay, 
for $i=0,1,2\,$: \ru{3}{0} }}
\\ \cline{1-2} \cline{4-5}
\tfrac{1}{\sqrt{2}} (X_{1+i} \mp \iu X_{3+i} ) & M && 
\mc{2}{|c|}{ \mbox{charge: } \mp \tfrac{1}{3} } \\
\tfrac{1}{\sqrt{2}} (Y_{1+i} \mp \iu Y_{3+i} ) & M && 
\mc{2}{|c|}{ \mbox{charge: } \pm \tfrac{2}{3} }
\\ \cline{1-2} \cline{4-5}
\end{array}$ \vs*{0.5ex}
\caption{The particle masses for the $\SU5 \times \U1$--model } \label{tab1}
\end{table}%
We recall that $m_t$ is the mass of the top quark, $m_N$ the mass scale of 
the right neutrinos and $M$ the grand unification scale, where we have assumed 
$m_N,M \gg m_t\,.$ Moreover, we have introduced the abbreviation
\[
\check{\lambda} = \sqrt{\lambda^2 +\tfrac{m_b^2 \nu_{10}^2}{m_t^2 m_n^2}} 
- \lambda \geq 0~.
\]

It is interesting to perform the transformation \rf[fy] in the Yukawa Larangian 
$\Lgr[m]$ of the fermionic action \rf[SFfin]. The contribution of the coupling 
of the $\phi_0'$ Higgs field to the fermions takes the form 
\eqa{rcl}{
\Lgr[\phi_0'] &=& \big( -\bsd_L^* ( \one_3 \ot ( M_d 
+ \tfrac{g_0}{\sqrt{8 (\mu_1+12 \mu_2)}} \phi_0' M_d ) ) \bsd_R 
- e_L^* ( M_e+ \tfrac{g_0} {\sqrt{8 (\mu_1 + 12 \mu_2)}} \phi_0' M_e ) e_R 
\npb \\
&-& \bsu_L^* (\one_3 \ot (M_u + \tfrac{g_0}{\sqrt{8 (\mu_1 + 12 \mu_2)}} 
\phi_0' M_u )) \bsu_R 
- \nu_L^* ( M_n^T + \tfrac{g_0} {\sqrt{8 (\mu_1+ 12 \mu_2)}} \phi_0' M_n^T ) 
\nu_R \big) \npb \\
&+& \mathrm{h.c} \npb \yn \label{f0} \\
&=& \big( -\bsd_L^* ( \one_3 \ot (1+ \tfrac{g_0}{m_t} \phi_0' ) M_d ) \bsd_R 
- e_L^* ( (1+ \tfrac{g_0}{m_t} \phi_0' )  M_e ) e_R \npb\\
&-& \bsu_L^* (\one_3 \ot (1+ \tfrac{g_0}{m_t} \phi_0' )  M_u )
) \bsu_R - \nu_L^* ((1+ \tfrac{g_0}{m_t} \phi_0' ) M_n^T ) \nu_R \big) 
+ \mathrm{h.c} ~.  
 }
Thus, the Higgs field $\phi_0'$ has the same properties as the standard 
model Higgs field. 

All other Higgs fields are too massive to observe. All Higgs and gauge 
fields with fractional--valued charge lead to proton decay. Without exception 
they receive a mass of the order of the grand unification scale $M\,,$ which 
must be chosen sufficiently large to ensure the observed stability of matter. 
The mass of the remaining Higgs fields with integer--valued charge is of the 
order $M,\lambda m_t, \lambda m_n, m_N$ or $\tfrac{m_N^2}{M}\,.$ These mass 
scales are situated somewhere between $m_t$ and $M\,.$ By assumption, $m_N$ and 
$\tfrac{m_N^2}{M}$ are very close to $M\,.$ Moreover, for generic choices of 
the mass matrices $M_{10}$ and $M_5\,,$ also $\lambda m_t$ and $\lambda m_n$ 
are close to $M\,.$ 

\section{The $\SU5$--Model}
\label{su5m}

Formally, we can derive the $\SU5$--model from our flipped 
$\SU5 \times \U1$--model discussed so far by the following restrictions and 
replacements: We put $\t{A}_{\mu} \equiv0$ ad hoc. Strictly speaking, this step 
is not allowed within our theory. However, one could imagine a formalism where 
the connection forms are not $\mbox{$(\Lambda^1 \ot \bbr{0})$} \op 
\mbox{($\Lambda^0 \g \ot \bbr{1})$}$--valued but 
$\mbox{$(\Lambda^1 \ot \h{\pi}(\f[a]))$} \op 
\mbox{($\Lambda^0 \g \ot \p{1}$)}$--valued. Now, taking the same L--cycle as 
before, however with $M_N \equiv 0\,,$ we obtain indeed a $\SU5$--GUT. The 
calculation is the same as before. However, since the graded centre $\cc{2}$ 
is not relevant in such a model, we must put $J_3 =0$ and $\zeta_{A,B,U,V}=0$ 
in the factorization procedure of Section~\ref{fact}. 
Moreover, instead of \rf[ort3] we perform the orthogonal transformation 
\eq[ort2]{
- \mb[\!\!]{c}{ Z_{\mu} \\ P_{\mu} } \!=\!
\mb[\!]{Z{18}}{
\cos \theta_W & {-} \sin \theta_W \\ \sin \theta_W & \cos \theta_W } \!\!
\mb[\!\!]{c}{ W^3_{\mu} \\  A'_{\mu} } ,\qquad 
\sin \theta_W=\sqrt{\tfrac{3}{8}}~.
  }
If we compute the electric charges we find that the labels are unconvenient 
now, because $u$ describes the $d$ quarks (and vice versa) and $\nu$ the 
electrons (and vice versa). Thus, we must permute the labels 
$u \leftrightarrow d$ and $\nu \leftrightarrow e\,.$ Then we obtain almost the 
same form \rf[SFfin] for the fermionic action, with the following 
modifications: 
\vs{-\topsep}
\begin{enumerate}
\item
We have $M_N \equiv0\,,$ in particular, the Lagrangian $\Lgr[h]''$ and the 
Higgs fields $\xi_E^0, \xi_F^0$ and $\xi_0$ are absent.
\vs{-\itemsep} \vs{-\parsep}

\item
In the Lagrangians $\Lgr[q]$ and $\Lgr[\ell]$ we have $Z_{\mu}' \equiv0\,,$ 
$\t{Z}_{\mu}' \equiv0$ and $\theta_W' \equiv0\,.$ 
\vs{-\itemsep} \vs{-\parsep}

\item 
In the Lagrangians $\Lgr[x]\,,$ $\Lgr[h]$ and $\Lgr[h]'\,,$ the 
fermion labels $\bsu$ and $\bsd$ are exchanged and $e$ and $\nu$ are exchanged. 
\vs{-\itemsep} \vs{-\parsep}

\item
The same exchanges occurs for the mass matrices: 
$M_u \leftrightarrow M_d\,,$ $\Mu \leftrightarrow M_{\t{d}}\,,$ 
$M_e \leftrightarrow M_n\,,$ $\Mn \leftrightarrow M_{\t{e}}\,.$

\end{enumerate}
\vs{-\topsep}
Compared with the $\SU5 \times \U1$--model one finds \cite{phd} identical 
tree--level predictions for the Weinberg angle, the masses of the $W$ and $Z$ 
boson and for the Higgs and gauge fields with fractional--valued charge. Again, 
there is precisely one standard model Higgs field $\phi_0'$ whose upper bound 
for the mass turns out to be $m_{\phi_0'} = 1.32\,m_t\,.$ Moreover, the masses 
of the remaining Higgs fields with integer--valued charge lie between $m_t$ and 
$M\,,$ generically close to $M\,.$ 

\section{Conclusion}
\label{cgu}

\settowidth\labelwidth{4.}%
	    \leftmargini\labelwidth
            \advance\leftmargini\labelsep
\begin{enumerate}

\item
We have succeeded in formulating the flipped $\SU5 \times \U1$--GUT within the 
framework of non--associative geometry. We have found interesting tree--level 
relations between fermionic and bosonic parameters: Given the fermionic 
parameters (fermion masses and Kobayashi--Maskawa mixing angles) and two 
$3 \times 3$--matrices determining the unification scale as input, we were able 
to compute all bosonic quantities: %\vs{-\topsep}
\begin{itemize}
\item the occurring multiplets of Higgs fields,
      \vs{-\itemsep} \vs{-\parsep}
\item the spontaneous symmetry breaking pattern,
      \vs{-\itemsep} \vs{-\parsep}
\item the masses of all Higgs fields,
      \vs{-\itemsep} \vs{-\parsep}
\item the masses of all Yang--Mills fields,
      \vs{-\itemsep} \vs{-\parsep}
\item the Weinberg angle.
\end{itemize} %\vs{-\topsep}
However, since not all input parameters are known, we were forced to be 
satisfied with estimations for some of the masses. 

\item 
The representation of the $\U1$--part of the $\SU5 \times \U1$--model is not an 
input but an algebraic consequence of the theory. This $\U1$--representation 
is unique and realized in nature. 

\item
In the $\SU5 \times \U1$--model there occur Higgs fields in complex $\ul{5}$--, 
complex \mbox{$\ul{50}$--,} complex $\ul{45}$-- and real $\ul{24}$--plets. 
After the spontaneous symmetry breaking, there survive $12$ Higgs fields of the 
$\ul{24}$--representation, $7$ Higgs fields of the $\ul{5}$--representation, 
$99$ Higgs fields of the $\ul{50}$--representation and $90$ Higgs fields of 
the $\ul{45}$--representation, and $16$ gauge fields become massive. 

\item
There occur three mass scales in the model under consideration: 
%\vs{-\topsep} 
\begin{itemize}
\item The lowest mass scale is the scale of the fermion masses reaching from 
      the neutrino masses to the mass of the top quark. Moreover, also the 
      electroweak gauge fields $Z,W^+,W^-$ belong to this scale, and~-- 
      remarkably~-- one Higgs field as well. 
      \vs{-\itemsep} \vs{-\parsep}

\item The mass of all fields leading to 
      proton decay is of the order of the grand unification scale $M\,.$ 
      \vs{-\itemsep} \vs{-\parsep}
      
\item The masses of Higgs fields which do not lead to proton decay lie 
      between the fermions scale and the grand unification scale, generically 
      close to $M\,.$ 
\end{itemize} %\vs{-\topsep}

\item 
There exists precisely one light Higgs field $\phi_0'\,,$ which has exactly the 
same properties as the standard model Higgs field. It couples to a fermion of 
the mass $m_f$ with the coupling constant $g_0 m_f/m_t\,.$ Moreover, it has the 
same couplings with the intermediate vector bosons $Z,W^+,W^-$ as the standard 
model Higgs field. The Higgs field $\phi_0'$ is a certain linear combination 
of the $\ul5$--representation and the $\ul{45}$--representation. 
This linear combination is the only one 
which corresponds to a zero mode of the grand unification sector. That the 
mass of $\phi_0'$ is generically different from zero is due to the fermion 
masses. Therefore, the Higgs field $\phi_0'$ receives a mass of the order of 
the mass $m_t$ of the top quark: For $m_t=176\,\mathrm{GeV}$ we have
$m_{\phi_0'} \leq 255\,\mathrm{GeV}\,.$
The reason that only an upper bound can be given is the incomplete knowledge of 
the input parameters. The upper bound is independent of all parameters related 
to grand unification. 

\item
The standard model is in perfect agreement with experiment. However, we have 
shown that the low energy sector of the $\SU5 \times \U1$--GUT is identical 
with the standard model. Thus, one must be careful with the extrapolation of 
the standard model to higher energies. 

\end{enumerate}

\begin{appendix}

\section{The Generation Space Matrices}
\label{appa} \vs*{-1ex}

\eqa{rclrcl}{
\h{M}^{10}_{aa} &:=& \tfrac{3}{10} M_{10}'{\!}^2 {+} (\tfrac{3}{5}\alpha_A 
{+} \zeta_A) \one_6  \;,  &
\h{M}^{10}_{cc} &:=& \tfrac{1}{10} M_N' M_N'{}^* {+} (\tfrac{3}{5}\alpha_U 
{+} \zeta_U) \one_6 \;,  \npb \\
\h{M}^{10}_{nn} &=& \tfrac{1}{10} \Mn' \Mn'{}^* 
{+} (\tfrac{3}{5} \alpha_V {+} \zeta_V) \one_6 \;, \npb \\ 
\h{M}^{10}_{bb} &:=& \tfrac{2}{5} \Mu' \Mu'{}^* {+} \tfrac{3}{5} M_d' M_d'{}^* 
{+} (\tfrac{3}{5} \alpha_B {+} \zeta_B) \one_6 \;, \hs*{-15em} 
\\
M^{10}_{aa} &:=& M_{10}'{\!}^2 {+} \beta_A \one_6 {+} 3 \delta_A \Mud^2 \;,
 & M^{10}_{nn} &:=& \Mn' \Mn'{}^* 
{+} \tfrac{1}{3} \beta_V \one_6 {+} \delta_V \Mud^2 \;,  \npb \\
M^{10}_{cc} &:=& M_N' M_N'{}^* {+} \tfrac{1}{3} \beta_U \one_6 
{+} \delta_U \Mud^2 \;, \qquad{} & 
\check{M}^{10}_{nn} &:=& \tfrac{1}{3} \check{\beta}_V \one_6 
{+} \check{\delta}_V \Mud^2 \;,  \\
M^{10}_{\{un\}} &:=& \th (\Mu' \Mn'{}^* {+} \Mn' \Mu'{}^* ) 
{+} \beta_W \one_6 {+} 3 \delta_W \Mud^2 \;, \hs*{-15em} \\
M^{10}_{[un]} &:=& \tfrac{1}{2 \iu} (\Mu' \Mn'{}^* {-} \Mn' \Mu'{}^* ) 
{+} \beta_W' \one_6 {+} 3 \delta_W' \Mud^2 \;, \hs*{-15em} \\ 
\t{M}^{10}_{nn} &:=& \Mn' \Mn'{}^* {+} \gamma_V \one_6 
{+} \epsilon_V \t{M}_V^2 \;, &
\t{M}^{10}_{cc} &:=& M_N' M_N'{}^* {+} \gamma_U \one_6 
{+} \epsilon_U \t{M}_V^2 \;, 
\npb \\ 
\t{M}^{10}_{\{cd\}} &:=& \th (M_N' M_d'{}^* {+} M_d' M_N'{}^*) 
{+} \t{\gamma}_U \one_6 {+} \t{\epsilon}_U \t{M}_V^2 \;, \hs*{-15em} \\
\t{M}^{10}_{[cd]} &:=& \tfrac{1}{2 \iu} (M_N' M_d'{}^* {-} M_d' M_N'{}^*) 
{+} \t{\gamma}_U' \one_6 {+} \t{\epsilon}_U' \t{M}_V^2 \;, \hs*{-15em} \npb \\ 
\t{M}^{10}_{\{un\}} &:=& \th (\Mu' \Mn'{}^* {+} \Mn' \Mu'{}^* ) 
{+} \gamma_W \one_6 {+} \epsilon_W \t{M}_V^2 \;, \hs*{-15em} \\
\t{M}_{[un]}^{10} &:=& \tfrac{1}{2 \iu} (\Mu' \Mn'{}^* {-} \Mn' \Mu'{}^* ) 
{+} \gamma_W' \one_6 {+} \epsilon_W' \t{M}_V^2 \;, \hs*{-15em} \yn \label{Mht}
\smallskip
\\
\h{M}^5_{aa} &:=& \tfrac{1}{5} M_5'{}^2 {+} (\tfrac{2}{5} \alpha_A 
{+} \zeta_A) \one_6 \;, \qquad{} & 
\h{M}^5_{cc} &:=& (\tfrac{2}{5} \alpha_U {+} \zeta_U) \one_6 \;, \npb \\
\h{M}^5_{nn} &:=& \tfrac{1}{5} \Mn'{}^* \Mn' {+} (\tfrac{2}{5} \alpha_V 
{+} \zeta_V) \one_6 \;, \\ 
\h{M}_{bb}^5 &:=& \tfrac{4}{5} \Mu'{}^* \Mu' 
{+} \tfrac{1}{5} \bar{M}_e' M_e'{}^T {+} (\tfrac{2}{5} \alpha_B 
{+} \zeta_B) \one_6 \;, \hs*{-15em} 
\npb \\
M^5_{aa} &:=& M_5'{}^2 {+} \beta_A \one_6 {+} \delta_A \Men^2 \;,  &
\check{M}_{nn}^5 &:=& \Mn'{}^* \Mn' {+} \check{\beta}_V \one_6 
{+} \check{\delta}_V \Men^2\;, \npb \\
M_{nn}^5 &:=& \beta_V \one_6 {+} \delta_V \Men^2\;, \qquad &
M_{cc}^5 &:=& \beta_U \one_6 {+} \delta_U \Men^2\;, \npb \\
M_{\{un\}}^5 &:=& \th (\Mn'{}^* \Mu' {+} \Mu'{}^* \Mn') {+} \beta_W \one_6 
{+} \delta_W \Men^2 \;, \hs*{-15em}  \npb \\ 
M^5_{[un]} &:=& \tfrac{1}{2 \iu} (\Mn'{}^* \Mu' {-} \Mu'{}^* \Mn') 
{+} \beta_W' \one_6 {+} \delta_W' \Men^2 \;, \hs*{-15em}
\smallskip
\\
\h{M}_{aa}^1 &:=& \zeta_A \one_6 \;,  &
\h{M}_{nn}^1 &:=& \zeta_V \one_6 \;, \npb \\
\h{M}_{cc}^1 &:=& \zeta_U \one_6 \;, & 
\h{M}_{bb}^1 &:=& M_e'{}^T \bar{M}_e' {+} \zeta_B \one_6 \;.
    }
The real constants $\alpha_A, \dots \zeta_V$ are determined by 
equation \rf[trt]. The solution is
\eqa{rclrcl}{
\alpha_A &=& {-} \tfrac{1}{8} \tr(M_{10}'{\!}^2) {+} \tfrac{1}{24} 
\tr (M_5'{}^2) \;, \quad{} &
\alpha_B &=& {-} \tfrac{1}{4} \tr(M_d' M_d'{}^*) 
{+} \tfrac{1}{4} \tr(M_e' M_e'{}^*)\;, \npb \eqnskip \\
\alpha_U &=& -\tfrac{1}{24} \tr( M_N' M_N'{}^*) \;, & \alpha_V &=& 0\;, \\
\zeta_A &=& \tfrac{1}{32} \tr(M_{10}'{\!}^2) - \tfrac{1}{32} \tr (M_5'{}^2) \;, 
\quad{} & \zeta_V &=& - \tfrac{1}{48} \tr(\Mn' \Mn'{}^*) \;, \eqnskip \\ 
\zeta_B &=& -\tfrac{1}{12} \tr(\Mu' \Mu'{}^*) 
+ \tfrac{1}{16} \tr(M_d' M_d'{}^*) - \tfrac{7}{48} \tr(M_e' M_e'{}^*) \;, 
\hs*{-15em} \\
\zeta_U &=& \tfrac{1}{96} \tr( M_N' M_N'{}^*) \;, 
\\
\beta_A &=& - \tfrac{1}{8} \tr(M_5'{}^2) - \tfrac{1}{24} \tr(M_{10}'{\!}^2) 
\;, \quad{}& \delta_A &=& - \dfrac{ \tr(M_{10}'{\!}^2 \Mud^2 
+ M_5'{}^2 \Men^2)} { \tr(3 (\Mud^2)^2 +(\Men^2)^2)}\;, \eqnskip \\
\beta_U &=& - \tfrac{1}{8} \tr(M_N' M_N'{}^*) \;, &
\delta_U &=& - 3 \dfrac{\tr(M_N' M_N'{}^* \Mud^2)}{ \tr(3 (\Mud^2)^2 + 
(\Men^2)^2)} \;, \\
\beta_V &=& - \tfrac{1}{8} \tr(\Mn' \Mn'{}^*) \;, \quad{}&
\check{\beta}_V &=& - \tfrac{1}{8} \tr(\Mn' \Mn'{}^*) \;, 
\yn \label{rpl2} 
\\
\delta_V &=& - \dfrac{\tr (3 \Mn' \Mn'{}^* \Mud^2)}{\tr (3(\Mud^2)^2 
+(\Men^2)^2)} \;, \quad{} & \check{\delta}_V &=& 
-\dfrac{\tr (\Mn'{}^* \Mn' \Men^2)}{\tr(3(\Mud^2)^2 + (\Men^2)^2 )} \;, \\
\beta_W &=& - \tfrac{1}{12} \tr(\Mu' \Mn'{}^* + \Mn' \Mu'{}^*) \;,\quad{} &
\beta_W' &=& - \tfrac{1}{12 \iu} \tr(\Mu' \Mn'{}^* - \Mn' \Mu'{}^*) \;, 
\eqnskip \\[0.5ex]
\delta_W &=& - \dfrac{ \tr((\Mu' \Mn'{}^* + \Mn' \Mu'{}^*) \Mud^2 
+ (\Mn'{}^* \Mu' + \Mu'{}^* \Mn') \Men^2)}
{ 2\,\tr(3 (\Mud^2)^2 +(\Men^2)^2)}\;, \hs*{-15em} \eqnskip \\[1.5ex]
\delta_W' &=& - \dfrac{ \tr((\Mu' \Mn'{}^* - \Mn' \Mu'{}^*) \Mud^2 
+ (\Mn'{}^* \Mu' - \Mu'{}^* \Mn') \Men^2)} 
{2 \iu \,\tr(3 (\Mud^2)^2 +(\Men^2)^2)}\;, \hs*{-15em} \\
\gamma_V &=& - \tfrac{1}{6} \tr(\Mn' \Mn'{}^*) \;, \quad{} &
\epsilon_V &=& - \dfrac{\tr(\Mn' \Mn'{}^* \t{M}_V^2)}{\tr((\t{M}_V^2)^2)}\;, 
\\
\gamma_W &=& - \tfrac{1}{12} \tr (\Mu' \Mn'{}^* + \Mn' \Mu'{}^*) \;,\quad{} &
\epsilon_W &=& - \dfrac{\tr((\Mu' \Mn'{}^* + \Mn' \Mu'{}^*) \t{M}_V^2)}{2\, 
\tr((\t{M}_V^2)^2)}\;, \npb \\[1ex]
\gamma_W' &=& - \tfrac{1}{12 \iu} \tr (\Mu' \Mn'{}^* - \Mn' \Mu'{}^*) \;, 
\quad{} & \epsilon_W' &=& - \dfrac{\tr((\Mu' \Mn'{}^* - \Mn' \Mu'{}^*) 
\t{M}_V^2)}{2 \iu\, \tr((\t{M}_V^2)^2)}\;,
\\
\gamma_U &=& - \tfrac{1}{6} \tr(M_N' M_N'{}^*) \;, \quad{} &
\epsilon_U &=& - \dfrac{\tr(M_N' M_N'{}^* \t{M}_V^2)}{\tr((\t{M}_V^2)^2)}\;, 
\\
\t{\gamma}_U &=& - \tfrac{1}{12} \tr (M_N' M_d'{}^* + M_d' M_N'{}^*) \;, 
\quad{} &
\t{\epsilon}_U &=& - \dfrac{\tr((M_N' M_d'{}^* + M_d' M_N'{}^*) 
\t{M}_V^2)}{2\, \tr((\t{M}_V^2)^2)}\;, \npb \\[1ex]
\t{\gamma}_U' &=& - \tfrac{1}{12 \iu} \tr (M_N' M_d'{}^* - M_d' M_N'{}^*) \;, 
\quad{} & \t{\epsilon}_U' &=& - \dfrac{\tr((M_N' M_d'{}^* - M_d' M_N'{}^*) 
\t{M}_V^2)}{2 \iu\, \tr((\t{M}_V^2)^2)}\;.
      }

\section{The Coefficients Occurring in the Higgs Potential} 
\label{appb} \vs*{-1ex}

\eqa{rclrcl}{
\mu_0	&=& \tfrac{1}{96} \tr(3 M_{10}'{\!}^2 + M_5'{}^2)~, \qquad{} &&& 
	   \hs*{2em} 
\mu_2 = \tfrac{1}{48} \tr( \Mn' \Mn'{}^* ) ~, 
	    \hs*{3em} \yn \label{nu012} \npb \\
\mu_1	&=& \tr (\tfrac{1}{16} M_d' M_d'{}^* + \tfrac{1}{12} \Mu' \Mu'{}^* 
	    + \tfrac{1}{48} M_e' M_e'{}^*) ~, \hs*{-15em}  &&& \hs*{2em} 
\mu_3 = \tfrac{1}{96} \tr( M_N' M_N'{}^* ) ~, \hs*{-15em}
\eqnskip \eqnskip
\smallskip
\\
\la[a] &=& \tr( 10  (\h{M}^{10}_{aa})^2 + 5 (\h{M}^5_{aa})^2 
       + (\h{M}^1_{aa})^2 ) ~, \hs*{-15em}  \yn \label{la} \npb \\
\la[b] &=& \tr(10 (\h{M}^{10}_{bb})^2 + 5 (\h{M}^5_{bb})^2 + (\h{M}^1_{bb})^2) 
       {}~,\hs*{-15em}	\\
\la[c] &=& \tr( 10 (\h{M}^{10}_{nn})^2 + 5 (\h{M}^5_{nn})^2 
       + (\h{M}^1_{nn})^2 ) ~, \hs*{-15em}  \\
\la[d] &=& \tr( 20 \h{M}^{10}_{aa} \h{M}^{10}_{bb} 
       + 10 \h{M}^5_{aa} \h{M}^5_{bb} + 2 \h{M}^1_{aa} \h{M}^1_{bb} ) ~,  
       \hs*{-15em}  \\
\la[e] &=& \tr( 20 \h{M}^1_{aa} \h{M}^{10}_{nn} 
       + 10 \h{M}^5_{aa} \h{M}^5_{nn} + 2 \h{M}^1_{aa} \h{M}^1_{nn} ) 
       \hs*{-15em}  \\
\la[f] &=& \tr( 20 \h{M}^{10}_{bb} \h{M}^{10}_{nn} 
       +10 \h{M}^5_{bb} \h{M}^5_{nn} +2 \h{M}^1_{bb} \h{M}^1_{nn} )~, 
       \hs*{-15em}  \\
\la[g] &=& \tr ( \tfrac{3}{2} (M_{10}' M_d' + M_d' M_{10}{\!}^T)
	   (M_{10}' M_d' + M_d' M_{10}{\!}^T)^* 
	   + 2 (M_5'{}^T)^2 M_e' M_e'{}^* ) ~, \hs*{-15em} \\
\la[h] &=& \tfrac{1}{4} \tr ( (M_{10}' M_d' - M_d' M_{10}'{\!}^T) 
        (M_{10}' M_d' - M_d' M_{10}'{\!}^T )^*)~, \hs*{-15em} \\
\la[i] &=& \tr ( 2 \Mu' \Mu'{}^* M_{10}'{\!}^2 ) ~, \hs*{5em} & 
\la[j] &=& \tr ( 2 \Mu'{}^* \Mu' M_5'{}^2 ) ~,  \\
\la[k] &=& \tr ( 2 \Mn' \Mn'{}^* M_{10}'{\!}^2 ) ~, & 
\la[l] &=& \tr ( 2 \Mn'{}^* \Mn' M_5'{}^2 ) ~, \\
\la[m] &=& \tr ( 4 \Mu' M_5' \Mu'{}^* M_{10}' ) ~, &
\la[n] &=& \mathrm{Re}(\tr ( 4 \Mu' \Mn'{}^* M_{10}'{\!}^2 )) ~, \\
\la[o] &=& \mathrm{Im}(\tr ( 4 \Mu' \Mn'{}^* M_{10}'{\!}^2 )) ~, & 
\la[p] &=& \mathrm{Re}(\tr ( 4 \Mu' M_5' \Mn'{}^* M_{10}' )) ~,  \\
\la[q] &=& \mathrm{Im}(\tr ( 4 \Mu' M_5' \Mn'{}^* M_{10}' )) ~, & 
\la[r] &=& \mathrm{Re}(\tr ( 4 \Mn'{}^* \Mu' M_5'{}^2 )) ~,  \\
\la[s] &=& \mathrm{Im}(\tr ( 4 \Mn'{}^* \Mu' M_5'{}^2 )) ~, &
\la[t] &=& \tr ( 4 \Mn' M_5' \Mn'{}^* M_{10}' ) ~,  \\
\la[u] &=& \tr ( 2 \Mn'{}^* \Mn' \bar{M}_e' M_e'{}^T ) ~, &
\la[v] &=& \tr ( 2 \Mn' \Mn'{}^* M_d' M_d'{}^* ) ~, \\
\la[w] &=& \tr ( 2 M_{d\t{n}}' M_{d\t{n}}'{}^* ) ~, 
\eqnskip \eqnskip
\smallskip
\\
\lac[a] &=& \tr( 10  (\h{M}^{10}_{cc})^2 + 5 (\h{M}^5_{cc})^2 
	+ (\h{M}^1_{cc})^2 ) ~, \hs*{-15em}  \yn \label{lac} \npb \\
\lac[b] &=& \tr( 20 \h{M}^{10}_{cc} \h{M}^{10}_{aa} 
	+ 10 \h{M}^5_{cc} \h{M}^5_{aa} + 2 \h{M}^1_{cc} \h{M}^1_{aa} ) ~,  
	\hs*{-15em}  \\
\lac[c] &=& \tr( 20 \h{M}^1_{cc} \h{M}^{10}_{bb} 
	+ 10 \h{M}^5_{cc} \h{M}^5_{bb} + 2 \h{M}^1_{cc} \h{M}^1_{bb} ) 
	\hs*{-15em}  \\
\lac[d] &=& \tr( 20 \h{M}^{10}_{cc} \h{M}^{10}_{nn} 
	+10 \h{M}^5_{cc} \h{M}^5_{nn} +2 \h{M}^1_{cc} \h{M}^1_{nn} )~, 
	\hs*{-15em}  \\
\lac[e] &=& \tfrac{1}{4} \tr ( (M_{10}' M_N' + M_N' M_{10}'{\!}^T) 
	(M_{10}' M_N' + M_N' M_{10}'{\!}^T )^*)~, \hs*{-15em} \\
\lac[f] &=& \tfrac{1}{4} \tr ( (M_{10}' M_N' - M_N' M_{10}'{\!}^T) 
        (M_{10}' M_N' - M_N' M_{10}'{\!}^T )^*)~, \hs*{-15em} \\
\lac[g] &=& \tfrac{1}{2} \mathrm{Re} ( 
        \tr ( (M_{10}' M_N' - M_N' M_{10}'{\!}^T) 
        (M_{10}' M_d' - M_d' M_{10}'{\!}^T )^*))~, \hs*{-15em} \\
\lac[h] &=& \tfrac{1}{2} \mathrm{Im} ( 
        \tr ( (M_{10}' M_N' - M_N' M_{10}'{\!}^T) 
	(M_{10}' M_d' - M_d' M_{10}'{\!}^T )^*))~, \hs*{-15em} \\
\lac[i] &=& \tr ( 2 \Mn' \Mn'{}^* M_N' M_N'{}^* ) ~, & 
\lac[j] &=& \mathrm{Re} (\tr ( 4 \Mn' \Mn'{}^* M_N' M_d'{}^* )) ~, \\
\lac[k] &=& \mathrm{Im} (\tr ( 4 \Mn' \Mn'{}^* M_N' M_d'{}^* )) ~, &
\lac[l] &=& \tr ( 2 M_{Nu}' M_{Nu}'{}^* ) ~, \\
\lac[m] &=& \mathrm{Re} (\tr ( 4 M_{d\t{n}}' M_{Nu}'{}^* )) ~, &
\lac[n] &=& \mathrm{Im} (\tr ( 4 M_{d\t{n}}' M_{Nu}'{}^* )) ~, 
\eqnskip \eqnskip
\smallskip
\\
\lat[a] &=& \tr( \tfrac{1}{3} (M^{10}_{aa})^2 + (M^5_{aa})^2 )~, & 
\lat[b] &=& \tr( 3 (\check{M}^{10}_{nn})^2 + (\check{M}^5_{nn})^2 )~,
	\label{lat} \yn \\
\lat[c] &=& \tr( 3 (M^{10}_{nn})^2 + (M^5_{nn})^2 )~,  &
\lat[d] &=& \tr( \tfrac{1}{3} (M^{10}_{\{un\}})^2 + (M^5_{\{un\}})^2 )~, \\
\lat[e] &=& \tr( \tfrac{1}{3} (M^{10}_{[un]})^2 + (M^5_{[un]})^2 )~, &
\lat[f] &=& \tr( 2 M^{10}_{aa} \check{M}^{10}_{nn} 
	+ 2 M^5_{aa} \check{M}^5_{nn}) ~, \\
\lat[g] &=& \tr( 2 M^{10}_{aa} M^{10}_{nn} + 2 M^5_{aa} M^5_{nn} )~, & 
\lat[h] &=& \tr( \tfrac{2}{3} M^{10}_{aa} M^{10}_{\{un\}} 
	+ 2 M^5_{aa} M^5_{\{un\}} )~, \\
\lat[i] &=& \tr( \tfrac{2}{3} M^{10}_{aa} M^{10}_{[un]} 
	+ 2 M^5_{aa} M^5_{[un]})~, &
\lat[j] &=& \tr( 6 \check{M}^{10}_{nn} M^{10}_{nn} 
	+ 2 \check{M}^5_{nn} M^5_{nn} )~, \\
\lat[k] &=& \tr( 2 \check{M}^{10}_{nn} M^{10}_{\{un\}} 
	+ 2 \check{M}^5_{nn} M^5_{\{un\}} )~, &
\lat[l] &=& \tr( 2 \check{M}^{10}_{nn} M^{10}_{[un]} 
	+ 2 \check{M}^5_{nn} M^5_{[un]} )~, \\
\lat[m] &=& \tr( 2 M^{10}_{nn} M^{10}_{\{un\}} 
	+ 2 M^5_{nn} M^5_{\{un\}} )~, &
\lat[n] &=& \tr( 2 M^{10}_{nn} M^{10}_{[un]} 
	+ 2 M^5_{nn} M^5_{[un]} )~, \\
\lat[o] &=& \tr( \tfrac{2}{3} M^{10}_{\{un\}} M^{10}_{[un]} 
	+ 2 M^5_{\{un\}} M^5_{[un]} )~, \\
\lat[p] &=& \tr( (\t{M}^{10}_{nn})^2) ~,  & 
\lat[q] &=& \tr( (\t{M}^{10}_{\{un\}})^2) ~, \\
\lat[r] &=& \tr( (\t{M}^{10}_{[un]})^2) ~,  &
\lat[s] &=& \tr( 2 \t{M}^{10}_{nn} \t{M}^{10}_{\{un\}} ) ~, \\
\lat[t] &=& \tr( 2 \t{M}^{10}_{nn} \t{M}^{10}_{[un]} ) ~,  &
\lat[u] &=& \tr( 2 \t{M}^{10}_{\{un\}} \t{M}^{10}_{[un]} ) ~,
\smallskip
\eqnskip \eqnskip
\\
\lah[a] &=& \tr( 3 (M^{10}_{cc})^2 + (M^5_{cc})^2 )~, & 
\lah[b] &=& \tr( 2 M^{10}_{cc} M^{10}_{aa} + 2 M^5_{cc} M^5_{aa} )~, \\
\lah[c] &=& \tr( 6 M^{10}_{cc} \check{M}^{10}_{nn} 
	    + 2 M^5_{cc} \check{M}^5_{nn} )~, &
\lah[d] &=& \tr( 6 M^{10}_{cc} M^{10}_{nn} + 2 M^5_{cc} M^5_{nn} )~, \\
\lah[e] &=& \tr( 2 M^{10}_{cc} M^{10}_{\{un\}} + 2 M^5_{cc} M^5_{\{un\}} )~, 
	    &
\lah[f] &=& \tr( 2 M^{10}_{cc} M^{10}_{[un]} + 2 M^5_{cc} M^5_{[un]} )~, 
\eqnskip \eqnskip
\smallskip 
\\
\lah[g] &=& \tr( (\t{M}^{10}_{cc})^2) ~,  & 
\lah[h] &=& \tr( (\t{M}^{10}_{\{cd\}})^2) ~,  \yn \label{lad} \\
\lah[i] &=& \tr( (\t{M}^{10}_{[cd]})^2) ~, &
\lah[j] &=& \tr( 2 \t{M}^{10}_{cc} \t{M}^{10}_{\{cd\}}) ~, \\
\lah[k] &=& \tr( 2 \t{M}^{10}_{cc} \t{M}^{10}_{[cd]}) ~, &
\lah[l] &=& \tr( 2 \t{M}^{10}_{\{cd\}} \t{M}^{10}_{[cd]}) ~, \\
\lah[m] &=& \tr( 2 \t{M}^{10}_{cc} \t{M}^{10}_{nn}) ~, &
\lah[n] &=& \tr( 2 \t{M}^{10}_{cc} \t{M}^{10}_{\{un\}}) ~, \\
\lah[o] &=& \tr( 2 \t{M}^{10}_{cc} \t{M}^{10}_{[un]}) ~, &
\lah[p] &=& \tr( 2 \t{M}^{10}_{\{cd\}} \t{M}^{10}_{nn}) ~, \\
\lah[q] &=& \tr( 2 \t{M}^{10}_{\{cd\}} \t{M}^{10}_{\{un\}}) ~, &
\lah[r] &=& \tr( 2 \t{M}^{10}_{\{cd\}} \t{M}^{10}_{[un]}) ~, \\
\lah[s] &=& \tr( 2 \t{M}^{10}_{[cd]} \t{M}^{10}_{nn}) ~, &
\lah[t] &=& \tr( 2 \t{M}^{10}_{[cd]} \t{M}^{10}_{\{un\}}) ~, \\
\lah[u] &=& \tr( 2 \t{M}^{10}_{[cd]} \t{M}^{10}_{[un]}) ~. 
         }

\section{The Quadratic Terms of the Higgs Potential}
\label{appc} \vs*{-1ex}

\begin{small}
\eqa{l}{
\Lgr[0] = \tfrac{1}{384} \big\{  \yn \label{lah}
\npb \\
\tfrac{1}{12 \mu_2 {+} \mu_1} \phi_0'{}^2 ( \begin{array}[t]{l} 
   8 \la[b] {+} 1152 \la[c] {+} 96 \la[f] 
   {+} \tfrac{3072}{5} \lat[b] {+} \tfrac{1024}{15} \lat[c] 
   {+} \tfrac{3072}{5} \lat[d] {+} \tfrac{1024}{5} \lat[j] 
   {-} \tfrac{3072}{5} \lat[k] {-} \tfrac{1024}{5} \lat[m] \\
   {+} 256 \lat[p] {+} 256 \lat[q] {-} 256 \lat[s] ) \end{array} \eqnskip \\
+ \sqrt{\tfrac{6}{5}} \tfrac{1}{\sqrt{\mu_0 (12 \mu_2 {+} \mu_1)}} 
   \psi_0 \phi_0' ( \begin{array}[t]{l} 
   8 \la[d] {+} 96 \la[e] {-} \tfrac{32}{3} \lah[c] {-} \tfrac{32}{9} \lah[d] 
   {+} \tfrac{32}{3} \lah[e] {-} \tfrac{80}{3} \lah[m] 
   {+} \tfrac{80}{3} \lah[n] \\
   {-} 512 \lat[b] {-} \tfrac{512}{9} \lat[c] {-} 512 \lat[d]
   {+} \tfrac{32}{5} \lat[f] {+} \tfrac{32}{15} \lat[g] 
   {-} \tfrac{32}{5} \lat[h] {-} \tfrac{512}{3} \lat[j] {+} 512 \lat[k] \\ 
   {+} \tfrac{512}{3} \lat[m] {-} \tfrac{1280}{3} \lat[p] 
   {-} \tfrac{1280}{3} \lat[q] {+} \tfrac{1280}{3} \lat[s] ) 
   \end{array} \eqnskip \\
+ \sqrt{3} \tfrac{1}{12 \mu_2 {+} \mu_1} \phi_0' \upsilon_0' 
   ( \begin{array}[t]{l} 
   \sqrt{\tfrac{\mu_1}{\mu_2}} (384 \la[c] {+} 16 \la[f] 
   {+} \tfrac{448}{5} \lat[b] {+} \tfrac{1216}{45} \lat[c] {+} 64 \lat[d] 
   {+} \tfrac{832}{15} \lat[j] {-} \tfrac{384}{5} \lat[k] 
   {-} \tfrac{256}{5} \lat[m] \\ 
   {+} \tfrac{256}{3} \lat[p] {+} \tfrac{128}{3} \lat[q] {-} 64 \lat[s] ) \\
   {+} \sqrt{\tfrac{\mu_2}{\mu_1}} (- 32 \la[b] {-} 192 \la[f] 
   {-} \tfrac{6912}{5} \lat[b] {+} \tfrac{256}{5} \lat[c] 
   {-} \tfrac{8448}{5} \lat[d] {-} \tfrac{768}{5} \lat[j] {+} 1536 \lat[k] \\ 
   {+} \tfrac{1024}{5} \lat[m] {-} 512 \lat[q] {+} 256 \lat[s] )) 
   \end{array} \eqnskip \\
+ \sqrt{3} \tfrac{1}{\sqrt{\mu_2(12 \mu_2 {+} \mu_1)}} \phi_0' \upsilon_{45} 
   ( -32 \lat[l] {-} \tfrac{32}{3} \lat[n] {+} 32 \lat[o] 
   {-} \tfrac{64}{3} \lat[t] {+} \tfrac{64}{3} \lat[u] ) \eqnskip \\
+ \sqrt{2} \tfrac{\sqrt{2}}{\sqrt{\mu_3(12 \mu_2 {+} \mu_1)}} \phi_0' \xi_0 
   ( 4 \lac[c] {+} 48 \lac[d] {+} \tfrac{64}{5} \lah[c] 
   {+} \tfrac{64}{15} \lah[d] {-} \tfrac{64}{5} \lah[e] {+} 16 \lah[m] 
   {-} 16 \lah[n] ) \eqnskip \\
+ \tfrac{1}{12 \mu_2 {+} \mu_1} \upsilon_0'{}^2 \big( \begin{array}[t]{l} 
   (\tfrac{9}{4} \lac[l] {-} 96 \la[f] {+} 24 \la[h] {+} 3 \la[i] {+} 3 \la[j] 
   {+} 9 \la[k] {+} 9 \la[l] {-} 3 \la[m] {-} 9 \la[t] 
   {+} \tfrac{1728}{5} \lat[b] {+} \tfrac{192}{5} \lat[c] \\ 
   {-} 192 \lat[d] {-} 192 \lat[j] {+} \tfrac{384}{5} \lat[k] 
   {+} \tfrac{896}{5} \lat[m] {-} 128 \lat[q] {+} 128 \lat[s] ) \\
   {+} \tfrac{\mu_1}{\mu_2} (96 \la[c] {+} \tfrac{3}{32} \lac[l] {+} \la[h] 
   {+} \tfrac{1}{8} \la[i] {+} \tfrac{1}{8} \la[j] {+} \tfrac{3}{8} \la[k] 
   {+} \tfrac{3}{8} \la[l] {-} \tfrac{1}{8} \la[m] {-} \tfrac{3}{8} \la[t] \\
   {+} \tfrac{184}{5} \lat[b] {+} \tfrac{376}{45} \lat[c] {+} 8 \lat[d] 
   {+} \tfrac{88}{15} \lat[j] {-} 16 \lat[k] {-} \tfrac{16}{3} \lat[m] 
   {+} \tfrac{64}{3} \lat[p] {+} \tfrac{16}{3} \lat[q] 
   {-} \tfrac{32}{3} \lat[s] ) \\
   {+} \tfrac{\mu_2}{\mu_1} (96 \la[b] {+} \tfrac{27}{2} \lac[l] {+} 144 \la[h] 
   {+} 18 \la[i] {+} 18 \la[j] {+} 54 \la[k] {+} 54 \la[l] {-} 18 \la[m] 
   {-} 54 \la[t] \\
   {+} \tfrac{31104}{5} \lat[b] {+} \tfrac{384}{5} \lat[c] 
   {+} \tfrac{19584}{5} \lat[d] {-} \tfrac{3456}{5} \lat[j] 
   {-} \tfrac{20736}{5} \lat[k] {+} \tfrac{2304}{5} \lat[m] 
   {+} 768 \lat[q] ) \big) \end{array} 
   \eqnskip \\
+ \sqrt{6} \tfrac{1}{\sqrt{\mu_0 (12 \mu_2 {+} \mu_1)}} \psi_3' \upsilon_0' 
   \begin{array}[t]{l} (\sqrt{\tfrac{\mu_1}{\mu_2}} 
   {+} 12 \sqrt{\tfrac{\mu_2}{\mu_1}} ) ( 
   2 \la[h] {+} \tfrac{1}{4} \la[i] {+} \tfrac{1}{4} \la[j] 
   {-} \tfrac{3}{4} \la[k] {-} \tfrac{3}{4} \la[l] {-} \tfrac{1}{4} \la[m] 
   {-} \tfrac{1}{4} \la[n] \\
   {+} \tfrac{1}{2} \la[p] {-} \tfrac{1}{4} \la[r] {+} \tfrac{3}{4} \la[t] 
   {+} 48 \lat[b] {-} \tfrac{16}{9} \lat[c] {+} 16 \lat[d] {-} 3 \lat[f] 
   {+} \tfrac{1}{3} \lat[g] {+} \lat[h] \\ 
   {+} \tfrac{16}{3} \lat[j] {-} 32 \lat[k] {+} \lah[c] 
   {-} \tfrac{1}{9} \lah[d] {-} \tfrac{1}{3} \lah[e] ) \end{array} \eqnskip \\
+ \tfrac{1}{\sqrt{\mu_2 (12 \mu_2 {+} \mu_1)}} \upsilon_0' \upsilon_{45} \big( 
   \sqrt{\tfrac{\mu_1}{\mu_2}} ( 
   {-} 16 \lat[l] {-} \tfrac{16}{3} \lat[n] {+} 8 \lat[o]
   {-} \tfrac{32}{3} \lat[t] {+} \tfrac{16}{3} \lat[u] ) 
   {+} \sqrt{\tfrac{\mu_2}{\mu_1}} ( {-} 96 \lat[o] {-} 64 \lat[u] ) \big) 
   \eqnskip \\
+ \sqrt{6} \tfrac{\sqrt{2}}{\sqrt{\mu_3 (12 \mu_2 {+} \mu_1)}} 
   \xi_0 \upsilon_0' \big(
   \begin{array}[t]{l} \sqrt{\tfrac{\mu_1}{\mu_2}} ( 
   8 \lac[d] {+} \tfrac{14}{15} \lah[c] {+} \tfrac{38}{45} \lah[d] 
   {-} \tfrac{2}{3} \lah[e] {+} \tfrac{8}{3} \lah[m] {-} \tfrac{4}{3} \lah[n] ) 
   \\
   {+} \sqrt{\tfrac{\mu_2}{\mu_1}} ( {-} 8 \lac[c] {-} \tfrac{72}{5} \lah[c] 
   {+} \tfrac{8}{5} \lah[d] {+} \tfrac{88}{5} \lah[e] {+} 16 \lah[n] ) \big) 
   \end{array}
   \eqnskip \\
+ \sqrt{\tfrac{2}{5}} \tfrac{1}{\sqrt{\mu_0 (12 \mu_2 {+} \mu_1)}} 
   \psi_0 \upsilon_0' \big( \begin{array}[t]{l} 
   \sqrt{\tfrac{\mu_1}{\mu_2}} ( 
   48 \la[e] {-} 10 \la[h] {-} \tfrac{5}{4} \la[i] {-} \tfrac{5}{4} \la[j] 
   {+} \tfrac{15}{4} \la[k] {+} \tfrac{15}{4} \la[l] {+} \tfrac{5}{4} \la[m] 
   {+} \tfrac{5}{4} \la[n] {-} \tfrac{5}{2} \la[p] \\ 
   {+} \tfrac{5}{4} \la[r] {-} \tfrac{15}{4} \la[t] {-} 112 \lat[b] 
   {-} \tfrac{304}{9} \lat[c] {-} 80 \lat[d] {+} \tfrac{7}{5} \lat[f] 
   {+} \tfrac{19}{15} \lat[g] {-} \lat[h] {-} \tfrac{208}{3} \lat[j] \\
   {+} 96 \lat[k] {+} 64 \lat[m] {-} \tfrac{640}{3} \lat[p] 
   {-} \tfrac{320}{3} \lat[q] {+} 160 \lat[s] \\
   {-} \tfrac{7}{3} \lah[c] {-} \tfrac{19}{9} \lah[d] 
   {+} \tfrac{5}{3} \lah[e] {-} \tfrac{40}{3} \lah[m] 
   {+} \tfrac{20}{3} \lah[n]  ) \\
   {+} \sqrt{\tfrac{\mu_1}{\mu_2}} ( 
   {-} 48 \la[d] {-} 120 \la[h] {-} 15 \la[i] {-} 15 \la[j] {+} 45 \la[k] 
   {+} 45 \la[l] {+} 15 \la[m] {+} 15 \la[n] \\ 
   {-} 30 \la[p] {+} 15 \la[r] {-} 45 \la[t] {+} 1728 \lat[b] {-} 64 \lat[c] 
   {+} 2112 \lat[d] {-} \tfrac{108}{5} \lat[f] {+} \tfrac{12}{5} \lat[g] \\ 
   {+} \tfrac{132}{5} \lat[h] {+} 192 \lat[j] 
   {-} 1920 \lat[k] {-} 256 \lat[m] {+} 1280 \lat[q] {-} 640 \lat[s] \\
   {+} 36 \lah[c] {-} 4 \lah[d] {-} 44 \lah[e] {-} 80 \lah[n] ) \big) 
   \end{array} \eqnskip \\
+ \tfrac{1}{\mu_2} \upsilon_{45}^2 ( \tfrac{3}{32} \lac[l] {+} \la[h] 
   {+} \tfrac{1}{8} \la[i] {+} \tfrac{1}{8} \la[j] {+} \tfrac{3}{8} \la[k] 
   {+} \tfrac{3}{8} \la[l] {-} \tfrac{1}{8} \la[m] {-} \tfrac{3}{8} \la[t] 
   {+} 8 \lat[e] {+} \tfrac{16}{3} \lat[r] ) \\
+ \tfrac{1}{\sqrt{\mu_0 \mu_2}} \sqrt{\tfrac{2}{5}} \psi_0 \upsilon_{45} 
   ( {-} \tfrac{7}{2} \la[o] {+} 4 \la[q] {-} \th \la[s] {-} \lat[i] 
   {+} 40 \lat[l] {+} \tfrac{40}{3} \lat[n] {-} 40 \lat[o] 
   {+} \tfrac{160}{3} \lat[t] {-} \tfrac{160}{3} \lat[u] 
   {+} \tfrac{5}{3} \lah[f] {+} \tfrac{20}{3} \lah[o]) \eqnskip \\
+  \sqrt{6} \tfrac{1}{\sqrt{\mu_0 \mu_2}} \psi_3' \upsilon_{45}
   ( -\th \la[o] {+} \th \la[s] {+} \lat[i] {-} 8 \lat[l] 
   {-} \tfrac{8}{3} \lat[n] {+} 8 \lat[o] {-} \tfrac{1}{3} \lah[f]) \eqnskip \\
+ \sqrt{6} \tfrac{\sqrt{2}}{\sqrt{\mu_2 \mu_3}} \xi_0 \upsilon_{45} 
   ( {-} \tfrac{2}{3} \lah[f] {-} \tfrac{4}{3} \lah[o] ) \eqnskip \\
+ \tfrac{1}{\mu_0} \psi_0^2 ( \begin{array}[t]{l}
   \tfrac{48}{5} \la[a] {+} \tfrac{3}{5} \la[g] {+} 10 \la[h] 
   {+} \tfrac{13}{5} \la[i] {+} \tfrac{7}{5} \la[j] {+} \tfrac{109}{5} \la[k] 
   {+} \tfrac{31}{5} \la[l] {-} \tfrac{4}{5} \la[m] {-} 7 \la[n] {+} 8 \la[p] 
   {-}	\la[r] {-} \tfrac{52}{5} \la[t] {+} \tfrac{2}{25} \lat[a] \\
   {+} 128 \lat[b] {+} \tfrac{128}{9} \lat[c] {+} 128 \lat[d] 
   {-} \tfrac{16}{5} \lat[f] {-} \tfrac{16}{15} \lat[g] 
   {+} \tfrac{16}{5} \lat[h] {+} \tfrac{128}{3} \lat[j] {-} 128 \lat[k] 
   {-} \tfrac{128}{3} \lat[m] {+} \tfrac{640}{3} \lat[p] \\
   {+} \tfrac{640}{3} \lat[q] {-} \tfrac{640}{3} \lat[s] 
   {+} \tfrac{48}{5} \lac[e] {+} \tfrac{2}{9} \lah[a]  
   {-} \tfrac{2}{15} \lah[b] {+} \tfrac{16}{3} \lah[c] 
   {+} \tfrac{16}{9} \lah[d] {-} \tfrac{16}{3} \lah[e] 
   {+} \tfrac{10}{3} \lah[g] {+} \tfrac{80}{3} \lah[m] 
   {-} \tfrac{80}{3} \lah[n] ) \end{array} \eqnskip \\
+ \tfrac{1}{\mu_0} \! \tsum_{i=1}^3 \! {\psi_i'}^2 ( \begin{array}[t]{l} 
   \la[g] {+} 6 \la[h] {+} 3 \la[i] {+} \la[j] {+} 3 \la[k] {+} 9 \la[l] 
   {+} 3 \la[n] {-} 3 \la[r] {+} 2 \lat[a] {+} 128 \lat[b] 
   {+} \tfrac{128}{9} \lat[c] {+} 128 \lat[d] {-} 16 \lat[f] \\ 
   {-} \tfrac{16}{3} \lat[g] {+} 16 \lat[h] 
   {+} \tfrac{128}{3} \lat[j] {-} 128 \lat[k] {-} \tfrac{128}{3} \lat[m] 
   {+} \tfrac{2}{9} \lah[a] {-} \tfrac{2}{3} \lah[b] {+} \tfrac{16}{3} \lah[c] 
   {+} \tfrac{16}{9} \lah[d] {-} \tfrac{16}{3} \lah[e] ) 
   \end{array} \eqnskip \\
+ \tfrac{1}{\mu_0} \tsum_{i=1}^8 \psi_i^2 ( \begin{array}[t]{l} 
   16 \la[h] {+} 2 \la[i] {+} 2 \la[j] {+} 2 \la[k] {+} 2 \la[l] {-} 2 \la[m] 
   {+}	2 \la[n] {-}  4 \la[p] +  2 \la[r] {-}	2 \la[t] {+} 2 \lat[a] 
   {+} 128 \lat[b] \\
   {+} \tfrac{128}{9} \lat[c] {+} 128 \lat[d] {+} 16 \lat[f] 
   {+} \tfrac{16}{3} \lat[g] {-} 16 \lat[h] {+} \tfrac{128}{3} \lat[j] 
   {-} 128 \lat[k] {-} \tfrac{128}{3} \lat[m] {+} \tfrac{256}{3} \lat[p] \\ 
   {+} \tfrac{256}{3} \lat[q] {-} \tfrac{256}{3} \lat[s] 
   {+} \tfrac{2}{9} \lah[a] {+} \tfrac{2}{3} \lah[b] {+} \tfrac{16}{3} \lah[c] 
   {+} \tfrac{16}{9} \lah[d] {-} \tfrac{16}{3} \lah[e] {+} \tfrac{4}{3} \lah[g] 
   {+} \tfrac{32}{3} \lah[m] {-} \tfrac{32}{3} \lah[n] )  
   \end{array} \eqnskip \\
+  \sqrt{\tfrac{3}{5}} \psi_0 \psi_3' ( 2 \la[g] {-} 20 \la[h] {+} 2 \la[i] 
   {-} 2 \la[j] {+} 2 \la[k] {-} 18 \la[l] {+} 4 \la[m] {+} 2 \la[n] 
   {-} 8 \la[p] {+} 6 \la[r] {+} 20 \la[t] ) \eqnskip \\
+ \sqrt{\tfrac{2}{\mu_0 \mu_3}} \sqrt{\tfrac{3}{5}} \psi_0 \xi_0 
   (8 \lac[b] {-} \tfrac{8}{9} \lah[a] 
   {+} \tfrac{4}{15} \lah[b] {-} \tfrac{32}{3} \lah[c] 
   {-} \tfrac{32}{9} \lah[d] {+} \tfrac{32}{3} \lah[e] 
   {-} \tfrac{20}{3} \lah[g] {-} \tfrac{80}{3} \lah[m] 
   {+} \tfrac{80}{3} \lah[n] ) \eqnskip \\
+ \tfrac{2}{\mu_3} \xi_0^2 ( 4 \lac[a] {+} \tfrac{8}{15} \lah[a] 
   {+} 2 \lah[g] ) \eqnskip \\
+ \tfrac{1}{\sqrt{\mu_0 \mu_2}} \sqrt{6} (\psi_1' \upsilon_{18} 
   {+} \psi_2' \upsilon_{63} ) ( \begin{array}[t]{l} 
   2 \la[h] {+} \tfrac{1}{4} \la[i] {+} \tfrac{1}{4} \la[j] 
   {-} \tfrac{3}{4} \la[k] {-} \tfrac{3}{4} \la[l] {-} \tfrac{1}{4} \la[m] 
   {-} \tfrac{1}{4} \la[n] {+} \tfrac{1}{2} \la[p] {-} \tfrac{1}{4} \la[r] 
   {+} \tfrac{3}{4} \la[t] \\
   {+} 48 \lat[b] {-} \tfrac{16}{9} \lat[c] {+} 16 \lat[d] {-} 3 \lat[f] 
   {+} \tfrac{1}{3} \lat[g] {+} \lat[h] {+} \tfrac{16}{3} \lat[j] 
   {-} 32 \lat[k] \\
   {+} \lah[c] {-} \tfrac{1}{9} \lah[d] {-} \tfrac{1}{3} \lah[e] ) 
   \end{array} \eqnskip \\
+ \tfrac{1}{\sqrt{\mu_0 \mu_2}} \sqrt{6} (\psi_1' \upsilon_{63} 
   {-} \psi_2' \upsilon_{18}) 
   ( -\th \la[o] {+} \th \la[s] {+} \lat[i] {-} 8 \lat[l] 
   {-} \tfrac{8}{3} \lat[n] {+} 8 \lat[o] {-} \tfrac{1}{3} \lah[f]) \eqnskip \\
+ \tfrac{1}{\sqrt{\mu_0 \mu_2}} \tsum_{i=1}^8 \psi_i \upsilon_i 
   \begin{array}[t]{l} 
   ( 8 \la[h] {+} \la[i] {+} \la[j] {-} 3 \la[k] {-} 3 \la[l] {-} \la[m] 
   {-} \la[n] {+} 2 \la[p] {-} \la[r] {+} 3 \la[t] {-} 64 \lat[b] 
   {-} \tfrac{64}{9} \lat[c] {+} 64 \lat[d] \\ 
   {-} 4 \lat[f] {-} \tfrac{4}{3} \lat[g] {-} 4 \lat[h] 
   {-} \tfrac{64}{3} \lat[j] {-} \tfrac{64}{3} \lat[p] 
   {+} \tfrac{64}{3} \lat[q] {-} \tfrac{4}{3} \lah[c] {-} \tfrac{4}{9} \lah[d] 
   {-} \tfrac{4}{3} \lah[e] {-} \tfrac{4}{3} \lah[m] {-} \tfrac{4}{3} \lah[n]) 
   \end{array} \eqnskip \\
+ \tfrac{1}{\sqrt{\mu_0 \mu_2}} \! \tsum_{i=1}^8 \! \psi_i \upsilon_{i+45} 
   ( {-} 2 \la[o] {+} 4 \la[q] {-} 2 \la[s] {-} 4 \lat[i] {-} 32 \lat[l] 
   {-} \tfrac{32}{3} \lat[n] {+} 32 \lat[o] {-} \tfrac{32}{3} \lat[t] 
   {+} \tfrac{32}{3} \lat[u] {-} \tfrac{4}{3} \lah[f] {-} \tfrac{4}{3} \lah[o]) 
   \eqnskip \\
+ \tfrac{\sqrt{2}}{\sqrt{\mu_0 \mu_3}} \tsum_{i=1}^8 \psi_i \xi_{i+32} 
   ( {-} 4 \lac[g] {+} 2 \lah[j] {+} 16 \lah[p] {-} 16 \lah[q] ) \\
+ \tfrac{\sqrt{2}}{\sqrt{\mu_0 \mu_3}} \tsum_{i=1}^8 \psi_i \xi_{i+81} 
   ( {-} 4 \lac[h] {+} 2 \lah[k] {+} 16 \lah[s] {-} 16 \lah[t] )  \eqnskip \\
+ \tfrac{2}{\mu_1} \tsum_{i=1}^6 \phi_i^2 ( \begin{array}[t]{l} 
   \la[g] {+} 8 \la[h] {+} \tfrac{13}{4} \la[i] {+} \tfrac{5}{4} \la[j] 
   {+} \tfrac{9}{4} \la[k] {+} \tfrac{9}{4} \la[l] {-} \tfrac{1}{4} \la[m] 
   {-} \tfrac{9}{4} \la[n] \\ 
   {+} \tfrac{3}{2} \la[p] {+} \tfrac{3}{4} \la[r] 
   {-} \tfrac{9}{4} \la[t] {+} \la[u] {+} \tfrac{1}{3} \la[v] 
   {+} \tfrac{32}{3} \la[w] {+} 162 \lat[b] {+} 2 \lat[c] \\ 
   {+} 98 \lat[d] {+} 2 \lat[e] {-} 18 \lat[j] {-} 126 \lat[k] {+} 14 \lat[m] 
   {+} \tfrac{64}{3} \lat[q] {+} \tfrac{64}{3} \lat[r] 
   {+} 2 \lah[h] {+} 2 \lah[i] )  \end{array} \eqnskip \\
+ \sqrt{2} \tfrac{2}{\sqrt{\mu_1\mu_3}} ( \begin{array}[t]{l} \phi_1 \xi_{44} 
   {+} \phi_2 \xi_{45} {+} \phi_3 \xi_{46} {+} \phi_4 \xi_{93} 
   {+} \phi_5 \xi_{94} {+} \phi_6 \xi_{95} )  \\
   ( -3 \lah[c] {+} \tfrac{1}{3} \lah[d] {+} \tfrac{7}{3} \lah[e] {+} 2 \lah[h] 
   {+} 2 \lah[i] {+} \tfrac{8}{3} \lah[n] ) \end{array} \eqnskip \\
+ \sqrt{2} \tfrac{2}{\sqrt{\mu_1\mu_3}} ( \phi_1 \xi_{93} {+} \phi_2 \xi_{94} 
   {+} \phi_3 \xi_{95} {-} \phi_4 \xi_{44} {-} \phi_5 \xi_{45} 
   {-} \phi_6 \xi_{46} ) 
   ( \tfrac{1}{3} \lah[f] {+} \tfrac{8}{3} \lah[o] ) \eqnskip \\
+ \sqrt{3} \tfrac{2}{\sqrt{\mu_1\mu_3}} ( \begin{array}[t]{l} \phi_1 \xi_{47} 
   {+} \phi_2 \xi_{48} {+} \phi_3 \xi_{49} {+} \phi_4 \xi_{96} 
   {+} \phi_5 \xi_{97} {+} \phi_6 \xi_{98} )  \\ 
   ( \tfrac{8}{3} \lac[g] {+} \tfrac{4}{9} \lac[j] {-} \tfrac{4}{9} \lac[m] 
   {+} \tfrac{2}{3} \lah[j] {+} \tfrac{8}{3} \lah[q] {-} \tfrac{8}{3} \lah[u] ) 
   \end{array} \eqnskip \\
+ \sqrt{3} \tfrac{2}{\sqrt{\mu_1\mu_3}} ( \begin{array}[t]{l} \phi_1 \xi_{96} 
   {+} \phi_2 \xi_{97} {+} \phi_3 \xi_{98} {-} \phi_4 \xi_{47} 
   {-} \phi_5 \xi_{48} {-} \phi_6 \xi_{49} ) \\
   ( \tfrac{8}{3} \lac[h] {+} \tfrac{4}{9} \lac[k] {+} \tfrac{4}{9} \lac[n] 
   {+} \tfrac{2}{3} \lah[k] {+} \tfrac{8}{3} \lah[r] {+} \tfrac{8}{3} \lah[t] ) 
   \end{array} \eqnskip \\
+ \sqrt{2} \tfrac{\sqrt{2}}{\sqrt{\mu_1\mu_2}} ( \begin{array}[t]{l} 
   \phi_1 \upsilon_9 {+} \phi_2 \upsilon_{10} {+} \phi_3 \upsilon_{11} 
   {+} \phi_4 \upsilon_{54} {+} \phi_5 \upsilon_{55} 
   {+} \phi_6 \upsilon_{56} )  \\
   ( {-} 4 \la[h] {-} \tfrac{1}{2} \la[i] {-}  \tfrac{1}{2} \la[j] 
   {-} \tfrac{3}{2} \la[k] {-} \tfrac{3}{2} \la[l] {+} \tfrac{1}{2} \la[m] 
   {+} 2 \la[n] {-} 2 \la[p] {+} \tfrac{3}{2} \la[t] {+} \la[u] 
   {+} \tfrac{1}{3} \la[v] {-} \tfrac{16}{3} \la[w] \\ 
   {+} 18 \lat[b] {-} \tfrac{10}{3} \lat[c] {-} 14 \lat[d] {+} 2 \lat[e] 
   {+} 14 \lat[j] {+} 2 \lat[k] {-} \tfrac{38}{3} \lat[m] 
   {+} \tfrac{32}{3} \lat[q] {-} \tfrac{32}{3} \lat[r] 
   {-} \tfrac{16}{3} \lat[s] ) \end{array} \eqnskip \\
+ \sqrt{2} \tfrac{\sqrt{2}}{\sqrt{\mu_1\mu_2}} (\phi_1 \upsilon_{54} 
   \begin{array}[t]{l} {+} \phi_2 \upsilon_{55} {+} \phi_3 \upsilon_{56} 
   {-} \phi_4 \upsilon_9 {-} \phi_5 \upsilon_{10} {-} \phi_6 \upsilon_{11} ) \\
    ( 2 \la[o] {-} 2 \la[q] {+} 8 \lat[l] {-} \tfrac{8}{3} \lat[n] 
   {-} 8 \lat[o] {-} \tfrac{16}{3} \lat[t] {+} \tfrac{32}{3} \lat[u] ) 
   \end{array} \eqnskip \\
+ \tfrac{\sqrt{2}}{\sqrt{\mu_1\mu_2}} ( \begin{array}[t]{l} 
   \phi_1 \upsilon_{12} {+} \phi_2 \upsilon_{13} {+} \phi_3 \upsilon_{14} 
   {+} \phi_4 \upsilon_{57} {+} \phi_5 \upsilon_{58} 
   {+} \phi_6 \upsilon_{59} ) \\  \hs*{-1em}
   ( 2 \la[u] {+} \tfrac{2}{3} \la[v] {+} \tfrac{16}{3} \la[w] {+} 36 \lat[b] 
   {+} 4 \lat[c] {-} 28 \lat[d] {+} 4 \lat[e] {-} 20 \lat[j] {+} 4 \lat[k] 
   {+} 12 \lat[m] {-} \tfrac{32}{3} \lat[q] {+} \tfrac{32}{3} \lat[r] 
   {+} 16 \lat[s] ) \end{array} \eqnskip \\
+ \tfrac{\sqrt{2}}{\sqrt{\mu_1\mu_2}} (\phi_1  \upsilon_{57} 
   {+} \phi_2 \upsilon_{58} {+} \phi_3 \upsilon_{59} {-} \phi_4 \upsilon_{12} 
   {-} \phi_5 \upsilon_{13} {-} \phi_6 \upsilon_{14} ) 
   ( 16 \lat[l] {-} 16 \lat[o] {+} 16 \lat[t] {-} \tfrac{32}{3} \lat[u] ) 
   \eqnskip \\
+ \sqrt{2} \tfrac{\sqrt{2}}{\sqrt{\mu_1\mu_2}} ( \phi_1 \upsilon_{39} 
   {+} \phi_2 \upsilon_{40} {+} \phi_3 \upsilon_{41} {+} \phi_4 \upsilon_{84} 
   {+} \phi_5 \upsilon_{85} {+} \phi_6 \upsilon_{86} ) 
   ( -6 \lah[p] {+} 2 \lah[q] {+} 2 \lah[u] )  \eqnskip \\
+ \sqrt{2} \tfrac{\sqrt{2}}{\sqrt{\mu_1\mu_2}} ( \phi_1 \upsilon_{84} 
   {+} \phi_2 \upsilon_{85} {+} \phi_3 \upsilon_{86} {-} \phi_4 \upsilon_{39} 
   {-} \phi_5 \upsilon_{40} {-} \phi_6 \upsilon_{41} ) 
   ( -2 \lah[r] {-} 6 \lah[s] {+} 2 \lah[t] )  \eqnskip \\
+ \tfrac{2}{\mu_3} (\tsum_{i=1}^6 \xi_i^2 {+} \tsum_{i=50}^{55} \xi_i^2) 
   (4 \lac[e] {+} \lac[l]) \eqnskip \\
+ \tfrac{2}{\mu_3} (\tsum_{i=7}^{12} \xi_i^2 {+} \tsum_{i=56}^{61} \xi_i^2) 
   (4 \lac[e] {+} 2 \lah[h] {+} 2 \lah[i]) \eqnskip \\ 
+ \tfrac{2}{\mu_3} (\tsum_{i=13}^{18} \xi_i^2 {+} \tsum_{i=62}^{67} \xi_i^2) 
   (16 \lac[e] {+} 2 \lah[h] {+} 2 \lah[i]) \eqnskip \\
+ \tfrac{2}{\mu_3} (\tsum_{i=19}^{24} \xi_i^2 {+} \tsum_{i=68}^{73} \xi_i^2) 
   (4 \lac[e] {+} \tfrac{1}{2} \lac[l] {+} \lah[h] {+} \lah[i]) \eqnskip \\
+ \tfrac{2}{\mu_3} (\tsum_{i=25}^{32} \xi_i^2 {+} \tsum_{i=74}^{81} \xi_i^2) 
   ( 9 \lac[e] {+} \lac[f] {+} \tfrac{1}{2} \lac[l] {+} \lah[h] {+} \lah[i]) 
   \eqnskip \\
+ \tfrac{2}{\mu_3} \tsum_{i=33}^{40} \xi_i^2  
   ( 9 \lac[e] {+} \lac[f] {+} 4 \lah[h]) 
+ \tsum_{i=82}^{89} \xi_i^2  ( 9 \lac[e] {+} \lac[f] {+} 4 \lah[i]) \eqnskip \\
+ \tfrac{2}{\mu_3} (\tsum_{i=41}^{43} \xi_i^2 {+} \tsum_{i=90}^{92} \xi_i^2) 
   ( \lac[e] {+} \lac[f] {+} \tfrac{8}{3} \lac[i] {+} \tfrac{1}{3} \lac[l] 
   {+} \tfrac{1}{9} \lah[a] {+} \tfrac{2}{3} \lah[g])	\eqnskip \\
+ \tfrac{2}{\mu_3} (\tsum_{i=44}^{46} \xi_i^2 {+} \tsum_{i=93}^{95} \xi_i^2) 
   ( \lac[e] {+} \lac[f] {+} \tfrac{1}{9} \lah[a] {+} \tfrac{2}{3} \lah[g]
   {+} \lah[h] {+} \lah[i])  \eqnskip \\
+ \tfrac{2}{\mu_3} (\tsum_{i=47}^{49} \xi_i^2 {+} \tsum_{i=96}^{98} \xi_i^2) 
   ( 4 \lac[e] {+} \tfrac{8}{3} \lac[f] {+} \tfrac{16}{9} \lac[i] 
   {+} \tfrac{1}{18} \lac[l] {+} \tfrac{2}{3} \lah[g] {+} \lah[h] {+} \lah[i])	
   \eqnskip \\
+ \tfrac{2}{\mu_3} \tsum_{i=1}^{6} (\xi_{i+6} \xi_{i+12} 
   {+} \xi_{i+55} \xi_{i+61}) ( 4 \lah[h] {-} 4 \lah[i]) \eqnskip \\
+ \tfrac{2}{\mu_3} \tsum_{i=1}^{6} (\xi_{i+6} \xi_{i+61} 
   {-} \xi_{i+55} \xi_{i+12}) ( {-} 4 \lah[l] ) \eqnskip \\
+ \tfrac{2}{\mu_3} \tsum_{i=1}^{8} \xi_{i+32} \xi_{i+81} 
   ( 4 \lah[l]) \eqnskip \\
+ \sqrt{6} \tfrac{2}{\mu_3} (\xi_{44} \xi_{47} {+} \xi_{45} \xi_{48} 
   {+} \xi_{46} \xi_{49} {+} \xi_{93} \xi_{96} {+} \xi_{94} \xi_{97} 
   {+} \xi_{95} \xi_{98}) ( \tfrac{2}{3} \lah[j] )  \eqnskip \\
+ \sqrt{6} \tfrac{2}{\mu_3} (\xi_{44} \xi_{96} {+} \xi_{45} \xi_{97} 
   {+} \xi_{46} \xi_{98} {-} \xi_{93} \xi_{47} {-} \xi_{94} \xi_{48} 
   {-} \xi_{95} \xi_{49}) ( \tfrac{2}{3} \lah[k] )  \eqnskip \\
+ \tfrac{\sqrt{2}}{\sqrt{\mu_2 \mu_3}} \tsum_{i=1}^{6} 
   (\xi_{i+18} \upsilon_{i+29} {+} \xi_{i+67} \upsilon_{i+74}) 
   ( {-} \lac[m] {-} 2 \lah[p] {-} 2 \lah[q] {-} 2 \lah[u] ) \eqnskip \\
+ \tfrac{\sqrt{2}}{\sqrt{\mu_2 \mu_3}} \tsum_{i=1}^{6} 
   (\xi_{i+18} \upsilon_{i+74} {-} \xi_{i+67} \upsilon_{i+29}) 
   ( {-} \lac[n] {-} 2 \lah[r] {+} 2 \lah[s] {+} 2 \lah[t] )  \eqnskip \\
+ \tfrac{\sqrt{2}}{\sqrt{\mu_2 \mu_3}} \tsum_{i=1}^{8} 
   (\xi_{i+24} \upsilon_{i+18} {+} \xi_{i+73} \upsilon_{i+63}) 
   ( \lac[g] {-} \lac[m] {-} 2 \lah[p] {-} 2 \lah[q] {-} 2 \lah[u] ) 
   \eqnskip \\
+ \tfrac{\sqrt{2}}{\sqrt{\mu_2 \mu_3}} \tsum_{i=1}^{8} 
   (\xi_{i+24} \upsilon_{i+63} {-} \xi_{i+73} \upsilon_{i+18}) 
   ( {-} \lac[h] {-} \lac[n] {-} 2 \lah[r] {+} 2 \lah[s] {+} 2 \lah[t] ) 
   \eqnskip \\
+ \tfrac{\sqrt{2}}{\sqrt{\mu_2 \mu_3}} \tsum_{i=1}^{8} (\xi_{i+32} \upsilon_i 
   {+} \xi_{i+81} \upsilon_{i+45}) ( {-} \lac[g] ) \eqnskip \\
+ \tfrac{\sqrt{2}}{\sqrt{\mu_2 \mu_3}} \tsum_{i=1}^{8} 
   (\xi_{i+32} \upsilon_{i+45} {-} \xi_{i+81} \upsilon_i) 
   ( \lac[h]) \eqnskip \\
+ \tfrac{\sqrt{2}}{\sqrt{\mu_2 \mu_3}} (\xi_{41} \upsilon_{15} 
   {+} \xi_{42} \upsilon_{16} {+} \xi_{43} \upsilon_{17} 
   {+} \xi_{90} \upsilon_{60} {+} \xi_{91} \upsilon_{61} 
   {+} \xi_{92} \upsilon_{62}) ( \tfrac{2}{3} \lah[c] {-} \tfrac{2}{3} \lah[d] 
   {+} \tfrac{2}{3} \lah[e] {-} 4 \lah[m] {+} \tfrac{4}{3} \lah[n] )  
   \eqnskip \\
+ \tfrac{\sqrt{2}}{\sqrt{\mu_2 \mu_3}} (\xi_{41} \upsilon_{60} 
   {+} \xi_{42} \upsilon_{61} {+} \xi_{43} \upsilon_{62} 
   {-} \xi_{90} \upsilon_{15} {-} \xi_{91} \upsilon_{16} 
   {-} \xi_{92} \upsilon_{17}) 
   ( \tfrac{2}{3} \lah[f] {+} \tfrac{4}{3} \lah[o] )  \eqnskip \\
+ \tfrac{\sqrt{2}}{\sqrt{\mu_2 \mu_3}} (\xi_{41} \upsilon_{36} 
   {+} \xi_{42} \upsilon_{37} {+} \xi_{43} \upsilon_{38} 
   {+} \xi_{90} \upsilon_{81} {+} \xi_{91} \upsilon_{82} 
   {+} \xi_{92} \upsilon_{83}) ( {-} \lac[g] {+} \tfrac{4}{3} \lac[j] 
   {-} \tfrac{1}{3} \lac[m] ) \eqnskip \\
+ \tfrac{\sqrt{2}}{\sqrt{\mu_2 \mu_3}} (\xi_{41} \upsilon_{81} 
   {+} \xi_{42} \upsilon_{82} {+} \xi_{43} \upsilon_{83} 
   {-} \xi_{90} \upsilon_{36} {-} \xi_{91} \upsilon_{37} 
   {-} \xi_{92} \upsilon_{38}) ( {-} \lac[h] {+} \tfrac{4}{3} \lac[k] 
   {+} \tfrac{1}{3} \lac[n] ) \eqnskip \\
+ \tfrac{\sqrt{2}}{\sqrt{\mu_2 \mu_3}} ( \begin{array}[t]{l} 
   \xi_{41} \upsilon_{42} {+} \xi_{42} \upsilon_{43} 
   {+} \xi_{43} \upsilon_{44} {+} \xi_{90} \upsilon_{87} 
   {+} \xi_{91} \upsilon_{88} {+} \xi_{92} \upsilon_{89}) \\
   ( -\tfrac{8}{3} \lac[i] {-} \tfrac{1}{3} \lac[l] 
   {+} 2 \lah[c] {-} \tfrac{2}{9} \lah[d] {-} \tfrac{2}{3} \lah[e] 
   {-} \tfrac{4}{3} \lah[m] {-} \tfrac{4}{3} \lah[n] ) \end{array} \eqnskip \\
+ \tfrac{\sqrt{2}}{\sqrt{\mu_2 \mu_3}} ( \xi_{41} \upsilon_{87} 
   {+} \xi_{42} \upsilon_{88} {+} \xi_{43} \upsilon_{89} 
   {-} \xi_{90} \upsilon_{42} 
   {-} \xi_{91} \upsilon_{43} {-} \xi_{92} \upsilon_{44}) 
   ( \tfrac{2}{3} \lah[f] {+} \tfrac{4}{3} \lah[o] ) \eqnskip \\
+ \tfrac{\sqrt{2}}{\sqrt{\mu_2 \mu_3}} (\xi_{44} \upsilon_9 
   {+} \xi_{45} \upsilon_{10} {+} \xi_{46} \upsilon_{11} 
   {+} \xi_{93} \upsilon_{54} {+} \xi_{94} \upsilon_{55} 
   {+} \xi_{95} \upsilon_{56}) 
   ( -\tfrac{1}{3} \lah[c] {-} \tfrac{5}{9} \lah[d] {-} \tfrac{1}{3} \lah[e] 
   {-} \tfrac{4}{3} \lah[m] {+} \tfrac{4}{3} \lah[n] )	\eqnskip \\
+ \tfrac{\sqrt{2}}{\sqrt{\mu_2 \mu_3}} (\xi_{44} \upsilon_{54} 
   {+} \xi_{45} \upsilon_{55} {+} \xi_{46} \upsilon_{56} 
   {-} \xi_{93} \upsilon_9 
   {-} \xi_{94} \upsilon_{10} {-} \xi_{95} \upsilon_{11}) 
   ( {-} \tfrac{1}{3} \lah[f] {+} \tfrac{4}{3} \lah[o] )  \eqnskip \\
+ \sqrt{2} \tfrac{\sqrt{2}}{\sqrt{\mu_2 \mu_3}} (\xi_{44} \upsilon_{12} 
  {+} \begin{array}[t]{l} 
  \xi_{45} \upsilon_{13} {+} \xi_{46} \upsilon_{14} {+} \xi_{93} \upsilon_{57} 
  {+} \xi_{94} \upsilon_{58} {+} \xi_{95} \upsilon_{59}) \\
   ( {-} \tfrac{1}{3} \lah[c] {+} \tfrac{1}{3} \lah[d] {-} \tfrac{1}{3} \lah[e] 
   {+} 2 \lah[m] {-} \tfrac{2}{3} \lah[n] ) \end{array} \eqnskip \\
+ \sqrt{2} \tfrac{\sqrt{2}}{\sqrt{\mu_2 \mu_3}} (\xi_{44} \upsilon_{57} 
   {+} \xi_{45} \upsilon_{58} {+} \xi_{46} \upsilon_{59} 
   {-} \xi_{93} \upsilon_{12} {-} \xi_{94} \upsilon_{13} 
   {-} \xi_{95} \upsilon_{14}) 
   ( {-} \tfrac{1}{3} \lah[f] {-} \tfrac{2}{3} \lah[o] )  \eqnskip \\
+ \tfrac{\sqrt{2}}{\sqrt{\mu_2 \mu_3}} (\xi_{44} \upsilon_{39} 
   {+} \xi_{45} \upsilon_{40} {+} \xi_{46} \upsilon_{41} 
   {+} \xi_{93} \upsilon_{84} {+} \xi_{94} \upsilon_{85} 
   {+} \xi_{95} \upsilon_{86}) 
   ( {-} \lac[g] {-} 6 \lah[p] {+} 2 \lah[q] {+} 2 \lah[u] )  \eqnskip \\
+ \tfrac{\sqrt{2}}{\sqrt{\mu_2 \mu_3}} (\xi_{44} \upsilon_{84} 
   {+} \xi_{45} \upsilon_{85} {+} \xi_{46} \upsilon_{86} 
   {-} \xi_{93} \upsilon_{39} {-} \xi_{94} \upsilon_{40} 
   {-} \xi_{95} \upsilon_{41}) 
   ( {-} \lac[h] {-} 2 \lah[r] {-} 6 \lah[s] {+} 2 \lah[t] )  \eqnskip \\
+ \sqrt{6} \tfrac{\sqrt{2}}{\sqrt{\mu_2 \mu_3}} (\xi_{47} \upsilon_9 
   \begin{array}[t]{l} 
   {+} \xi_{48} \upsilon_{10} {+} \xi_{49} \upsilon_{11} 
   {+} \xi_{96} \upsilon_{54} {+} \xi_{97} \upsilon_{55} 
   {+} \xi_{98} \upsilon_{56})	\\
   ( {-} \tfrac{2}{3} \lac[g] {+} \tfrac{2}{9} \lac[j] {+} \tfrac{1}{9} \lac[m] 
   {-} \tfrac{2}{3} \lah[p] {+} \tfrac{2}{3} \lah[q] {+} \tfrac{2}{3} \lah[u] ) 
   \end{array} \eqnskip \\
+ \sqrt{6} \tfrac{\sqrt{2}}{\sqrt{\mu_2 \mu_3}} (\xi_{47} \upsilon_{54} 
   \begin{array}[t]{l}	{+} \xi_{48} \upsilon_{55} {+} \xi_{49} \upsilon_{56} 
   {-} \xi_{96} \upsilon_9 {-} \xi_{97} \upsilon_{10} 
   {-} \xi_{98} \upsilon_{11})	\\
   ( \tfrac{2}{3} \lac[h] {-} \tfrac{2}{9} \lac[k] {+} \tfrac{1}{9} \lac[n] 
   {+} \tfrac{2}{3} \lah[r] {+} \tfrac{2}{3} \lah[s] {-} \tfrac{2}{3} \lah[t] )  
   \end{array} \eqnskip \\
+ \sqrt{3} \tfrac{\sqrt{2}}{\sqrt{\mu_2 \mu_3}} (\xi_{47} \upsilon_{12} 
   {+} \xi_{48} \upsilon_{13} {+} \xi_{49} \upsilon_{14} 
   {+} \xi_{96} \upsilon_{57} 
   {+} \xi_{97} \upsilon_{58} {+} \xi_{98} \upsilon_{59}) 
   ( \tfrac{4}{9} \lac[j] {-} \tfrac{1}{9} \lac[m] {+} 2 \lah[p] 
   {-} \tfrac{2}{3} \lah[q] {-} \tfrac{2}{3} \lah[u] )	\eqnskip \\
+ \sqrt{3} \tfrac{\sqrt{2}}{\sqrt{\mu_2 \mu_3}} (\xi_{47} \upsilon_{57} 
   {+} \begin{array}[t]{l} 
   \xi_{48} \upsilon_{58} {+} \xi_{49} \upsilon_{59} {-} \xi_{96} \upsilon_{12} 
   {-} \xi_{97} \upsilon_{13} {-} \xi_{98} \upsilon_{14}) \\
   ( -\tfrac{4}{9} \lac[k] {-} \tfrac{1}{9} \lac[n] {-} \tfrac{2}{3} \lah[r] 
   {-} 2 \lah[s] {+} \tfrac{2}{3} \lah[t] ) \end{array} \eqnskip \\
+ \sqrt{6} \tfrac{\sqrt{2}}{\sqrt{\mu_2 \mu_3}} (\xi_{47} \upsilon_{39} 
   {+} \xi_{48} \upsilon_{40} {+} \xi_{49} \upsilon_{41} 
   {+} \xi_{96} \upsilon_{84} 
   {+} \xi_{97} \upsilon_{85} {+} \xi_{98} \upsilon_{86}) 
   ( {-} 2 \lah[m] {+} \tfrac{2}{3} \lah[n] ) \eqnskip \\
+ \sqrt{6} \tfrac{\sqrt{2}}{\sqrt{\mu_2 \mu_3}} (\xi_{47} \upsilon_{84} 
   {+} \xi_{48} \upsilon_{85} {+} \xi_{49} \upsilon_{86} 
   {-} \xi_{96} \upsilon_{39} 
   {-} \xi_{97} \upsilon_{40} {-} \xi_{98} \upsilon_{41}) 
   ( {-} \tfrac{2}{3} \lah[o] ) \eqnskip \\
+ \tfrac{1}{\mu_2} \tsum_{i=1}^8 \upsilon_i^2 ( \begin{array}[t]{l} 
   \la[h] {+} \tfrac{1}{8} \la[i] {+} \tfrac{1}{8} \la[j] 
   {+} \tfrac{9}{8} \la[k] {+} \tfrac{9}{8} \la[l] {-} \tfrac{1}{8} \la[m] 
   {-} \tfrac{3}{8} \la[n] {+} \tfrac{3}{4} \la[p] {-} \tfrac{3}{8} \la[r] 
   {-} \tfrac{9}{8} \la[t] \\
   {+} 8 \lat[b] {+} \tfrac{8}{9} \lat[c] {+} 8 \lat[d] 
   {+} \tfrac{8}{3} \lat[j] {+} 8 \lat[k] {+} \tfrac{8}{3} \lat[m] 
   {+} \tfrac{16}{3} \lat[p] {+} \tfrac{16}{3} \lat[q] 
   {+} \tfrac{16}{3} \lat[s] ) \end{array} \eqnskip \\
+ \tfrac{1}{\mu_2} \tsum_{i=46}^{53} \upsilon_i^2 ( \begin{array}[t]{l} 
   \la[h] {+} \tfrac{1}{8} \la[i] {+} \tfrac{1}{8} \la[j] 
   {+} \tfrac{9}{8} \la[k] {+} \tfrac{9}{8} \la[l] {-} \tfrac{1}{8} \la[m] 
   {-} \tfrac{3}{8} \la[n] {+} \tfrac{3}{4} \la[p] {-} \tfrac{3}{8} \la[r] 
   {-} \tfrac{9}{8} \la[t] {+} 8 \lat[e] {+} \tfrac{16}{3} \lat[r] ) 
   \end{array} \eqnskip \\
+ \tfrac{1}{\mu_2}  \tsum_{i=1}^8 \upsilon_i \upsilon_{i+45} 
   ( 8 \lat[l] {+} \tfrac{8}{3} \lat[n] {+} 8 \lat[o] {+} \tfrac{16}{3} \lat[t] 
   {+} \tfrac{16}{3} \lat[u] ) \eqnskip \\
+ \tfrac{1}{\mu_2} (\tsum_{i=9}^{11} \upsilon_i^2 
   {+} \tsum_{i=54}^{56} \upsilon_i^2 )( \begin{array}[t]{l} 
   \la[h] {+} \tfrac{1}{8} \la[i] {+} \tfrac{1}{8} \la[j] 
   {+} \tfrac{29}{8} \la[k] {+} \tfrac{5}{8} \la[l] {-} \tfrac{1}{8} \la[m] 
   {-} \tfrac{5}{8} \la[n] {+} \tfrac{3}{4} \la[p] {-} \tfrac{1}{8} \la[r] 
   {-} \tfrac{9}{8} \la[t] \\
   {+} \th \la[u] {+} \tfrac{1}{6} \la[v] {+} \tfrac{4}{3} \la[w] 
   {+} \lat[b] {+} \tfrac{25}{9} \lat[c] {+} \lat[d] {+} \lat[e] 
   {+} \tfrac{5}{3} \lat[j] {+} \lat[k] {+} \tfrac{5}{3} \lat[m] \\ 
   {+} \tfrac{8}{3} \lat[p] {+} \tfrac{8}{3} \lat[q] {+} \tfrac{8}{3} \lat[r] 
   {-} \tfrac{8}{3} \lat[s] ) \end{array} \eqnskip \\
+ \tfrac{1}{\mu_2} (\tsum_{i=12}^{14} \upsilon_i^2 
   {+} \tsum_{i=57}^{59} \upsilon_i^2 )( \begin{array}[t]{l} 
   \la[h] {+}  \tfrac{1}{8} \la[i] {+} \tfrac{1}{8} \la[j] 
   {+} \tfrac{9}{8} \la[k] {+} \tfrac{1}{8} \la[l] {-} \tfrac{1}{8} \la[m] 
   {-} \tfrac{3}{8} \la[n] {+} \tfrac{1}{4} \la[p] {+} \tfrac{1}{8} \la[r] 
   {+} \tfrac{3}{8} \la[t] {+} \la[u]\\
   {+} \tfrac{1}{3} \la[v] {+} \tfrac{2}{3} \la[w] {+} 2 \lat[b] {+} 2 \lat[c] 
   {+} 2 \lat[d] {+} 2 \lat[e] {-} 2 \lat[j] {+} 2 \lat[k] {-} 2 \lat[m] \\
   {+} 12 \lat[p] {+} \tfrac{4}{3} \lat[q] {+} \tfrac{4}{3} \lat[r] 
   {-} 4 \lat[s] ) \end{array} \eqnskip \\
+ \tfrac{1}{\mu_2} (\tsum_{i=15}^{17} \upsilon_i^2 
   {+} \tsum_{i=60}^{62} \upsilon_i^2 )( \begin{array}[t]{l} 
   \la[h] {+} \tfrac{1}{8} \la[i] {+} \tfrac{1}{8} \la[j] 
   {+} \tfrac{9}{8} \la[k] {+} \tfrac{1}{8} \la[l] {-} \tfrac{1}{8} \la[m] 
   {-} \tfrac{3}{8} \la[n] {+} \tfrac{1}{4} \la[p] {+} \tfrac{1}{8} \la[r] 
   {+} \tfrac{3}{8} \la[t] {+} 4 \lat[b] \\
   {+} 4 \lat[c] {+} 4 \lat[d] {+} 4 \lat[e] {-} 4 \lat[j] 
   {+} 4 \lat[k] {-} 4 \lat[m] {+} 24 \lat[p] {+} \tfrac{8}{3} \lat[q] 
   {+} \tfrac{8}{3} \lat[r] {-} 8 \lat[s] )  \end{array} \eqnskip \\
+ \tfrac{1}{\mu_2} ( \upsilon_{18}^2 {+} \upsilon_{63}^2 ) \begin{array}[t]{l}
   ( \tfrac{3}{32} \lac[l] {+} \la[h] {+} \tfrac{1}{8} \la[i] 
   {+} \tfrac{1}{8} \la[j] {+} \tfrac{3}{8} \la[k] {+} \tfrac{3}{8} \la[l] 
   {-} \tfrac{1}{8} \la[m] {-} \tfrac{3}{8} \la[t] {+} \tfrac{3}{2} \la[u] 
   {+} \th \la[v] \\ 
   {+} 27 \lat[b] {+} \tfrac{1}{3} \lat[c] {+} 3 \lat[d] {+} 3 \lat[e] 
   {-} 3 \lat[j] {-} 9 \lat[k] {+} \lat[m] ) \end{array} \eqnskip \\
+ \tfrac{1}{\mu_2} (\tsum_{i=19}^{26} \upsilon_i^2 
   {+} \tsum_{i=64}^{71} \upsilon_i^2) \begin{array}[t]{l}
   (\la[h] {+} \tfrac{1}{8} \la[i] {+} \tfrac{1}{8} \la[j] 
   {+} \tfrac{9}{8} \la[k] {+} \tfrac{9}{8} \la[l] {-} \tfrac{1}{8} \la[m] 
   {-} \tfrac{3}{8} \la[n] {+} \tfrac{3}{4} \la[p] {-} \tfrac{3}{8} \la[r] 
   {-} \tfrac{9}{8} \la[t] {+} 2 \la[w] \\ 
   {+} 4 \lat[p] {+} 4 \lat[q] {+} 4 \lat[r] {+} 4 \lat[s] ) 
   \end{array} \eqnskip \\
+ \tfrac{1}{\mu_2} (\tsum_{i=27}^{29} \upsilon_i^2 
   {+} \tsum_{i=72}^{74} \upsilon_i^2 ) 
   \begin{array}[t]{l} (\la[h] {+} \tfrac{1}{8} \la[i] {+} \tfrac{1}{8} \la[j] 
   {+} \tfrac{9}{8} \la[k] {+} \tfrac{1}{8} \la[l] {-} \tfrac{1}{8} \la[m] 
   {-} \tfrac{3}{8} \la[n] {+} \tfrac{1}{4} \la[p] \\ 
   {+} \tfrac{1}{8} \la[r] {+} \tfrac{3}{8} \la[t] 
   {+} 2 \la[u] {+} \tfrac{2}{3} \la[v] {+} \tfrac{4}{3} \la[w]) \end{array} 
   \eqnskip \\
+ \tfrac{1}{\mu_2} (\tsum_{i=30}^{35} \upsilon_i^2 
   {+} \tsum_{i=75}^{80} \upsilon_i^2) \begin{array}[t]{l}
   (\la[h] {+} \tfrac{1}{8} \la[i] {+} \tfrac{1}{8} \la[j] 
   {+} \tfrac{49}{8} \la[k] {+} \tfrac{9}{8} \la[l] {-} \tfrac{1}{8} \la[m] 
   {-} \tfrac{7}{8} \la[n] {+} \tfrac{5}{4} \la[p] {-} \tfrac{3}{8} \la[r] 
   {-} \tfrac{21}{8} \la[t] \\
   {+} 2 \la[w] {+} 4 \lat[p] {+} 4 \lat[q] {+} 4 \lat[r] {+} 4 \lat[s] ) 
   \end{array} \eqnskip \\
+ \tfrac{1}{\mu_2} (\tsum_{i=36}^{38} \upsilon_i^2 
   {+} \tsum_{i=82}^{84} \upsilon_i^2 ) \begin{array}[t]{l}
   (\la[h] {+} \tfrac{1}{8} \la[i] {+} \tfrac{1}{8} \la[j] 
   {+} \tfrac{49}{8} \la[k] {+} \tfrac{1}{8} \la[l] {-} \tfrac{1}{8} \la[m] 
   {-} \tfrac{7}{8} \la[n] \\
   {+} \tfrac{3}{4} \la[p] {+} \tfrac{1}{8} \la[r] {+} \tfrac{7}{8} \la[t] 
   {+} 2 \la[u] {+} \tfrac{2}{3} \la[v] {+} \tfrac{4}{3} \la[w]) 
   \end{array} \eqnskip \\
+ \tfrac{1}{\mu_2} (\tsum_{i=39}^{41} \upsilon_i^2 
   {+} \tsum_{i=84}^{86} \upsilon_i^2) \begin{array}[t]{l}
   (\la[h] {+} \tfrac{1}{8} \la[i] {+} \tfrac{1}{8} \la[j] 
   {+} \tfrac{49}{8} \la[k] {+} \tfrac{1}{8} \la[l] {-} \tfrac{1}{8} \la[m] 
   {-} \tfrac{7}{8} \la[n] {+} \tfrac{3}{4} \la[p] {+} \tfrac{1}{8} \la[r] 
   {+} \tfrac{7}{8} \la[t] \\
   {+} 2 \la[w] {+} 36 \lat[p] {+} 4 \lat[q] {+} 4 \lat[r] {-} 12 \lat[s] ) 
   \end{array} \eqnskip \\
+ \tfrac{1}{\mu_2} (\tsum_{i=42}^{44} \upsilon_i^2 
   {+} \tsum_{i=87}^{89} \upsilon_i^2 )( \begin{array}[t]{l} 
   \tfrac{2}{3} \lac[i] {+} \tfrac{1}{12} \lac[l] {+} \la[h] 
   {+} \tfrac{1}{8} \la[i] {+} \tfrac{1}{8} \la[j] {+} \tfrac{1}{8} \la[k] 
   {+} \tfrac{9}{8} \la[l] {-} \tfrac{1}{8} \la[m] {+} \tfrac{1}{8} \la[n] \\
   {+} \tfrac{1}{4} \la[p] {-} \tfrac{3}{8} \la[r] {+} \tfrac{3}{8} \la[t] 
   {+} 36 \lat[b] {+} \tfrac{4}{9} \lat[c] {+} 4 \lat[d] {+} 4 \lat[e] 
   {-} 4 \lat[j] \\
   {-} 12 \lat[k] {+} \tfrac{4}{3} \lat[m] {+} \tfrac{8}{3} \lat[p] 
   {+} \tfrac{8}{3} \lat[q] {+} \tfrac{8}{3} \lat[r] {+} \tfrac{8}{3} \lat[s] ) 
   \end{array} \eqnskip \\
+ \sqrt{2} \tfrac{1}{\mu_2} ( \begin{array}[t]{l} 
   \upsilon_9 \upsilon_{12} {+} \upsilon_{10} \upsilon_{13} 
   {+} \upsilon_{11} \upsilon_{14} {+} \upsilon_{54} \upsilon_{57} 
   {+} \upsilon_{55} \upsilon_{58} {+} \upsilon_{56} \upsilon_{59})  \\ \!\!
   ( \la[u] {+} \tfrac{1}{3} \la[v] {-} \tfrac{4}{3} \la[w] {+} 2 \lat[b] 
   {-} \tfrac{10}{3} \lat[c] {+} 2 \lat[d] {+} 2 \lat[e] 
   {+} \tfrac{2}{3} \lat[j] {+} 2 \lat[k] {+} \tfrac{2}{3} \lat[m] 
   {-} 8 \lat[p] {-} \tfrac{8}{3} \lat[q] {-} \tfrac{8}{3} \lat[r] 
   {+} \tfrac{16}{3} \lat[s] ) \end{array} \eqnskip \\
+ \sqrt{2} \tfrac{1}{\mu_2} ( \upsilon_9 \upsilon_{57} 
   {+} \upsilon_{10} \upsilon_{58} {+} \upsilon_{11} \upsilon_{59} 
   {-} \upsilon_{12} \upsilon_{54} {-} \upsilon_{13} \upsilon_{55} 
   {-} \upsilon_{14} \upsilon_{56} ) 
   ( \tfrac{8}{3} \lat[n] {-} \tfrac{8}{3} \lat[t] ) \eqnskip \\
+ \tfrac{1}{\mu_2} ( \upsilon_{15} \upsilon_{42} \begin{array}[t]{l} 
   {+} \upsilon_{16} \upsilon_{43} {+} \upsilon_{17} \upsilon_{54} 
   {+} \upsilon_{60} \upsilon_{87} 
   {+} \upsilon_{61} \upsilon_{88} {+} \upsilon_{62} \upsilon_{89}) \\
   ( 24 \lat[b] {+} \tfrac{8}{3} \lat[c] {-} 8 \lat[d] {+} 8 \lat[e] 
   {-} \tfrac{40}{3} \lat[j] {+} 8 \lat[k] {+} \tfrac{8}{3} \lat[m] 
   {+} 16 \lat[p] {-} \tfrac{16}{3} \lat[q] {+} \tfrac{16}{3} \lat[r] 
   {+} \tfrac{16}{3} \lat[s] ) \end{array} \eqnskip \\
+ \tfrac{1}{\mu_2} ( \upsilon_{15} \upsilon_{87} 
   {+} \upsilon_{16} \upsilon_{88} {+} \upsilon_{17} \upsilon_{89} 
   {-} \upsilon_{60} \upsilon_{42} {-} \upsilon_{61} \upsilon_{43} 
   {-} \upsilon_{62} \upsilon_{44} ) 
   ( {-} 8 \lat[l] {-} \tfrac{8}{3} \lat[n] {+} 8 \lat[o] 
   {-} \tfrac{16}{3} \lat[t] {+} \tfrac{16}{3} \lat[u] ) \eqnskip \\
+ \tfrac{1}{\mu_2} ( \upsilon_{36} \upsilon_{42} 
   {+} \upsilon_{37} \upsilon_{43} {+} \upsilon_{38} \upsilon_{54} 
   {+} \upsilon_{81} \upsilon_{87} {+} \upsilon_{82} \upsilon_{88} 
   {+} \upsilon_{83} \upsilon_{89}) 
   (- \tfrac{2}{3} \lac[j] {+} \tfrac{1}{6} \lac[m]) \eqnskip \\
+ \tfrac{1}{\mu_2} ( \upsilon_{36} \upsilon_{87} 
   {+} \upsilon_{37} \upsilon_{88} 
   {+} \upsilon_{38} \upsilon_{89} {-} \upsilon_{81} \upsilon_{42} 
   {-} \upsilon_{82} \upsilon_{43} {-} \upsilon_{83} \upsilon_{44} ) 
   ( \tfrac{2}{3} \lac[k] {+} \tfrac{1}{6} \lac[n] ) 
   \big\} {+} I.T~.
  }
\end{small} \vs*{-2ex}

\end{appendix}

\end{document}